\documentclass[
reprint,
superscriptaddress,
nofootinbib,
amsmath,amssymb,
aps,
rmp,
floatfix,longbibliography
]{revtex4-2}
\usepackage{graphicx}
\usepackage{dcolumn}
\usepackage[hypertexnames, breaklinks=true, bookmarksnumbered=true, bookmarksopen=true, colorlinks=true, linktocpage=true, allcolors=magenta]{hyperref}
\usepackage{amsmath,amssymb,amsfonts,amsthm}
\usepackage{mathtools} 
\usepackage[T1]{fontenc}
\usepackage[latin9]{inputenc}
\usepackage{times}
\usepackage{txfonts}
\usepackage{bbm} 
\usepackage{bm} 
\usepackage{mathrsfs} 
\usepackage{xcolor}
\usepackage{natbib}
\usepackage{braket}

\DeclareMathOperator{\Tr}{Tr}
\allowdisplaybreaks

\newtheorem{theorem}{Theorem}

\newtheorem{conjecture}{Conjecture}

\begin{document}

\title{Catalysis of entanglement and other quantum resources}

\author{Chandan Datta}
\email{dattachandan10@gmail.com}
\affiliation{Centre for Quantum Optical Technologies, Centre of New Technologies,
University of Warsaw, Banacha 2c, 02-097 Warsaw, Poland}
\affiliation{Institute for Theoretical Physics III, Heinrich Heine University D\"{u}sseldorf, Universit\"{a}tsstra{\ss}e 1, D-40225 D\"{u}sseldorf, Germany}
\author{Tulja Varun Kondra}
\author{Marek Miller}
\author{Alexander Streltsov}
\email{a.streltsov@cent.uw.edu.pl}

\affiliation{Centre for Quantum Optical Technologies, Centre of New Technologies,
University of Warsaw, Banacha 2c, 02-097 Warsaw, Poland}

\begin{abstract}
In chemistry, a catalyst is a substance which enables a chemical reaction or increases its rate, while remaining unchanged in the process. Instead of chemical reactions, \emph{quantum catalysis} enhances our ability to convert quantum states into each other under physical constraints. The nature of the constraints depends on the problem under study and can arise, e.g., from energy preservation. This article reviews the most recent developments in quantum catalysis and gives a historical overview of this research direction. We focus on the catalysis of quantum entanglement and coherence, and also discuss this phenomenon in quantum thermodynamics and general quantum resource theories. We review applications of quantum catalysis and also discuss the recent efforts on universal catalysis, where the quantum state of the catalyst does not depend on the states to be transformed. Catalytic embezzling is also considered, a phenomenon that occurs if the catalyst's state can change in the transition.
\end{abstract}

\maketitle

\tableofcontents

\section{Introduction}
\label{sec:Introduction}

In 1835, Jacob Berzelius observed a common feature among several chemical reactions: they all required an additional substance, which remained unchanged in the process. He described this feature with the word ``catalysis'', which goes back to  Greek, and can be translated as ``down'' or ``loosen''~\cite{Berzelius1835,vanSanten2000catalysis}. A systematic study of catalytic processes in chemistry has been performed later by Wilhelm Ostwald, who gave a precise definition which is also used today: ``Catalysis is the acceleration of a slow chemical process by the presence of a foreign substance''~\cite{Ostwald1894}. First known examples of chemical catalysis used by humans are fermentation processes, which have been used for thousands of years~\cite{vanSanten2000catalysis}. Today, chemical catalysis is broadly used in the chemical industry and is essential for many industrial processes.
In addition, the importance of catalytic reactions in biochemistry cannot be overstated:
to the point that the ability of living organisms to ``catalyse chemical reactions efficiently and selectively'' via metabolic pathways could be called one of the ``fundamental conditions for life'' \cite{nelson2008lehninger}.

Quantum catalysis is conceptually similar to chemical catalysis but differs from it in several important details. A simple analogy between quantum and chemical catalysis can be established by replacing ``chemical reaction'' with ``quantum state transition''. With this, a quantum catalyst is a quantum system which enables otherwise impossible transitions between quantum states.  First investigations of quantum catalysis go back to the work of \citet{JonathanPhysRevLett.83.3566}, who studied transitions between entangled states shared by two remote parties (Alice and Bob). It is commonly assumed that the remote parties have the complete freedom to manipulate their quantum systems locally, including all kinds of transformations and measurements which are compatible with the laws of quantum mechanics. Additionally, Alice and Bob can exchange the outcomes of their measurements via a classical communication channel. This allows each party to adjust their local setups to the outcomes of the previous measurements obtained by the other party. This setup describes everything Alice and Bob can do to their shared quantum state if they cannot exchange other quantum particles~\cite{BennettPhysRevA.54.3824, HorodeckiRevModPhys.81.865}. 

In the first example of entanglement catalysis given by~\citet{JonathanPhysRevLett.83.3566}, Alice and Bob aim to convert their quantum state $\ket{\psi}^{AB}$ into another state $\ket{\phi}^{AB}$ under the limitations described above. A few months earlier, \citet{NielsenPhysRevLett.83.436} presented conditions for such state transformations, which allow to check whether $\ket{\psi}^{AB}$ can be converted into $\ket{\phi}^{AB}$ for any pair of states. \citet{JonathanPhysRevLett.83.3566} found that for some states an otherwise impossible transition can be enabled by using an additional pair of particles $A'$ and $B'$ held by Alice and Bob, respectively. 
More precisely, by choosing an appropriate entangled state $\ket{\eta}^{A'B'}$, they showed that $\ket{\psi}^{AB} \otimes \ket{\eta}^{A'B'}$ can be converted into $\ket{\phi}^{AB} \otimes \ket{\eta}^{A'B'}$, even though a direct conversion from $\ket{\psi}^{AB}$ into $\ket{\phi}^{AB}$ is not possible. As a result the state $\ket{\eta}^{A'B'}$ remains unchanged and the two-particle system $A'B'$ can be seen as a quantum catalyst.

In recent years, it became clear that entanglement is not the only quantum feature to exhibit catalytic behaviour, mostly thanks to the development of quantum resource theories~\cite{ChitambarRevModPhys.91.025001}, which goes back to the systematic study of quantum entanglement~\cite{HorodeckiRevModPhys.81.865} as a resource for the emerging quantum technology. Every quantum resource theory is based on a set of free states and free operations, corresponding to states and transformations which are easy to prepare and implement within well-defined physical constraints. Apart from quantum entanglement, the framework of quantum resource theories gave significant insight into the thermodynamical properties of quantum systems~\cite{BrandaoPhysRevLett.111.250404, Goold_2016,GOUR20151}, and the role of purity~\cite{PhysRevA.67.062104,GOUR20151} and quantum coherence~\cite{Baumgratz_2014,StreltsovRevModPhys.89.041003} for technological applications.  Catalytic effects have been found in most resource theories studied in recent literature, allowing to overcome certain constraints of the theory by adding a particle which remains unchanged in the procedure. 

The original formulation of quantum catalysis proposed by \citet{JonathanPhysRevLett.83.3566} did not allow for any correlations between the catalyst and the primary system as a result of a catalytic process. This constraint has been lifted in recent literature, allowing the procedure to build up and preserve correlations. This fruitful approach has significantly improved our understanding of quantum catalysis, with applications in the resource theories of quantum coherence~\cite{AbergPhysRevLett.113.150402}, quantum thermodynamics~\cite{Wilminge19060241,PhysRevX.8.041051,shiraishi2020quantum,ShiraishiPhysRevLett.128.089901}, purity~\cite{BoesPhysRevLett.122.210402,wilming2020entropy}, and entanglement~\cite{Kondra2102.11136,Datta2022entanglement}. In many cases, it has significantly simplified the analysis of catalytic state transformations within the corresponding resource theory~\cite{PhysRevX.8.041051,shiraishi2020quantum,ShiraishiPhysRevLett.128.089901,BoesPhysRevLett.122.210402,wilming2020entropy,Kondra2102.11136,Datta2022entanglement,datta_complete_monotone}.

In this article, we review recent developments in the catalysis of quantum entanglement and other quantum resource theories. The article is structured as follows: In Section~\ref{sec:QuantumResourceTheories} we recall the main features of quantum resource theories. Exact and approximate catalysis are discussed in detail in Sections~\ref{sec:ExactCatalysis} and~\ref{sec:ApproxCatalysis}, respectively. In Section~\ref{sec:Applications} we review applications of catalysis for quantum information processing. Section~\ref{sec:UniversalCatalysis} presents recent results on universal catalysis. Catalytic embezzling phenomena are discussed in Section~\ref{sec:CatalyticEmbezzling}, and perspectives and open problems are presented in Section~\ref{sec:Perspectives}.

\section{Quantum resource theories} \label{sec:QuantumResourceTheories}

The study of quantum resource theories~\cite{ChitambarRevModPhys.91.025001} goes back to the early research in quantum entanglement~\cite{HorodeckiRevModPhys.81.865}.
After the first observation of quantum entanglement in nature~\cite{AspectPhysRevLett.49.1804,AspectPhysRevLett.47.460},
it was suggested that entanglement might eventually find interesting technological applications.
Among others, \citet{BennettPhysRevLett.70.1895} established that maximally entangled singlet states could be regarded as a vehicle for quantum teleportation.
And even before that,
a cryptographic scheme with singlets was described in~\cite{EkertPhysRevLett.67.661}.  
Thus,
from the point of view of various technological applications,
an important problem emerged,
how to find an optimal method of transforming arbitrary quantum states into maximally entangled singlets:
\begin{equation}
    \label{eq:singlet}
    \ket{\phi_{2}^{+}} = \frac{1}{\sqrt{2}} \left( \ket{00} + \ket{11} \right).
\end{equation}
This question is highly relevant also in the context of mixed,
or noisy, quantum states,
which are the right model of a quantum system in any real experiment.
One of the goals of quantum resource theories is to develop optimal strategies for transforming arbitrary quantum states,
states perhaps not so useful for some applications into ones that are deemed more ``resourceful''.

Any quantum resource theory specifies a context in which certain states and transformations are considered \emph{free}.
Free states are always available at no cost --
an additional subsystem as a free state can be added or discarded at will.
For example, 
the resource theory of entanglement regards all separable states~\cite{WernerPhysRevA.40.4277} as free:
\begin{equation}
    \rho_\mathrm{sep} = \sum_ip_i \rho_i^A \otimes \sigma_i^B,
\end{equation}
where $\rho_i^A$ and $\sigma_i^B$ are local quantum states.
A composite quantum system can be brought to a separable state without the requirement of having shared quantum entanglement in advance,
using only local operations and classical communication (LOCC). For a general resource theory,
free operations are those transformations of quantum systems that can be implemented without the need of having access to resourceful, non-free states.
In the resource theory of entanglement, free operations are precisely the class of LOCC. 

It was the resource theory of quantum entanglement that was historically the first to undergo a systematic study~\cite{HorodeckiRevModPhys.81.865}.
Intuitively, 
entanglement is a valuable resource in systems that allow for clear spatial separation into several discrete subsystems.
Not all prominent applications of quantum technologies, however,
emphasise the role of such spatial separation.
For example,
an instance of quantum computation is typically assumed to be a local process. The role of entanglement in quantum computation has been the subject of numerous publications, and it has been shown that under certain assumptions, a quantum computer requires entanglement in order to exhibit an exponential speed-up compared to classical computation~\cite{Jozsa2003}.
As of today, it remains unclear, however,
whether it is quantum entanglement that is a crucial component of quantum computation, 
especially when noisy quantum states are considered.
An interesting example to the contrary is the one clean qubit model~\cite{KnillPhysRevLett.81.5672},
which performs exponentially faster than the best-known classical algorithm,
with vanishing entanglement~\cite{DattaPhysRevA.72.042316,Naseri2205.13610}. Moreover, given certain quantum computational tasks,
large amounts of entanglement can have a detrimental effect
on the performance of a quantum processor~\cite{GrossPhysRevLett.102.190501,Naseri2205.13610}.
One should also note that universal quantum computation is proven to be possible with arbitrary little entanglement~\cite{VanDenNestPhysRevLett.110.060504}.

The framework of the general quantum resource theory allows for investigating quantum systems beyond the constraints of particular resource manipulation scenarios.
Since not all quantum information processing tasks require entanglement,
a considerable amount of effort has been put in recent years into the research of other resource theories~\cite{ChitambarRevModPhys.91.025001}, among others the resource theories of purity~\cite{PhysRevA.67.062104,GOUR20151,Streltsov_2018},
quantum thermodynamics~\cite{BrandaoPhysRevLett.111.250404,Goold_2016,GOUR20151},
coherence~\cite{Baumgratz_2014,StreltsovRevModPhys.89.041003,Aberg2006,Wu2021},
asymmetry~\cite{GourPhysRevA.80.012307,Gour_2008,VaccaroPhysRevA.77.032114,Marvian2014,MarvianPhysRevA.90.062110},
and imaginarity~\cite{WuPhysRevLett.126.090401,WuPhysRevA.103.032401,Hickey_2018}.
As it has been shown that for every quantum resource theory there exists a channel discrimination task showing a quantitative advantage of utilising that particular resource~\cite{TakagiPhysRevLett.122.140402,TakagiPhysRevX.9.031053}, the newly developed theories may prove to be of considerable importance to future applications of processing quantum information.

\begin{figure*}
\includegraphics[width=0.65\paperwidth]{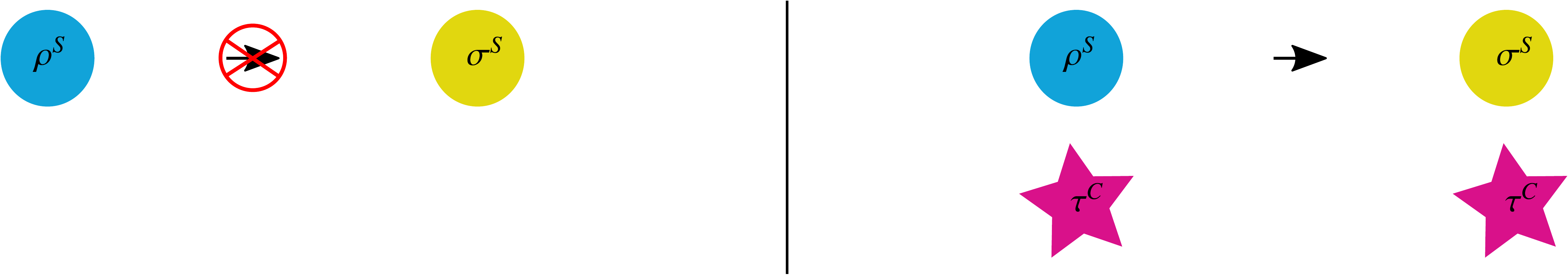}

\caption{Quantum catalysis: Consider two system states $\rho^S$ and $\sigma^S$ such that a conversion $\rho^S \rightarrow \sigma^S$ cannot be achieved with free operations (left part of the figure). In many resource theories, a conversion might still be enabled for some states by a catalyst in the state $\tau^C$, allowing for the transformation $\rho^S \otimes \tau^C \rightarrow \sigma^S \otimes \tau^C$ (right part of the figure).}
\label{fig:Catalysis}
\end{figure*}

\section{Exact catalysis}
\label{sec:ExactCatalysis}

Due to the rapid growth of quantum information science in recent years,
we can identify and quantify the quantum resources required for different quantum information protocols.
One of the important studies in this direction is to learn how quantum states can be transformed into each other under some physical constraints. More precisely, we are interested in knowing whether a state $\sigma$ can be achieved from a given resourceful state $\rho$ as a result of some physical operation allowed by the theory.  Sometimes the transformation is possible deterministically, i.e., there exists a free operation $\Lambda_f$ such that $\sigma = \Lambda_f[\rho]$. When the transformation from $\rho$ to $\sigma$ is not possible deterministically, we consider the possibility of probabilistic conversion. Nonetheless, there exist states which cannot be converted into each other even probabilistically \cite{Dur_PhysRevA.62.062314, Fang_2018, Sauerwein_PhysRevX.8.031020, Regula_2022,Wu2020}. In that case, there may exist an additional state that works as a catalyst enabling the transformation while remaining unchanged in the process, see also Fig.~\ref{fig:Catalysis}. 

In this section, we will discuss exact catalysis, where no error is allowed
in the transformation. Historically, exact catalysis was first introduced
for quantum entanglement \cite{JonathanPhysRevLett.83.3566}, and extended to other quantum resource theories more recently. In the following, we will review recent results on quantum catalysis for different resource theories. Before presenting the results, we introduce the concept of exact catalysis. An exact catalytic transformation between two states $\rho^S$ and $\sigma^S$ of a system $S$ is possible when there exists a catalyst $C$ in a state $\tau^C$ and a free operation $\Lambda_f$ acting on the system $S$ and the catalyst $C$ such that \cite{JonathanPhysRevLett.83.3566}
\begin{equation}
    \Lambda_f(\rho^S\otimes\tau^C)=\sigma^S\otimes\tau^C. 
\end{equation}
Exact catalytic transformation is very stringent, and over time relaxations of it have been considered in the literature, allowing the catalyst to build up correlations with the primary system \cite{AbergPhysRevLett.113.150402,LostaglioPhysRevLett.123.020403,Boes_Cat_rand,Wilde_classical,CatGauss_2021}. In particular, an exact correlated catalytic transformation between two states $\rho^S$ and $\sigma^S$ is possible if and only if there exists a catalyst state $\tau^C$, a correlated state $\sigma^{SC}$ and a free operation $\Lambda_f$ such that
\begin{equation}\label{eq:CorrCat}
    \Lambda_f(\rho^S\otimes\tau^C)=\sigma^{SC} \,\, \mbox{with}\,\, \Tr_C\sigma^{SC}=\sigma^S\,\, \mbox{and}\,\,\Tr_S\sigma^{SC}=\tau^C.
\end{equation}

Having defined exact catalysis for general quantum resource theories, we will focus on the catalysis of entanglement in the following.

\subsection{Entanglement}
\label{subsec:ExactCatalysis:Entanglement}

Historically, 
\citet{JonathanPhysRevLett.83.3566}
introduced the notion of exact catalysis of quantum states first in the context of the resource theory of entanglement. 
The idea is based on the following fundamental result,
which characterises fully the necessary and sufficient conditions to be met by all admissible transformations of pure bipartite entangled states. 
Consider two states $\ket{\psi}^{AB}=\sum_{i}\sqrt{\alpha_i}\ket{ii}$ and $\ket{\phi}^{AB}=\sum_{i}\sqrt{\beta_i}\ket{ii}$,
with their respective Schmidt coefficients $\{\alpha_i\}$ and $\{\beta_i\}$ sorted in non-increasing order.

\begin{theorem}[Nielsen's theorem, see \cite{NielsenPhysRevLett.83.436}]
\label{thm:EntMajo}
The transformation $\ket{\psi}^{AB} \rightarrow \ket{\phi}^{AB}$ is possible via LOCC if and only if the reduced states $\psi^A$ and $\phi^A$ fulfill the majorization relation $\psi^A \prec \phi^A$, i.e., their Schmidt coefficients satisfy 
\begin{equation}\label{Nielsen criteria}
    \sum_{i=0}^n\alpha_i\leq  \sum_{i=0}^n\beta_i \,\,\, \mathrm{for \, all} \,\,\, 0\leq n\leq d-1,
\end{equation}
with $d=\min \{d_A,d_B\}$, where $d_A$ and $d_B$ are the dimensions of the Hilbert space of Alice and Bob, respectively. 
\end{theorem}
\noindent The same result as in Theorem \ref{thm:EntMajo} has been derived independently by \citet{HardyPhysRevA.60.1912}.

It is not difficult to find such states which do not satisfy the above relations, and hence an LOCC transition between them is not possible. 
As an example, consider the following two states~\cite{JonathanPhysRevLett.83.3566}:
\begin{align}
\ket{\psi} & =\sqrt{0.4}\ket{00}+\sqrt{0.4}\ket{11}+\sqrt{0.1}\ket{22}+\sqrt{0.1}\ket{33}\label{example1},\\
\ket{\phi} & =\sqrt{0.5}\ket{00}+\sqrt{0.25}\ket{11}+\sqrt{0.25}\ket{22}.\label{example2}
\end{align}
It can be easily verified that the above states do not satisfy Nielsen's theorem, and as a result, a deterministic conversion is not possible. 
When a deterministic transformation is not possible, one may consider probabilistic transformation with maximum success probability given by \cite{PhysRevLett.83.1046}:
\begin{equation}\label{EntOptProb}
    P_{\max}(\ket{\psi}\rightarrow\ket{\phi})=\min_{n\in[0,d-1]}\frac{\sum_{i=n}^{d-1}\alpha_i}{\sum_{i=n}^{d-1}\beta_i}.
\end{equation}
For the states given in Eqs. (\ref{example1}) and (\ref{example2}), the optimal probability of transforming $\ket{\psi}$ into $\ket{\phi}$ is $0.8$.
Surprisingly, if the two parties have access to an additional entangled state $\ket{\eta}=\sqrt{0.6}\ket{00}+\sqrt{0.4}\ket{11}$,
then they can perform the transformation $\ket{\psi}\otimes\ket{\eta}\rightarrow \ket{\phi}\otimes\ket{\eta}$ with certainty~\cite{NielsenPhysRevLett.83.436,JonathanPhysRevLett.83.3566}.
After the transformation the state $\ket{\eta}$ remains unchanged and can be used again for the same purpose.
We can say that the state $\ket{\eta}$ acts as a catalyst and enables a transformation that otherwise would not be possible.
Note that in the context of quantum entanglement, 
what is nowadays predominantly called ``catalysis'', 
used to be described as ``entanglement-assisted LOCC operation'' (ELOCC)
\cite{JonathanPhysRevLett.83.3566, EisertPhysRevLett.85.437}.

A maximally entangled state of any dimension cannot catalyse a transformation between two arbitrary incomparable bipartite pure states \cite{JonathanPhysRevLett.83.3566}. A surprising result that already suggests a rich structure of catalytic transformations of quantum resources. Moreover, if for a pair of states $\ket{\psi}$ and $\ket{\phi}$,
there exist catalytic transformations $\ket{\phi}\rightarrow\ket{\psi}$ and
$\ket{\psi}\rightarrow\ket{\phi}$,
i.e., if two states are interconvertible under catalytic LOCC, 
then there is a \emph{local} unitary $U$ such that $\ket{\phi} = U \ket{\psi}$ \cite{JonathanPhysRevLett.83.3566}.
As a consequence, 
if a direct, non-catalytic transformation from $\ket{\psi}$ to $\ket{\phi}$ is not possible,
whereas a catalytic transformation exists in that case,
then the reverse transition $\ket{\phi}\rightarrow\ket{\psi}$ cannot be realised even as a catalytic one.
Hence, 
catalysis can have an advantage over the direct transformation of quantum states only in the case when the two states are incomparable: $\ket{\phi}\nleftrightarrow\ket{\psi}$,
i.e., there is neither a direct transformation $\ket{\psi}\rightarrow\ket{\phi}$ nor $\ket{\phi}\rightarrow\ket{\psi}$. However, note that the above result is based on assuming pure state catalysts, and how the result behaves with mixed catalyst states is not known.
The result suggests that, 
as far as transformations of two-qubit pure entangled states are concerned,
there is no advantage with a catalyst in a pure state,
since it is always true that either $\ket{\psi}\rightarrow\ket{\phi}$ or $\ket{\phi}\rightarrow\ket{\psi}$ in this case.

A catalytic LOCC transformation with catalysts in pure states has no advantage in the case of two-qutrit pure states as well. 
In order to see this, 
\citet{JonathanPhysRevLett.83.3566} prove that a bipartite state $\ket{\psi}$
of local dimension $d$ can be converted catalytically into another bipartite state $\ket{\phi}$,
only if the following inequalities hold 
\begin{equation}\label{d_catalyst_cond}
\alpha_0\leq\beta_0\,\, \mbox{and}\,\, \alpha_{d-1}\geq\beta_{d-1},
\end{equation} 
where the Schmidt coefficients $\{\alpha_i\}$ and $\{\beta_i\}$,
$i=0,1,\ldots,d-1$,
of respectively $\ket{\psi}$ and $\ket{\phi}$ are sorted in non-increasing order.
In the case of $d=3$, the above necessary condition for catalysis can be stated as
\begin{equation}\label{d_catalyst_cond_for_qutrits}
\alpha_0\leq\beta_0\,\, \mbox{and}\,\, \alpha_{0} + \alpha_{1} \leq \beta_{0} + \beta_{1},
\end{equation} 
since the numbers $\{\alpha_i\}$ and $\{\beta_i\}$ sum up to identity.
But this is exactly the condition \eqref{Nielsen criteria} of Nielsen's majorization theorem!
Hence, as in the two-qubit case, 
there is no advantage in catalysis for bipartite systems with $d=3$:
the existence of a catalytic conversion of entangled states implies that these states can be converted directly via LOCC. A genuine catalytic conversion of incomparable states occurs first in systems of dimension $d=4$, as we have already seen in Eqs. (\ref{example1}) and (\ref{example2}). 

When a catalytic transition from  $\ket{\psi}$ to $\ket{\phi}$ is not possible with certainty,
it can happen that by adding a catalyst state we will increase the probability of the conversion.
As an example,
consider two incomparable states \cite{JonathanPhysRevLett.83.3566}:
\begin{eqnarray}
   &&\ket{\psi}=\sqrt{0.6}\ket{00}+\sqrt{0.2}\ket{11}+\sqrt{0.2}\ket{22},\\
   &&\ket{\phi}=\sqrt{0.5}\ket{00}+\sqrt{0.4}\ket{11}+\sqrt{0.1}\ket{22}. 
\end{eqnarray}
The optimal probability to convert $\ket{\psi}$ into $\ket{\phi}$ is $P_{\max}(\ket{\psi}\rightarrow\ket{\phi})=0.8$ \cite{PhysRevLett.83.1046},
whereas with the help of another state $\ket{\eta}=\sqrt{0.65}\ket{00}+\sqrt{0.35}\ket{11}$,
we have $P_{\max}(\ket{\psi}\otimes\ket{\eta}\rightarrow\ket{\phi}\otimes\ket{\eta})=0.904$. Of course, there are situations when even adding a catalyst in an arbitrary pure state cannot increase the optimal probability of transformation.
This is the case, e.g. when the optimal probability of converting a bipartite state $\ket{\psi}$ into $\ket{\phi}$ is $P_{\max}(\ket{\psi}\rightarrow\ket{\phi})=\alpha_{d-1}/\beta_{d-1}$.
Indeed,
let $\{\gamma_j\}_{j=0}^{d'-1}$ be the ordered, non-zero Schmidt coefficients of $\ket{\eta}$.
Since the Schmidt coefficients,
not necessarily ordered,
of the state $\ket{\psi} \otimes \ket{\eta}$ are given by $\{ \alpha_i \gamma_j \}_{i=0,j=0}^{d-1,d'-1}$,
and similarly: $\{ \beta_i \gamma_j \}_{i=0,j=0}^{d-1,d'-1}$
for the state $\ket{\phi} \otimes \ket{\eta}$,
from Eq.~\eqref{EntOptProb} we obtain \cite{JonathanPhysRevLett.83.3566}
\begin{align}
P_{\max}(\ket{\psi}\otimes\ket{\eta}\rightarrow\ket{\phi}\otimes\ket{\eta}) & =\min\limits _{\substack{l=0,1,\ldots,d-1 \label{prob_cat} \\
k=0,1,\ldots,d'-1
}
}\frac{\sum\limits _{i\geq l,j\geq k}\alpha_{i}\gamma_{j}}{\sum\limits _{i\geq l,j\geq k}\beta_{i}\gamma_{j}}\\
 & \leq\frac{\alpha_{d-1}}{\beta_{d-1}}, \nonumber
\end{align}
where the inequality is obtained by setting $l=d-1, k=d'-1$,
instead of minimising over all possible pairs $l,k$.

Of course,
a pure catalyst state can also increase the transformation probability.
\citet{PhysRevA.69.062310,1397943} have studied the advantage of pure catalysts in probabilistic entangled state transformations. Here, the goal is to find a pure catalyst state $\ket{\eta}$ increasing the optimal conversion probability, i.e.,
\begin{equation}
P_{\max}(\ket{\psi}\otimes\ket{\eta}\rightarrow\ket{\phi}\otimes\ket{\eta})>P_{\max}(\ket{\psi}\rightarrow\ket{\phi}).
\end{equation}
The following necessary and sufficient condition has been derived for the existence of a catalyst $\ket{\eta}$ for such an enhancement \cite{PhysRevA.69.062310}:
\begin{equation}
P_{\max}(\ket{\psi}\rightarrow\ket{\phi})<\min\left\{\frac{\alpha_{d-1}}{\beta_{d-1}},1\right\},
\end{equation}
where $\alpha_{d-1}$ and $\beta_{d-1}$ are the smallest Schmidt coefficient of $\ket{\psi}$ and $\ket{\phi}$ respectively. Furthermore, they also discussed the scenario where the probability can be increased with the help of a two-qubit pure catalyst. In addition, an efficient algorithm has been proposed to find the most economical catalyst in such cases \cite{1397943}.

As mentioned earlier, if the local dimension of the given state is smaller than or equal to $3$, a catalyst does not help in accomplishing a forbidden transformation~\cite{JonathanPhysRevLett.83.3566}. Now one may wonder that for a given final state $\ket{\phi}$ of dimension at least 4 there exists an initial state $\ket{\psi}$ with $\psi^A\nprec\phi^A$, such that $\ket{\psi}$ can be converted catalytically into $\ket{\phi}$. It turns out that almost for all final states $\ket{\phi}$ there exists a catalytic conversion from $\ket{\psi}$ \cite{PhysRevA.64.042314}. To be precise, the following has been shown: For a given final state $\ket{\phi}$, there exists a pure state catalyst and an initial state $\ket{\psi}$ with $\psi^A\nprec\phi^A$, such that the transformation $\ket{\psi}\rightarrow\ket{\phi}$ is possible catalytically if and only if $\beta_0\neq\beta_l$ and $\beta_m\neq\beta_{d-1}$ for some $l,m$ with $0<l<m<d-1$ \cite{PhysRevA.64.042314}. Here, $\beta_i$ are Schmidt coefficients of the state $\ket{\phi}$. This result emphasises that for almost all final states of dimension 4 or more, the sets of initial states to achieve a final state $\ket{\phi}$ using a pure state catalyst and without a catalyst are not the same, i.e., $T(\phi)\neq S(\phi)$. Here $T(\phi)$ and $S(\phi)$ represent the sets of possible initial states $\ket{\psi}$ that can be converted into $\ket{\phi}$ with a catalyst in an arbitrary pure state and without a catalyst, respectively. Note that no restriction has been imposed on the dimension of the catalyst. For a restriction on the dimension of the catalyst, the set of initial states is represented by $T_k(\phi)$, where $k$ is the maximum dimension of the catalyst. Then it has been shown that there exists no $k$ such that $T_k(\phi)=T(\phi)$ when $T(\phi)\neq S(\phi)$ \cite{PhysRevA.64.042314}. Although the set $T_k(\phi)$ is closed, the set $T(\phi)$ is not closed in general \cite{PhysRevA.64.042314}. Therefore, the dimension of the catalyst is unbounded in general to determine which states can be catalytically converted into a given final state. Furthermore, as we have previously discussed maximally entangled state cannot be a catalyst. Naturally, one may ask which states are useful to be catalysts. As shown by~\citet{PhysRevA.64.042314}, it turns out that almost all bipartite pure states are potential catalysts. Precisely the following has been shown: For a bipartite state $\ket{\eta}=\sum_{i=0}^{d-1}\sqrt{\xi_i}\ket{ii}$ of local dimension $d$ with a nonuniform distribution of Schmidt coefficients there exist states $\ket{\psi}$ and $\ket{\phi}$ in dimension $4$ with $\psi^A\nprec\phi^A$, such that $\psi^A\otimes\eta^A\prec\phi^A\otimes\eta^A$ \cite{PhysRevA.64.042314}. 

\label{par:trumping_cond}
Since the discovery of entanglement catalysis by~\citet{JonathanPhysRevLett.83.3566}, an important problem was to characterize all pairs of states $\ket{\psi}^{AB}$ and $\ket{\phi}^{AB}$ allowing for a catalytic LOCC transformation. Results in this direction have been presented in \cite{Nechita2008,Aubrun2008,Sun2005,Turgut_2007,Klimesh0709.3680}, focusing on pure catalyst states. When a catalyst enables a forbidden transformation between two bipartite pure states, the respective Schmidt coefficients satisfy a trumping condition. More precisely, a state $\ket{\psi}$ is trumped by another state $\ket{\phi}$, i.e., $\psi^A \prec_T \phi^A$ if there exists a catalyst in the state $\ket{\eta}$ such that $\psi^A\otimes\eta^A\prec\phi^A\otimes\eta^A$ holds \cite{PhysRevA.64.042314}. The trumping condition is an extension of Nielsen's majorization relation presented earlier in Theorem \ref{thm:EntMajo}. 

A detailed characterization of the trumping condition is mathematically very complex \cite{PhysRevA.64.042314}. \citet{Sun2005} provided a necessary and sufficient condition for the existence of a pure two-qubit catalyst state for $4\otimes4$ incomparable pure states. Furthermore, they have given an efficient algorithm to decide whether a $k\otimes k$ pure catalyst state exists for a pair of $n\otimes n$ incomparable pure states. \citet{Aubrun2008,Nechita2008} provided some more partial results in this direction by determining the closure of the set $T(\phi)$ consisting of initial states $\ket{\psi}$ for a given final state $\ket{\phi}$. Precisely, they showed that the closure of $T(\phi)$ in $l_1$ norm can be completely characterized by the following inequalities: $||\psi^A||_p\leq||\phi^A||_p$ for all $p\geq1$, where $||\cdot||_p$ represents $l_p$ norm. Despite these efforts, the characterization was not complete and a full characterization has been provided by \citet{Turgut_2007} and \citet{Klimesh0709.3680}. To be precise, it has been shown that the trumping condition is equivalent to \cite{Turgut_2007,Klimesh0709.3680}
\begin{equation}
    f_\alpha(s(\phi))>f_\alpha(s(\psi)) \,\,\forall \,\, \alpha\in(-\infty,\infty),
\end{equation}
where $s(\psi)$ and $s(\phi)$ represent the Schmidt coefficients of the states $\ket{\psi}$ and $\ket{\phi}$, respectively. The function $f_\alpha(x)$ for a $d$-dimensional probability vector $x$ is expressed as
\begin{eqnarray}\label{eq:falpha}
 f_{\alpha}(x)=\begin{cases}
 \ln \sum_{i=1}^{d} x_i^\alpha & (\alpha >1);\\
\sum_{i=1}^{d} x_i \ln x_i &  (\alpha =1);\\
 -\ln \sum_{i=1}^{d} x_i^\alpha & (0<\alpha<1);\\
 -\sum_{i=1}^{d} \ln x_i &  (\alpha =0);\\
\ln \sum_{i=1}^{d} x_i^\alpha & (\alpha <0). 
 \end{cases}
\end{eqnarray}

If one allows an infinitesimally small error in the initial state, i.e., $||\psi^A -\psi_{\varepsilon}^A||_1\leq \varepsilon$ with the trace norm $||M||_{1}=\Tr\sqrt{M^{\dagger}M}$, then the trumping condition $\psi_\varepsilon^A\prec_T\phi^A$ is equivalent to the monotonicity of the R\'enyi entropies \cite{Brandao3275}. 
\begin{theorem}[See Proposition 4 of the Supplementary Information in \cite{Brandao3275}]
Consider two states $\ket{\psi}$ and $\ket{\phi}$ with Schmidt coefficients $s(\psi)$ and $s(\phi)$ respectively. Then for arbitrary $\varepsilon>0$, there exists a state $\psi_\varepsilon^A$ with $||\psi^A -\psi_{\varepsilon}^A||_1\leq \varepsilon$ and $\psi_\varepsilon^A\prec_T\phi^A$ if and only if 
\begin{eqnarray}\label{catalyst_renyi}
 H_{\alpha}(s(\psi))\geq H_{\alpha}(s(\phi)) 
\end{eqnarray}
is true for all $\alpha\in (-\infty,\infty)$.
\end{theorem}
 Here, $H_{\alpha}$ represent the R\'enyi entropies which can be expressed as
\begin{equation}
    H_{\alpha}(x)=\frac{\mathrm{sgn}(\alpha)}{1-\alpha}\ln\left(\sum_{i=1}^{d} x_i^\alpha\right) \,\, \mbox{for} \,\,\alpha \in \mathbb{R} \setminus \{0,1\}. 
\end{equation}
$H_{\alpha}(x)$ for $\alpha \in \{-\infty, 0, 1, \infty\}$ are defined by taking limits and expressed as 
\begin{eqnarray}
 &H_0(x) &= \ln \mbox{rank}(x),\,\, H_1(x)=-\sum_{i=1}^d x_i \ln x_i\\
 &H_{-\infty}(x) &= \ln x_{\min}, \,\, H_{\infty}(x)=-\ln x_{\max},
\end{eqnarray}
where ${x_{\max}}$ and $x_{\min}$ are the maximum and minimum elements of the vector $x$, respectively.

In real life, noise is inevitable, and as a result, we are forced to deal with mixed states rather than pure states. Mixed state transformations are much more complicated \cite{PhysRevA.61.040301} as majorization relations described in Theorem \ref{thm:EntMajo} are not applicable in general. Consequently, the conversion of mixed states with the help of a catalyst gets convoluted. The reason is that we need to first check whether a mixed state transformation $\rho\rightarrow\sigma$ is possible. In general, it is a formidable task. Nevertheless, the problem has been solved for some classes of mixed states. \citet{EisertPhysRevLett.85.437} explored transformations between a special class of rank two mixed states of the form
\begin{eqnarray}
 &\rho &= \nu \ket{\psi}\!\bra{\psi}+(1-\nu)\ket{\omega}\!\bra{\omega}\,\,\mbox{and}\label{mixed_example1}\\
 &\sigma &= \gamma\ket{\phi}\!\bra{\phi}+(1-\gamma)\ket{\omega}\!\bra{\omega},\label{mixed_example2}
\end{eqnarray}
where $\gamma=\nu\mbox{Tr}[\Omega]$, $\Omega=\Pi\ket{\psi}\!\bra{\psi}\Pi$, $\Pi=\openone-\ket{\omega}\!\bra{\omega}$ and $\ket{\omega}$ is a product state. Moreover, we take $|\bra{\phi}\omega\rangle|^2=0$. With this an LOCC transformation from $\rho$ to $\sigma$ implies that \cite{EisertPhysRevLett.85.437} 
\begin{equation}\label{mixed_transform}
    \frac{\mbox{Tr}_A\Omega}{\mbox{Tr}\Omega}\prec \phi^A. 
\end{equation}
Now let us choose
\begin{eqnarray}
 \ket{\psi}&=&\sqrt{0.38}\ket{00}+\sqrt{0.38}\ket{11}+\sqrt{0.095}\ket{22}\nonumber\label{pure_example1}\\
 &+&\sqrt{0.095}\ket{33} + \sqrt{0.05}\ket{44}, \\ 
 \ket{\phi}&=&\sqrt{0.5}\ket{00}+\sqrt{0.25}\ket{11}+\sqrt{0.25}\ket{22},\label{pure_example2}\\ \ket{\omega}&=&\ket{44},\label{pure_example3}
\end{eqnarray}
 and $\gamma=0.95\nu$, where $1>\nu>0$ \cite{EisertPhysRevLett.85.437}. With this one can easily verify that Eq.~(\ref{mixed_transform}) is not satisfied, and hence the transformation $\rho\rightarrow\sigma$ is not possible. 
 
One may ask whether the forbidden transformation can be accomplished with the help of a catalyst. The answer is yes and the protocol is as follows \cite{EisertPhysRevLett.85.437}. Alice performs a von Neumann measurement on her subsystem with Kraus operators $K_1=\sum_{i=0}^3\ket{i}\!\bra{i}$ and $K_2=\ket{4}\!\bra{4}$. When she gets an outcome associated with $K_2$, no further operations are needed as we have state $\ket{\omega}$ with probability $(1-\gamma)$. In the case of other outcomes, the final state is
\begin{equation}
    \ket{\varphi}=\sqrt{0.4}\ket{00}+\sqrt{0.4}\ket{11}+\sqrt{0.1}\ket{22}+\sqrt{0.1}\ket{33}.
\end{equation}
Note that, the reduced state of $\ket{\varphi}$ is not majorized by the reduced state of $\ket{\phi}$. Therefore, a direct transformation is not possible and a catalyst could help in the transformation $\ket{\varphi}\rightarrow\ket{\phi}$. In fact, by taking a catalyst in state $\ket{\zeta}=\sqrt{0.4}\ket{00}+\sqrt{0.6}\ket{11}$, we obtain 
\begin{equation}
    \varphi^A\otimes\zeta^A\prec\phi^A\otimes\zeta^A.
\end{equation}
Finally, discarding the classical information, we achieve the desired state $\sigma$.

\label{par:Li_MixedCatalysisTwoQubits}
In \cite{li2011catalysis},
the authors analyse the catalytic transformation of mixed states of two qubits.
Building on the results from \cite{Gour_PhysRevA.72.022323,li2011sufficient},
their work sheds some light on the question of whether a mixed catalyst state could provide an advantage for entanglement catalysis. For a given mixed state $\rho$, the authors define a family of entanglement monotones:
\begin{equation}
    \label{eq:Li_EntanglmentMonotonesMixed}
    \hat{E}_{k,l}(\rho) = \min \limits_{\{ p_j, \ket{\psi_j} \}} \, 
        \sum \limits_{j} p_j \, \hat{E}_{k,l}(\ket{\psi_j}), 
\end{equation}
with $l=2,3,\ldots,d$ and $k \in (0, 1]$,
where the minimum is taken over all ensembles $\{p_j, \ket{\psi_j} \}$ of probability distributions $p_j$ and bipartite pure states $\ket{\psi_j}$
such that $\rho = \sum_j p_j \ket{\psi_j}$,
and
\begin{equation}
    \label{eq:Li_EntanglmentMonotonesPure}
    \hat{E}_{k,l}(\ket{\psi}) = \min \left \{ 1, \sum \limits_{i=l}^{d} \frac{\alpha_i}{k} \right \},
\end{equation}
for any bipartite pure state  $\ket{\psi}=\sum_{i}^{d}\sqrt{\alpha_i}\ket{ii}$
with ordered Schmidt coefficients.
The ensemble that minimises Eq. \eqref{eq:Li_EntanglmentMonotonesMixed} is called \emph{optimal}.  

In another paper, \cite{li2011sufficient}, the same authors give a non-standard definition of an LOCC transformation between mixed states.
According to Definition 2.6 in \cite{li2011sufficient}, a mixed state $\rho$
can be converted into another mixed state $\sigma$ via LOCC if there is an LOCC operation between optimal ensembles of $\rho$ and $\sigma$.  
For a precise definition of a transformation between ensembles of pure states,
see \cite{li2011sufficient}, p.~97.  
For our discussion here,
it is enough to notice that if there is an LOCC map $\Lambda$, such that $\Lambda(\rho) = \sigma$ in the ordinary sense,
then $\rho$ can be converted to $\sigma$ via LOCC according to Definition 2.6 in \cite{li2011sufficient} -- but not necessarily the other way round.

The main result from \cite{li2011catalysis},
namely Proposition~2.2 therein,
can be now stated as follows:
\label{prop:Li_GeneralisationTrump}
Suppose $\rho$, $\sigma$, and $\omega$ are rank 2 mixed bipartite states of a two-qubit system.  
Then the state $\rho \otimes \omega$ can be converted into $\sigma \otimes \omega$ via LOCC according to Definition~2.6 in \cite{li2011sufficient}
if and only if
\begin{equation}
\label{eq:Li_MixedTrump}
    \hat{E}_{k,l}(\rho \otimes \omega) \geq \hat{E}_{k,l}(\sigma \otimes \omega),
\end{equation}
for all $k \in (0,1]$ and $l=2,3,4$.
In other words, 
as long as we consider rank 2 mixed states, 
a catalytic transformation between two-qubit systems is governed by the family of entanglement monotones defined in Eq.~\eqref{eq:Li_EntanglmentMonotonesMixed}.
The result is a generalisation of the trumping condition for pure states,
see the discussion above, p.~\pageref{par:trumping_cond}.

Another important result in \cite{li2011catalysis},
Proposition 2.5,
follows quickly from the  proof of the Proposition~2.2,
and gives us a partial answer to the question of whether mixed catalyst states could provide an advantage over pure states:
\label{prop:Li_NoPureCatalyst}
If $\rho$, $\sigma$ are rank 2 mixed bipartite states of a two-qubit system
that \emph{cannot} be converted into each other via LOCC,
but for which a catalyst's state $\omega$ exists as in Eq.~\eqref{eq:Li_MixedTrump},
then $\omega$ \emph{cannot} be pure,
even if we allowed for $\omega$ to have an arbitrary Schmidt rank.
Note that this statement assumes that all LOCC transformations between mixed states are according to Definition~2.6 in \cite{li2011sufficient},
but as we have already discussed above, 
that implies that there is no catalytic LOCC transformation in the ordinary sense as well. 

Lastly, the authors give an example of a mixed catalyst state 
that facilitates a transformation between rank 2 mixed two-qubit states as above,
see Sec.~3 in \cite{li2011catalysis}.
However, again,
since the authors' definition of a LOCC transformation between mixed states
describes a larger family of maps than standard LOCC on the convex set of mixed states,
the question of whether an ordinary map and a mixed catalyst state exist remains open. 
Given that we know that no pure two-qubit state can be a catalyst in this case, 
a positive answer to that question would be a strong argument for the advantages of mixed-state catalysis. 

Next, we will discuss the advantage of a catalyst for small transformations where we want to change the state slightly. In particular, the goal is to convert a state $\rho$ into another state $\sigma$ with $F(\rho,\sigma) > 1-\varepsilon$, where $F(\rho,\sigma)=\left(\mbox{Tr}[\sqrt{\sigma}\rho\sqrt{\sigma}]^{1/2}\right)^2$ is the fidelity \cite{UHLMANN1976273}. In such a scenario, there exist states $\rho$ and $\sigma$ such that the transformation $\rho\rightarrow\sigma$ is impossible without a catalyst and can be accomplished with a suitable catalyst. Indeed, such states can be designed using Eqs.~(\ref{mixed_example1}) and~(\ref{mixed_example2})~\cite{EisertPhysRevLett.85.437}. On the contrary, a surprising result exists when we want to transform a pure state slightly. For all pure states $\ket{\psi}$ there exists an $\varepsilon>0$, such that the following statement holds for all states $\ket{\phi}$ with $|\braket{\psi|\phi}|^2>1-\varepsilon$: the catalytic transformation $\ket{\psi}\rightarrow\ket{\phi}$ is impossible with a pure catalyst if it is impossible without the catalyst \cite{EisertPhysRevLett.85.437}.

The structure of a catalyst for a successful transformation between two incomparable pure states was studied by~\citet{ZHOU200070}. To be precise, the authors showed that successful transformation of $\ket{\psi}=\sum_{i=0}^{d-1}\sqrt{\alpha_i}\ket{ii}$ into $\ket{\phi}=\sum_{i=0}^{d-1}\sqrt{\beta_i}\ket{ii}$ can be accomplished with the help of a pure catalyst state having $m$ non-zero Schmidt coefficients, i.e., $\ket{\eta}=\sum_{i=0}^{m-1}\sqrt{\gamma_i}\ket{ii}$, only if \cite{ZHOU200070}
\begin{equation}
    m \gamma_{m-1}\leq P_{\max}(\ket{\psi}\rightarrow\ket{\phi}),
\end{equation}
where $P_{\max}(\ket{\psi}\rightarrow\ket{\phi})$ is described in Eq. (\ref{EntOptProb}), represents the optimal probability of transforming $\ket{\psi}$ into $\ket{\phi}$ \cite{PhysRevLett.83.1046}. Furthermore, they extended the above result to stochastic scenarios. As a matter of fact, they showed that the probability of the transformation $P_{\max}(\ket{\psi}\rightarrow\ket{\phi})$ can be increased with the help of a pure state catalyst $\ket{\eta}$, only if \cite{ZHOU200070}
\begin{equation}
   m \gamma_{m-1}\leq P_{\max}(\ket{\psi}\rightarrow\ket{\phi})/P', 
\end{equation}
where $P'\geq P_{\max}(\ket{\psi}\rightarrow\ket{\phi})$ represents the enhancement in the probability for catalytic transformation. Their results provide some insight into the structure of the catalyst. 

In \cite{PhysRevA.65.022307}, the concept of mutual catalysis has been introduced. For this, suppose there are two pairs of states $(\ket{\psi}, \ket{\phi})$ and $(\ket{\eta}, \ket{\zeta})$. The transformations $\ket{\psi}\rightarrow\ket{\phi}$ and $\ket{\eta}\rightarrow\ket{\zeta}$ are not possible deterministically. However, sometimes it is possible that joint LOCC can transform $\ket{\psi}\otimes\ket{\eta}$ to $\ket{\phi}\otimes\ket{\zeta}$. Therefore, $\ket{\psi}$ and $\ket{\eta}$ help each other to achieve the desired target states. As an example, consider the following states \cite{PhysRevA.65.022307}
\begin{align}
\ket{\psi} & =\sqrt{0.4}\ket{00}+\sqrt{0.36}\ket{11}+\sqrt{0.14}\ket{22}\\
 & \quad+\sqrt{0.1}\ket{33},\nonumber \\
\ket{\phi} & =\sqrt{0.5}\ket{00}+\sqrt{0.25}\ket{11}+\sqrt{0.25}\ket{22},\\
\ket{\eta} & =\sqrt{0.6}\ket{00}+\sqrt{0.4}\ket{11},\\
\ket{\zeta} & =\sqrt{0.55}\ket{00}+\sqrt{0.45}\ket{11}.
\end{align}
It is easy to check that $\ket{\psi} \nleftrightarrow \ket{\phi}$ and $\ket{\eta}\nrightarrow\ket{\zeta}$ \cite{NielsenPhysRevLett.83.436}. However, they can be transformed jointly into the desired target states. Necessary conditions for such transformations have been derived in \cite{PhysRevA.65.022307}. Furthermore, the authors have discussed the scenario where a conversion from $\ket{\psi}$ to $\ket{\phi}$ is not possible even with the help of a catalyst but can be realized with mutual catalysis \cite{PhysRevA.65.022307}.

According to Theorem \ref{thm:EntMajo}, there exist states $\ket{\psi}$ and $\ket{\phi}$, such that $\ket{\psi}\rightarrow\ket{\phi}$ cannot be realized with certainty. However, in some cases, if we possess multiple copies of $\ket{\psi}$, they can be transformed to the same number of $\ket{\phi}$ \cite{PhysRevA.65.052315}. Along the same line, a concept of multicopy catalysis has been introduced by~\citet{PhysRevA.71.062306}. A state $\ket{\eta}$ is called a multiple-copy catalyst when it cannot catalyze the transformation $\ket{\psi}\rightarrow\ket{\phi}$, however the same can be accomplished with the help of $\ket{\eta}^{\otimes k}$ for some $k>1$. As an example~\citet{PhysRevA.71.062306} consider the following case. Consider the initial state $ \ket{\psi}$ as given in Eq. (\ref{example1}) and the target state is
\begin{equation}
    \ket{\phi}=\sqrt{0.5}\ket{00}+\sqrt{0.25}\ket{11}+\sqrt{0.22}\ket{22}+\sqrt{0.03}\ket{33}.
\end{equation}
The transformation $\ket{\psi}\rightarrow\ket{\phi}$ cannot be realized with a pure two-qubit catalyst state \cite{Sun2005}. However, note that the above transformation can be realized with a pure two-qutrit catalyst state. Now consider a situation where Alice and Bob only have access to some copies of a two-qubit pure state. One may wonder whether the above transformation can be realized with multiple copies of a two-qubit catalyst. Indeed there exists a two-qubit state $\ket{\eta}=\sqrt{0.6}\ket{00}+\sqrt{0.4}\ket{11}$, such that $\psi^A\otimes(\eta^A)^{\otimes 5}\prec\phi\otimes(\eta^A)^{\otimes 5}$ \cite{PhysRevA.71.062306}. Furthermore, they have derived some necessary conditions for an arbitrary pure state to be a catalyst for a given transformation $\ket{\psi}\rightarrow\ket{\phi}$. In addition, an advantage in probabilistic transformation with multiple copy catalysis has also been shown in \cite{PhysRevA.71.062306}. Instead of a single copy of the initial state, if we have multiple copies of it and want to transform it into the same number of copies of the final state with the help of a multicopy catalyst, then a study has been done on the tradeoff between the number of copies of the source state and that of the catalyst state. To be precise, they have shown that the number of copies required for a catalyst state decreases with the number of copies of the source state \cite{PhysRevA.71.062306}. In \cite{DuanPhysRevA.71.042319,Duanarxiv2004} it has been shown that any multiple copy entangled state transformation \cite{PhysRevA.65.052315} can be realized in single copy level with the help of a suitable pure catalyst.
It is natural to wonder whether the opposite is also true. But the answer is negative. As shown in \cite{PhysRevA.74.042312}, the catalytic assisted transformation is strictly more powerful than multiple copy transformations. Now, it is natural to ask whether they are equivalent in the asymptotic scenario. Indeed it has been shown that catalytically assisted transformation is asymptotically equivalent to the multiple-copy transformation when the dimension of the catalyst and number of copies tend to infinity \cite{PhysRevA.72.024306}. More precisely, these two kinds of transformations are equivalent in the sense that they can simulate each other's ability to transform a given initial state to a given target state with the same optimal success probability. Optimal probability for multiple-copy state transformation from $\ket{\psi}$ to $\ket{\phi}$ can be defined as \cite{PhysRevA.72.024306}
\begin{equation}
    P_M(\ket{\psi}\rightarrow\ket{\phi})=\sup_{m}[P_{\max}(\ket{\psi}^{\otimes m}\rightarrow \ket{\phi}^{\otimes m})]^{1/m}.
\end{equation}
On the other hand, the optimal conversion probability for catalytically assisted transformation is given by \cite{PhysRevA.72.024306}
\begin{equation}
    P_C(\ket{\psi}\rightarrow\ket{\phi})=\sup_{\ket{\eta}} P_{\max}(\ket{\psi}\otimes\ket{\eta}\rightarrow \ket{\phi}\otimes\ket{\eta}).
\end{equation}
In \cite{PhysRevA.72.024306}, it has been shown that $P_M(\ket{\psi}\rightarrow\ket{\phi})=P_C(\ket{\psi}\rightarrow\ket{\phi})$.
Furthermore, \citet{DuanPhysRevA.71.042319} have shown that multiple copy transformation, with the aid of a pure catalyst, is equivalent to single copy catalytic transformation with an arbitrary pure catalyst state. Let $V(\phi)$ be the set of all initial states $\{\ket{\psi}\}$ such that $\ket{\psi}\otimes\ket{\eta}\rightarrow\ket{\phi}\otimes\ket{
\eta}$ for some $\ket{\eta}$, and $V_M(\phi)$ is the set of all initial states $\{\ket{\psi}\}$ such that $\ket{\psi}^{\otimes m}\otimes\ket{\eta}\rightarrow\ket{\phi}^{\otimes m}\otimes\ket{
\eta}$ for some $m\geq 1$ and $\ket{\eta}$. Obviously $V(\phi)\subseteq V_M(\phi)$. In fact, it has been shown that they are equal $V(\phi)= V_M(\phi)$ \cite{DuanPhysRevA.71.042319}. Therefore, the power of catalytic assisted transformation (assuming pure catalyst states) cannot be elevated by increasing the number of copies of the original state. 

While the previous discussion showed when a catalytic transformation is possible, it does not provide any information about the catalyst. To be precise, it does not give information about the entanglement content or the dimension of the catalyst required for a transformation. \citet{SandersPhysRevA.79.054302} provided a lower bound on the dimension of a possible catalyst state $\ket{\eta}$ for a transformation $\ket{\psi}\otimes\ket{\eta}\rightarrow\ket{\phi}\otimes\ket{\eta}$. Suppose $\ket{\psi}$, $\ket{\phi}$ are two incomparable pure states with $d$ non-zero Schmidt coefficients and the transformation $\ket{\psi}\otimes\ket{\eta}\rightarrow\ket{\phi}\otimes\ket{\eta}$ is possible with the help of a pure catalyst state $\ket{\eta}$ that has $r$ non-zero Schmidt coefficients. Then it has been shown that the minimum number of non-zero Schmidt coefficients for the catalyst should be \cite{SandersPhysRevA.79.054302}
\begin{equation}\label{eq:SandersCataLower}
    r\geq 1+\frac{d-1}{d}\frac{\log_2 C_{d-1}(\phi)-\log_2 C_{d-1}(\psi)}{\log_2 C_{d}(\psi)-\log_2 C_{d}(\phi)},
\end{equation}
where $C_k$ represents generalized $k$-th concurrence and can be expressed as
\begin{equation}
    C_k(\psi)=\left(\frac{f_k[s(\psi)]}{f_k[s(\psi^+)]}\right)^{1/k},
\end{equation}
where $\psi^+$ is a maximally entangled state of dimension $d$ and $f_k(x)$ represents $k$th elementary symmetric function of $d$ variables: $f_k(x)=\sum_{i_1<\cdots<i_k}x_{i_1}\cdots x_{i_k}$ \cite{Gour_PhysRevA.71.012318}.

So far, we have not discussed the amount of entanglement required for a catalyst to enable a transformation between a pair of incomparable states. However, the answer is not fully known and only some partial results are available in the literature. \citet{PhysRevA.71.062306} derived some necessary conditions for a bipartite pure entangled state to be a catalyst. In fact, the following has been proved: Suppose $\ket{\psi}$ and $\ket{\phi}$ are two incomparable states of Schmidt rank $d$ with Schmidt coefficients $\{p\}$ and $\{q\}$ respectively, and a catalyst state $\ket{\eta}$ that has $k$ number of non-zero Schmidt coefficients represented by $\{t\}$ enables the transformation from $\ket{\psi}$ to $\ket{\phi}$. Then for any $l\in L_{\psi,\phi}$ and $0\leq i \leq k-2$, the Schmidt coefficients of the catalyst state satisfy the following three conditions:
\begin{subequations}\label{eq:DuanCondCata}
\begin{eqnarray}
 &&\frac{t_0}{t_{k-1}}>\frac{q_l}{q_{l+1}},\\
 &&\frac{t_0}{t_i}>\frac{q_l}{q_{l+1}}\,\,\,\, \mbox{or}\,\,\,\, \frac{t_i}{t_{i+1}}<\frac{q_0}{q_{l}},\\
 &&\frac{t_{i+1}}{t_{d-1}}>\frac{q_l}{q_{l+1}}\,\,\,\, \mbox{or}\,\,\,\,
 \frac{t_i}{t_{i+1}}<\frac{q_{l+1}}{q_{d-1}},
\end{eqnarray}
\end{subequations}
where $L_{\psi,\phi}$ represents the set of all $l$ such that  
\begin{equation}
   L_{\psi,\phi}=\left\{l: 0\leq l \leq d-1\,\, \mbox{and}\,\, \sum_{j=0}^l p_j> \sum_{j=0}^l q_j\right\}.  
\end{equation}
These restrictions on the Schmidt coefficients represent the deviation of state $\ket{\eta}$ from the product and maximally entangled states. 

Very recently, \citet{GraboweckyPhysRevA.99.052348} presented some stronger conditions than Eqs.~\eqref{eq:DuanCondCata}.
They showed that for any two incomparable states $\ket{\psi}$, $\ket{\phi}$ with $d$ non-zero Schmidt coefficients and a catalyst state $\ket{\eta}$ having $k$ non-zero Schmidt coefficients, if $\psi^A\otimes\eta^A\prec\phi^A\otimes\eta^A$,
then \cite{GraboweckyPhysRevA.99.052348}
\begin{subequations}\label{eq:GourCondCata}
\begin{align}
\max_{i\in\{0,\ldots,k-2\}}\frac{t_{i}}{t_{i+1}} & <\min\left\{ \frac{q_{0}}{q_{m}},\frac{q_{n+1}}{q_{d-1}}\right\} \equiv a,\label{condition1}\\
\frac{t_{0}}{t_{k-1}} & >\max_{l\in L_{\psi,\phi}}\frac{q_{l}}{q_{l+1}}\equiv b,\label{condition2}
\end{align}
\end{subequations}
where $m=\min L_{\psi,\phi}$ and $n=\max L_{\psi,\phi}$. These two conditions show how much a catalyst state can differ from both a product state and a maximally entangled state. Previously in Eq.~\eqref{eq:SandersCataLower}, we have seen a lower bound on the dimension of a catalyst in terms of the generalised $k$-th concurrence~\cite{SandersPhysRevA.79.054302}. However, calculating $k$-th concurrence for a higher-dimensional state is not easy in general. In that context, \citet{GraboweckyPhysRevA.99.052348} provided the following lower bound on the dimension of a catalyst, which can be easily calculated: 
\begin{equation}\label{eq:GourDimCata}
    k>\frac{\ln b}{\ln a}+1.
\end{equation}

The constraints on the Schmidt coefficients of the catalyst state presented in Eqs. (\ref{eq:DuanCondCata}) and (\ref{eq:GourCondCata}) depend only on the Schmidt coefficients of the target state. Recently, \citet{Guo_2021} studied a similar scenario and showed that for $\psi^A\otimes\eta^A\prec\phi^A\otimes\eta^A$ the following result holds:
\begin{equation}
    \frac{t_0}{t_{k-1}}>\max_{l\in L_{\psi,\phi}} \left(\min\left\{\frac{p_l}{p_{l+1}},\frac{q_l}{q_{l+1}}\right\}\right)\equiv c,
\end{equation}
where $k$ is the number of non-zero Schmidt coefficients of state $\ket{\eta}$ and also represents the dimension of the catalyst. Furthermore, they put a bound on the minimum dimension of the catalyst state that can be expressed as \cite{Guo_2021}
\begin{equation}
    k>\frac{\ln c}{\ln (a\sqrt{b})}+1.
\end{equation}
Unlike Eq. (\ref{eq:GourDimCata}), the above bound depends on both $\{p\}$ and $\{q\}$. One can compare and check which bound provides a good lower bound.

\citet{Duarte_2016} explored an interesting question about self-catalysis. Given two incomparable states $\ket{\psi}$ and $\ket{\phi}$, they studied whether $\ket{\psi}$ can catalyze the transformation $\ket{\psi}\otimes\ket{\psi}\rightarrow\ket{\phi}\otimes\ket{\psi}$. 
As an example, consider the following incomparable states \cite{Duarte_2016}:
\begin{align}
\ket{\psi} & =\sqrt{0.9}\ket{00}+\sqrt{0.081}\ket{11}+\sqrt{0.01}\ket{22}\nonumber \\
 & +\sqrt{0.009}\ket{33},\\
\ket{\phi} & =\sqrt{0.95}\ket{00}+\sqrt{0.03}\ket{11}+\sqrt{0.02}\ket{22}. \label{multicopy_self_catalysis_example}
\end{align}
The transformation of $\ket{\psi}$ into $\ket{\phi}$ can be accomplished with the help of $\ket{\psi}$ as catalyst, i.e., $\ket{\psi}\otimes\ket{\psi}\rightarrow\ket{\phi}\otimes\ket{\psi}$. 
One can modify it further by considering multi-copy self catalysis, where $\ket{\psi}\otimes\ket{\psi}\nrightarrow\ket{\phi}\otimes\ket{\psi}$ is not possible but $\ket{\psi}\otimes\ket{\psi}^{\otimes N}\rightarrow\ket{\phi}\otimes\ket{\psi}^{\otimes N}$ is possible for some $N>1$ \cite{Duarte_2016}.
For instance, consider the initial state to be
\begin{equation}
   \ket{\psi_1}=\sqrt{0.918}\ket{00}+\sqrt{0.07}\ket{11}+\sqrt{0.006}\ket{22}+\sqrt{0.006}\ket{33} 
\end{equation}
and the target state to be $\ket{\phi}$ in Eq. (\ref{multicopy_self_catalysis_example}). Then the transformation of $\ket{\psi_1}$ into $\ket{\phi}$ can be achieved using $\ket{\psi_1}^{\otimes 4}$ as a catalyst. 
\citet{Duarte_2016} tried to answer the question of how typical the self-catalysis phenomenon is using numerical methods. 
Although they found that self-catalysis occurs in systems of any dimension, the phenomenon is relatively rare \cite{Duarte_2016}. 
They also explored the self-catalysis of stochastic transformations and concluded that self-catalysis in this case occurs more often than in the deterministic one \cite{Duarte_2016}. In a recent paper, \citet{Duarte_machine} discussed in detail whether neural network models can detect catalysis or self-catalysis for bipartite pure entangled state conversions.

The practical implementation of majorization relations was first studied by~\citet{Gagatsos2013}. In fact, they showed that beam splitter can be used to achieve several classes of majorization relations. Furthermore, they constructed incomparable states and showed that these states can be catalytically converted. Such catalyst states can be easily implemented by using single-photon states or two-mode squeezed vacuum states \cite{Gagatsos2013}. 

Recently, \citet{Santra2021} studied the catalytic entanglement concentration of two-qubit pure states. They derived a formula for the maximum probability of converting a finite number of copies of a general two-qubit pure state into a singlet with the help of a two-qubit pure catalyst state. Furthermore, they have shown that any pure two-qubit entangled state can be a potential catalyst. Based on the approach given in \cite{Santra2021}, the authors further discussed the advantage of catalysis in the long-range entanglement distribution over a quantum network in \cite{PhysRevA.103.012407}. 

In a recent numerical study, the entanglement properties of pure catalyst states have been explored \cite{gupta2022statistics}. The average amount of entanglement required to catalyze a forbidden transformation decreases with the increment in entanglement difference between the pure initial and target states. Furthermore, for multiple copy state transformations, the amount of entanglement required to catalyze the transformation is lower compared to the single copy catalytic transformation.

An intriguing role of entangled catalysts in the context of implementing some nonlocal operations has been discussed in \cite{PhysRevLett.88.167903}. In fact, \citet{PhysRevLett.88.167903} have shown that certain nonlocal tasks cannot be implemented without using additional entanglement. Interestingly, this additional entanglement does not need to be consumed and can be recovered exactly, thus serving as a catalyst in the process. More precisely, there exists a certain two-qubit unitary gate $\tilde{U}_{AB}$ that cannot be implemented from the available two-qubit unitary gate $U_{AB}$ by means of LOCC only \cite{PhysRevLett.88.167903}. However, the same can be implemented by means of local unitary operations and using a maximally entangled state as a catalyst \cite{PhysRevLett.88.167903}.

\citet{Kay_2006} explored the cloning of non-maximally entangled states with LOCC and a copy of a maximally entangled state. The authors completely characterised the set of clonable states. Furthermore, it has been shown that catalysts do not provide any advantage in the cloning process \cite{Kay_2006}.

In general, multipartite entangled state transformation is more complex in nature
\cite{walter2016multipartite} 
and that makes catalytic transformation even more complex. Unlike catalytic transformation in bipartite scenarios, very few results are known for catalytic state transformations in multipartite scenarios. One of the first results in this direction was provided by \citet{Chen_2010}. More precisely, they studied the advantage of catalysis in stochastic LOCC (SLOCC) transformation of multipartite pure states. In fact, it has been shown that there exist state transformations that cannot be accomplished by SLOCC but can be realised by both catalytic SLOCC and multicopy SLOCC. Catalytic SLOCC represents the probabilistic transformation of $\ket{\psi}$ to $\ket{\phi}$ with the help of a catalyst state $\ket{\eta}$, such that the transformation $\ket{\psi}\otimes\ket{\eta}\rightarrow\ket{\phi}\otimes\ket{\eta}$ has non-vanishing probability. On the other hand, a transformation is possible via multicopy SLOCC if there exists a $k$, such that $\ket{\psi}^{\otimes k}$ can be converted into $\ket{\phi}^{\otimes k}$ with non-vanishing probability. As specific examples, consider the following transformations~\cite{Chen_2010}: \\
i) $\ket{\psi}_{ABC}\rightarrow \ket{\phi^+_2}_{AB}\otimes\ket{\phi^+_2}_{BC}\otimes\ket{\phi^+_2}_{CA}$, where $\ket{\psi}_{ABC}$ is any generalised GHZ-type state of tensor rank 6. \\
ii) $\ket{GHZ_{n}^{n-1}}\rightarrow\ket{W_n}$, where $\ket{GHZ_n^{d}}=\sum_{i=0}^{d-1}\ket{i}^{\otimes n}/\sqrt{d}$ and $\ket{W_n}$ is $n$-partite generalisation of $\ket{W_3}=(\ket{001}+\ket{010}+\ket{100})/\sqrt{3}$.  \\
Both of the above state transformations are not possible even stochastically but can be accomplished via both catalytic SLOCC and multicopy SLOCC. Furthermore, it has been shown that if $\ket{\psi}$ can be converted to $\ket{\phi}$ via multicopy SLOCC then the same can be realised with catalytic SLOCC  \cite{Chen_2010}.

Recently \citet{Neven_2021} found a majorization condition for a class of GHZ-like states, which completely characterises LOCC transformations. For this, suppose $g$ and $h$ are two invertible, complex and diagonal matrices satisfy $\Tr(g^\dagger g)=\Tr(h^\dagger h)$. Then the transformation 
\begin{equation}\label{ghz}
    \openone \otimes \cdots \otimes \openone\otimes g \ket{\mathrm{GHZ}_n^d}\rightarrow \openone \otimes \cdots \otimes \openone\otimes h \ket{\mathrm{GHZ}_n^d}
\end{equation}
 is possible by LOCC if and only if 
\begin{equation}\label{multipartite_majo}
  \left(|g_1|^2,\ldots,|g_d|^2\right)^T  \prec \left(|h_1|^2,\ldots,|h_d|^2\right)^T
\end{equation}
holds. Here, $\left(g_1,\ldots,g_d\right)$ and $\left(h_1,\ldots,h_d\right)$ are diagonal elements of the matrices $g$ and $h$, respectively. Note that $g$ and $h$ correspond to the GHZ-like states of the form $\sum_{i=0}^{d-1}\sqrt{a_i}\ket{i}^{\otimes n}$. As a direct consequence of this we see that $\ket{\psi}$ and  $\ket{\phi}$ characterised by 
\begin{align}
g & =\mathrm{diag}(\sqrt{0.45},\sqrt{0.35},\sqrt{0.12},\sqrt{0.08}),\\
h & =\mathrm{diag}(\sqrt{0.56},\sqrt{0.21},\sqrt{0.17},\sqrt{0.06}),
\end{align}
respectively, are incomparable as they fail to satisfy Eq. (\ref{multipartite_majo}). However, the transformation can be accomplished with another GHZ-like state $\ket{\varphi}$ characterised by
\begin{equation}
f=\mathrm{diag}(\sqrt{0.63},\sqrt{0.27},\sqrt{0.07},\sqrt{0.03}),
\end{equation}
that acts as a catalyst in the process. In fact, the transformation $\ket{\psi}\otimes\ket{\varphi}\rightarrow\ket{\phi}\otimes\ket{\varphi}$ is equivalent to the following transformation:
\begin{equation}
    \openone \otimes \cdots \otimes \openone\otimes (g\otimes f) \ket{\mathrm{GHZ}_n^{16}}\rightarrow \openone \otimes \cdots \otimes \openone\otimes (h\otimes f) \ket{\mathrm{GHZ}_n^{16}}.
\end{equation}

This completes our discussion of exact entanglement catalysis, and we will consider the exact catalysis of coherence in the following.

\subsection{Quantum coherence}
\label{subsec:ExactCatalysis:Coherence}

Recent studies suggest that quantum coherence might be another useful resource for many information processing tasks that are otherwise impossible without it \cite{StreltsovRevModPhys.89.041003}.  There are two different notions of coherence in the literature: \emph{unspeakable} and \emph{speakable} \cite{Marvian_PhysRevA.94.052324}.  In the case of unspeakable coherence, the levels of eigenspaces are important for the quantification of coherence, for example, energy. Whereas for the latter one, it is irrelevant which eigenspaces are responsible for the coherence. While the former type of coherence is characterized using the resource theory of asymmetry \cite{Marvian_2013}, the latter is characterized using the usual notion of resource theory of coherence introduced by \citet{Baumgratz_2014}. In the following, we will discuss catalysis in both cases.

Quantum catalysis in the resource theory of coherence was 
introduced in \cite{AbergPhysRevLett.113.150402}.  In this paper, the author studied the question of how to use a catalyst to create superposition between different energy levels under energy-preserving unitary operations. As energy must remain conserved, a system in a definite energy level cannot be changed to a superposition of different energy levels. However, an additional system or reservoir with a high degree of coherence can help to achieve this.  Due to this, the amount of coherence in the reservoir can decrease. In contrast, \citet{AbergPhysRevLett.113.150402} showed that the coherence in the reservoir can be used to turn the state of the system into a superposition of energy levels without consuming it in the process. Hence, coherence in this reservoir acts like a catalyst in the process. The advantage has been shown by constructing two examples: the doubly-infinite ladder and the half-infinite ladder \cite{AbergPhysRevLett.113.150402}. Furthermore, as a direct consequence, it has been shown that this catalytic procedure has application in work extraction \cite{AbergPhysRevLett.113.150402}. Note that in \r{A}berg's protocol the state of the reservoir does not remain the same. Nonetheless, the argument for catalysis is that the coherence resource in the reservoir does not degrade in the process and can be reused on another system to apply the same transformation. 

The change in the reservoir's state is unusual to the definition of catalysis.
Hence, an alternative description to \cite{AbergPhysRevLett.113.150402} has been provided by \citet{Korzekwa_2016}, who preferred the term ``repeatability'' over catalysis as the state of the reservoir changes.  They studied work extraction from the coherence of the given system while the reservoir can be used repeatedly. The amount of work that can be extracted from a system in the state $\rho$ is equal to the difference in free energy $\Delta F(\rho)$ when the allowed operations conserve the energy only on average \cite{PopescuNatComm}. Here $\Delta F(\rho)=F(\rho)-F(\gamma)$ with the Helmholtz free energy
\begin{equation}
    F(\rho)=\Tr[\rho H]-k T S(\rho), \label{eq:HelmholtzFreeEnergy}
\end{equation}
where $H$ and $\gamma$ correspond to the system Hamiltonian and thermal state respectively, and $S(\rho) = -\Tr[\rho \log \rho]$ is the von Neumann entropy.\footnote{If not stated otherwise, we define the von Neumann entropy with the natural logarithm.} However, in \cite{Korzekwa_2016}, the authors were interested in the work extraction only from the coherence of the system, and in that case, 
\begin{equation}
    F(\rho)-F(D[\rho])=k T S(\rho||D[\rho])
\end{equation}
is the amount of free energy available for the extraction from the coherence of state $\rho$, where $D$ is a dephasing operation in the energy eigenbasis and $S(\rho||D[\rho])$ represents a measure of coherence \cite{GourPhysRevA.80.012307} with the quantum relative entropy $S(\rho||\sigma) = \Tr[\rho \log \rho] - \Tr[\rho \log \sigma]$. \citet{Korzekwa_2016} showed that the amount of work extracted from the coherence of state $\rho$ is arbitrarily close to the free energy extraction limit, i.e., $k T S(\rho||D[\rho])$. However, all the coherence cannot be converted to work, as some of the work will be needed to repump the reservoir so that it can be reused. Nevertheless, they showed that the amount of extracted work can be arbitrarily close to the coherence provided when the coherence of the reference system is sufficiently large.   

Recently, \citet{Vaccaro_2018} analyzed the proposal of \citet{AbergPhysRevLett.113.150402} to check whether coherence can be really used repeatedly without any degradation in its performance. In particular, \citet{Vaccaro_2018} showed that if the reservoir is used repeatedly then the final systems can have correlations that have not been considered by~\citet{AbergPhysRevLett.113.150402}. The correlations between the final systems actually affect the efficiency of repeated operations. To be precise, the coherence drawn from the reservoir by uncorrelated systems is completely different compared to the coherence drawn by correlated systems. Furthermore, they showed that neglecting these correlations among the final systems could lead to unphysical phenomena, for example, two non-orthogonal reservoir states can be discriminated almost perfectly \cite{Vaccaro_2018}.

In the context of whether coherence is catalytic, an important study was done by~\citet{LostaglioPhysRevLett.123.020403}. In fact, it has been shown that coherence can neither be cloned nor broadcasted for finite dimensional systems. The coherence of a finite-dimensional system $\rho^E$ can be broadcast if there exists a time-translation covariant channel $\Lambda$ and an incoherent state $\rho^S$, such that $[\rho^S,H^S]=0$ with
\begin{equation}\label{broadcasting}
    \Lambda(\rho^E\otimes\rho^S)=\sigma^{ES},
\end{equation}
where $\rho^E=\Tr_S(\sigma^{ES})$, $\sigma^S=\Tr_E(\sigma^{ES})$ and $[\sigma^S,H^S]\neq0$. The action of a time-translation covariant channel $\Lambda$ on a state $\rho$ with Hamiltonian $H$ is expressed as $\Lambda(\rho)=\Tr_A[U(\rho\otimes\sigma^A)U^\dagger]$, where $\sigma^A$ is some ancillary state with Hamiltonian $H_A$ and $U$ represents some global unitary with $[U,H+H_A]=0$. On the other hand, the coherence of the system $\rho^S$ can be cloned when there is no correlation between the sub-systems, i.e.,
\begin{equation}
    \Lambda(\rho^E\otimes\rho^S)=\sigma^{ES}=\rho^E\otimes\sigma^S.
\end{equation}
Here, the system $E$ acts like a catalyst. In such a scenario, correlated catalysis under covariant operations cannot broadcast or clone coherence in an initially incoherent state $\rho^S$ \cite{LostaglioPhysRevLett.123.020403}. However, the questions of coherence cloning and broadcasting remain open for infinite dimensional systems. Note that the results of \citet{AbergPhysRevLett.113.150402} suggest that a weaker version of coherence broadcasting, where the state $\rho^E$ is allowed to change in the procedure, is possible for infinite dimensional systems. Hence, it seems like broadcasting as defined earlier in Eq. (\ref{broadcasting}), where the state $\rho^E$ remains unchanged may not be possible. 

In the above result, we see that broadcasting of coherence to an initially incoherent state is not possible from a finite-dimensional system \cite{LostaglioPhysRevLett.123.020403}. On the contrary, \r{A}berg's result suggests that it could be possible with infinite-dimensional systems \cite{AbergPhysRevLett.113.150402}. Recently, \citet{Ding2021} tried to create a connection between the two results by studying the role of a finite-dimensional catalyst in enlarging the set of accessible states under time-translational covariant operations. In their protocol they allow correlation between the system and the catalyst after the transformation, however, the reduced state of the catalyst remains unchanged. As correlation is allowed, they first showed that pure state catalysts do not help in any state transformations under covariant operations. Precisely, if $\rho$ cannot be transformed into $\rho'$ under covariant operations, then there exists no pure catalyst state $\ket{\phi}$ that can catalyse the transformation $\rho\otimes\phi\rightarrow\rho'\otimes\phi$ under covariant catalytic transformations. This result is a generalisation of the previously studied result by \citet{Marvian_2013}, where $\rho$ and $\rho'$ are considered to be pure. Now if one can allow correlation between the system and the catalyst then there exist coherent states such that the achievable set of states from it is strictly larger under covariant operations. This has been shown explicitly for qubit states \cite{Ding2021}. Furthermore, with the dimension of the catalyst the accessible set of states increases \cite{Ding2021}. 

In the same spirit as Nielsen's theorem in entanglement theory (see Theorem~\ref{thm:EntMajo}), a counterpart has been derived in coherence theory to address the conversions among pure states \cite{PhysRevA.91.052120,PhysRevLett.116.120404,PhysRevA.96.032316}. Before presenting the result, we introduce incoherent states and incoherent operations. To introduce these, we first fix a basis $\{\ket{i}\}$ (called as an incoherent basis) in a finite-dimensional Hilbert space. The incoherent states are diagonal in the incoherent basis and can be represented as $\sum_i  p_i \ket{i}\!\bra{i}$, where $p_i\geq 0$ and $\sum_i{p_i}=1$. The set of all such incoherent states is represented by $\mathcal{I}$. A quantum operation characterised by a set of Kraus operators $\{K_n\}$ satisfying $\sum_n K_n^\dagger K_n=\openone$, is called incoherent operation if the Kraus operators satisfy $K_n \rho K_n^\dagger/ \Tr\left(K_n \rho K_n^\dagger\right)\in \mathcal{I}$ for all $n$ and $\rho\in\mathcal{I}$ \cite{Baumgratz_2014}. Let us consider a pair of states: $\ket{\psi}=\sum_{i=0}^{d-1} \alpha_i \ket{i}$ and $\ket{\phi}=\sum_{i=0}^{d-1} \beta_i \ket{i}$, where $\alpha_i$ and $\beta_i$ are arranged in the following way $|\alpha_0|^2\geq\cdots\geq|\alpha_{d-1}|^2$ and $|\beta_0|^2\geq\cdots\geq|\beta_{d-1}|^2$.
Then $\ket{\psi}$ can be converted to $\ket{\phi}$ via incoherent operations iff $\Delta[\psi]\prec\Delta[\phi]$, or equivalently the following holds \cite{PhysRevA.91.052120,PhysRevLett.116.120404,PhysRevA.96.032316}
\begin{equation}\label{coherence_majorization}
   \sum_{i=0}^k |\alpha_i|^2 \leq \sum_{i=0}^k |\beta_i|^2 \,\,\,\,\forall k=\{0,\ldots,d-1\},
\end{equation}
where $\psi = \ket{\psi}\!\bra{\psi}$ and $\Delta[\rho]=\sum_{i}\braket{i|\rho|i}\ket{i}\!\bra{i}$ represents complete dephasing in the incoherent basis. Note that the above result also holds true for other classes of free operations considered in coherence theory \cite{PhysRevLett.116.120404}. 

Following similar ideas as for entanglement, it is not difficult to find examples of incomparable states where the above conditions are not satisfied. However, in some cases, an additional coherent state can help in accomplishing a forbidden transformation without being changed in the process \cite{PhysRevA.91.052120}. The above result in Eq. (\ref{coherence_majorization}) sheds some light on the problem of coherence catalysis, similar to the previously discussed cases for entanglement \cite{JonathanPhysRevLett.83.3566}. In fact, maximally coherent states cannot catalyze a forbidden transformation between two arbitrary incomparable pure states \cite{PhysRevA.91.052120}. Furthermore, if a forbidden transformation $\ket{\psi}\nrightarrow\ket{\phi}$ can be accomplished with the help of a pure catalyst state, then the reverse transformation $\ket{\phi}\nrightarrow\ket{\psi}$ cannot be catalyzed with a pure catalyst state \cite{PhysRevA.91.052120}. Hence, pure catalysts can only catalyse transformations between incomparable pure states. Another important consequence of Eq.~\eqref{coherence_majorization} is that conversion of $\ket{\psi}$ into $\ket{\phi}$ is possible with the help of a pure catalyst if $|\alpha_0|\leq |\beta_0|$ and $|\alpha_{d-1}|\geq |\beta_{d-1}|$ \cite{PhysRevA.91.052120}.

\citet{BuPhysRevA.93.042326} studied the necessary and sufficient conditions for catalytic coherence transformations considering pure catalyst states. They derived necessary and sufficient conditions for the existence of a catalyst for a pair of incomparable pure states. Suppose $\ket{\psi}$ and $\ket{\phi}$ are two incomparable states of dimension $d$. Then there exists a catalyst state $\ket{\eta}$ accomplishing the transformation $\ket{\psi}\otimes\ket{\eta}\rightarrow \ket{\phi}\otimes\ket{\eta}$,
if and only if \cite{BuPhysRevA.93.042326} 
\begin{align}
H_{\alpha}(\Delta[\psi]) & >H_{\alpha}(\Delta[\phi])\,\,\,\forall\alpha\in(-\infty,\infty)/0\,\,\, \mbox{and}\label{coherence_renyi1}\\
\Tr(\log\Delta[\psi]) &> \Tr(\log\Delta[\phi]). \label{coherence_renyi2}
\end{align}
Here 
\begin{equation}\label{eq:QuRenEnt}
    H_{\alpha}(\rho)=\frac{\mbox{sgn}(\alpha)}{1-\alpha}\log \Tr [\rho^\alpha]
\end{equation}
represents R\'enyi entropy. The strict inequalities in the above equations can be made less strict by allowing an infinitesimally small error in the initial or in the final state. In fact, for arbitrary $\varepsilon>0$, there exists a state $\ket{\psi_\varepsilon}$ with $||\psi-\psi_\varepsilon||_1\leq\varepsilon$ such that $\ket{\psi_\varepsilon}$ can be converted into $\ket{\phi}$ via catalytic incoherent operations (assuming pure catalyst states), if and only if \cite{BuPhysRevA.93.042326}
\begin{align}
H_{\alpha}(\Delta[\psi]) & \geq H_{\alpha}(\Delta[\phi])\,\,\,\forall\alpha\in(-\infty,\infty)/0\,\,\,\mbox{and} \label{coherence_renyi3}\\
\Tr(\log\Delta[\psi]) & \geq\Tr(\log\Delta[\phi]). \label{coherence_renyi4}
\end{align}
Note the analogy between this result and the case of entanglement as discussed in Eq. (\ref{catalyst_renyi}). Furthermore, \citet{BuPhysRevA.93.042326} explored the structure of qubit catalyst states for four-dimensional state transformations. The necessary and sufficient conditions for the existence of a qubit catalyst $\ket{\eta}=\sqrt{a}\ket{0}+\sqrt{1-a}\ket{1}$ for two four dimensional incomparable states $\ket{\psi}=\sum_{i=0}^3\sqrt{\alpha_i}\ket{i}$ and $\ket{\phi}=\sum_{i=0}^3\sqrt{\beta_i}\ket{i}$ are
\begin{align}
\alpha_{0} & \leq\beta_{0},\\
\alpha_{0}+\alpha_{1} & >\beta_{0}+\beta_{1},\\
\alpha_{0}+\alpha_{1}+\alpha_{2} & \leq\beta_{0}+\beta_{1}+\beta_{2},
\end{align}
and additionally
\begin{eqnarray}\label{range_catalyst_coherence}
&& \max\left\{\frac{\alpha_0+\alpha_1-\beta_0}{\beta_1+\beta_2}, 1-\frac{\alpha_3-\beta_3}{\beta_2-\alpha_2}\right\}\nonumber\\
 && \leq a \leq \min \left\{\frac{\beta_0}{\alpha_0+\alpha_1}, \frac{\beta_0-\alpha_0}{\alpha_1-\beta_1}, 1-\frac{\beta_3}{\alpha_2+\alpha_3}\right\}.
\end{eqnarray}
Here, without loss of generality we assume that $\{\alpha_i\}$ and $\{\beta_i\}$ are real and they are arranged in descending order. Hence, Eq.~(\ref{range_catalyst_coherence}) provides a range of the catalyst for the above transformation. Similar to the result in entanglement theory \cite{JonathanPhysRevLett.83.3566}, catalytic conversions (assuming pure catalysts) of incomparable pure states are not possible for two and three-dimensional systems, and the first nontrivial result can be found in dimension four as shown in Eq. (\ref{range_catalyst_coherence}), see \cite{BuPhysRevA.93.042326}.

So far we have discussed deterministic state transformations. If a pair of states cannot be transformed deterministically, i.e, they violate Eq. (\ref{coherence_majorization}), they still can be transformed stochastically with some optimal probability $P_{\max}$ \cite{DuQIQC2015}. \citet{BuPhysRevA.93.042326} studied the advantage of a catalyst in stochastic state transformations. In fact, they showed that for two incomparable pure states $\ket{\psi}=\sum_{i=0}^{d-1}\alpha_i\ket{i}$ and $\ket{\phi}=\sum_{i=0}^{d-1}\beta_i\ket{i}$,
there exists a catalyst state $\ket{\eta}$ such that $P_{\max}(\ket{\psi}\rightarrow\ket{\phi})< P_{\max}(\ket{\psi}\otimes\ket{\eta}\rightarrow\ket{\phi}\otimes\ket{\eta})$, if and only if
\begin{equation}
    P_{\max}(\ket{\psi}\rightarrow\ket{\phi})<\min\left\{\frac{|\alpha_{d-1}|^2}{|\beta_{d-1}|^2},1\right\}.
\end{equation}
Note that a similar result is also present in entanglement theory \cite{JonathanPhysRevLett.83.3566}. By analogy to self-catalysis in entanglement \cite{Duarte_2016}, in \cite{BuPhysRevA.93.042326}, the authors also explored self-catalysis in the context of coherence theory. As an example, consider the states
\begin{align}
\ket{\psi} & =\sqrt{0.9}\ket{0}+\sqrt{0.081}\ket{1}+\sqrt{0.01}\ket{2}+\sqrt{0.009}\ket{3},\\
\ket{\phi} & =\sqrt{0.95}\ket{0}+\sqrt{0.03}\ket{1}+\sqrt{0.02}\ket{2}.
\end{align}
It is easy to check that $\ket{\psi}\nrightarrow\ket{\phi}$, while at the same time $\ket{\psi}\otimes\ket{\psi}\rightarrow\ket{\phi}\otimes\ket{\psi}$ \cite{BuPhysRevA.93.042326}.

Earlier, we discussed the advantage of catalysis-assisted probabilistic transformation of pure states. Naturally, one may wonder whether a catalyst exists such that it can enhance the probability of transforming a mixed state into a pure state. Recently, \citet{Liu_2020} explored this scenario for strictly incoherent operations and derived necessary and sufficient conditions for the existence of a pure catalyst such that the optimal transformation probability is enhanced. A quantum operation characterised by Kraus operators $\{K_n\}$ is called a strictly incoherent operation if all $K_n$ and $K_n^\dagger$ are incoherent in the given incoherent basis \cite{PhysRevLett.116.120404,PhysRevX.6.041028}. For such operations \citet{Liu_2020} showed that $P_{\max}(\rho\otimes\eta\rightarrow\psi\otimes\eta)>P_{\max}(\rho\rightarrow\psi)$ is possible, if and only if the following holds at least for one $\mu$:
\begin{equation}
     P_{\max}(\phi^\mu\rightarrow\psi)<\min\left\{\frac{\phi^{\mu}_n}{\psi_n},1\right\},
 \end{equation}
where $\{\ket{\phi^\mu}\}$ correspond to the maximally dimensional pure coherent state subspaces of $\rho$, $\ket{\phi^\mu}=\sum_{i=0}^{d_1-1} \phi^\mu_i\ket{i}$, $\ket{\psi}=\sum_{i=0}^{d_2-1} \psi_i\ket{i}$,  $n=\max\{d_1-1,d_2-1\}$, and $d_1$ and $d_2$ denote the dimensions of $\ket{\phi}^\mu$ and $\ket{\psi}$ respectively. The coefficients of the pure states are arranged in descending order. A state $\rho$ has a pure coherent state subspace of dimension $n$ if there exists an incoherent projector $\mathbb{P}$ such that $\frac{\mathbb{P}_\mu\rho\mathbb{P}_\mu}{\mbox{Tr}\mathbb{P}_\mu\rho\mathbb{P}_\mu}=\phi^\mu$ and $\phi^\mu$ has coherence rank $n$ \cite{Liu_2020}. Coherence rank corresponds to the number of nonzero terms $\phi_i^\mu$ in $\ket{\phi}^\mu$. The subspace is maximally coherent when the rank of $\phi^\mu$ cannot be increased.  

Catalytic coherence transformation of some classes of mixed states has been studied by~\citet{Xing_2020}. Consider the following rank-2 mixed states of the forms
\begin{eqnarray}
    &&\rho=\alpha\ket{\psi}\!\bra{\psi}+(1-\alpha)\ket{4}\!\bra{4}\,\, \mbox{and}\\
  && \sigma=\beta\ket{\phi}\!\bra{\phi}+(1-\beta)\ket{4}\!\bra{4},
\end{eqnarray}
where 
\begin{align}
\ket{\psi} & =\sqrt{0.38}\ket{0}+\sqrt{0.38}\ket{1}+\sqrt{0.095}\ket{2}+\sqrt{0.095}\ket{3}\nonumber \\
 & +\sqrt{0.05}\ket{4},\\
\ket{\phi} & =\sqrt{0.5}\ket{0}+\sqrt{0.25}\ket{1}+\sqrt{0.25}\ket{2}.
\end{align}
The transformation of $\rho$ into $\sigma$ is not possible via incoherent transformation \cite{Xing_2020}. However, as shown in \cite{Xing_2020}, the transformation can be achieved with the help of a catalyst in state $\ket{\varphi}=\sqrt{0.6}\ket{0}+\sqrt{0.4}\ket{1}$. Note that the above states and the catalytic conversion protocol are similar to the entanglement case discussed in \cite{EisertPhysRevLett.85.437}. The above transformation of rank-2 mixed states can be translated to some special class of higher-rank states as well. In addition, for general mixed states catalytic coherence transformation is also discussed in \cite{Xing_2020}. 

\citet{XING2020} also studied \emph{supercatalytic} coherence transformation. Suppose we have two incomparable states $\ket{\psi}$ and $\ket{\phi}$, and we want to achieve the transformation $\ket{\psi}\rightarrow\ket{\phi}$. We already see that such transformations can be realized with the help of some auxiliary state if Eqs. (\ref{coherence_renyi1}) and (\ref{coherence_renyi2}) are satisfied. In contrast to this, the transformation can be realised in another way where the auxiliary state changes in the process, but the coherence of the final auxiliary system is increased. More precisely, $\ket{\psi}\otimes\ket{\varphi_1}\rightarrow\ket{\phi}\otimes\ket{\varphi_2}$, where $\ket{\varphi_1}$ and $\ket{\varphi_2}$ represent auxiliary states with $C_r(\varphi_2)\geq C_r(\varphi_1)$. Here $C_r$ represents the relative entropy of coherence \cite{Baumgratz_2014}.  Examples and some interesting results of such supercatalytic coherence transformation are discussed in \cite{XING2020}.

\subsection{Quantum thermodynamics} \label{subsec:ExactCatalysis:Thermodynamics}

Catalysis has also been broadly explored in quantum thermodynamics, investigating transformations of quantum systems under thermal operations in the presence of catalysts. Thermal operations model transformations of a quantum system $S$ with Hamiltonian $H_S$ interacting with a heat reservoir $B$ with Hamiltonian $H_B$ and in a thermal state  $\gamma^{B} = e^{-\beta H_B} / Z_B$,
where $\beta =  1/kT $ is the inverse temperature 
and $Z_B = \Tr e^{-\beta H_B}$ is the partition function.
The interaction is described by a unitary $U_{SB}$ that conserves the total energy: $[U_{SB}, H_S + H_B] = 0$.
A thermal operation $\Lambda$ on a state $\rho^S$ is defined as~\cite{Janzing2000}
\begin{equation}
    \Lambda(\rho^S)=\Tr_B\left(U_{SB}\left[\rho^S\otimes\gamma^B\right]U_{SB}^\dagger\right).
\end{equation}
See also \cite{horodecki2013fundamental} for a detailed discussion on thermal operations. 

To describe \textit{catalytic} thermal operations,~\citet{Brandao3275} allow for an additional system $\sigma^C$ as a catalyst. Precisely, a transformation from $\rho^S$ to $\mu^S$ is possible via catalytic thermal operations, if and only if there exists a thermal operation $\Lambda$ and a state $\sigma^C$ taking $\rho\otimes \sigma$ to $\mu \otimes\sigma$~\cite{Brandao3275}, i.e., 
\begin{equation}\label{CatTheOp}
     \Lambda(\rho^S\otimes\sigma^C)=\Tr_B\left(U_{SCB}\left[\rho^S\otimes\sigma^C\otimes\gamma^B\right]U_{SCB}^\dagger\right)=\mu^S\otimes\sigma^C,
\end{equation}
where the unitary $U_{SCB}$ satisfies $[U_{SCB},H_S+H_C+H_B]=0$ and $H_C$ represents the Hamiltonian of the catalyst system $C$. Moreover, the authors find necessary and sufficient conditions for such transformation to be possible when the initial and the final states are block-diagonal in energy eigenbasis \cite{Brandao3275}. 
\begin{theorem}[See family of second laws in \cite{Brandao3275}]
There exists a catalytic thermal operation mapping a diagonal state $\rho$ to another diagonal state $\rho'$, if and only if
\begin{equation}\label{secondlaw}
     F_{\alpha}(\rho, \gamma_{\beta})\geq  F_{\alpha}(\rho',\gamma_{\beta})\,\,\textrm{for}\,\,\alpha\geq 0.
\end{equation}
\end{theorem}
\noindent Here, the generalized free energies are defined as
\begin{equation}
\label{eq:generalisedFreeEnergies}
    F_{\alpha}(\rho,\gamma_\beta) = k T S_{\alpha}(\rho|| \gamma_\beta) -k T \log Z,
\end{equation}
where 
\begin{equation}
\label{eq:RenyiRelEntropy}
     S_{\alpha}(\rho  ||  \sigma) =\begin{cases} 
     \frac{1}{\alpha-1}  \log\Tr \left(\rho^{\alpha} \sigma^{1-\alpha} \right), \,\, \alpha \in [0,1) \\
     \frac{1}{\alpha-1}  \log\Tr \left[\left(\sigma^{\frac{1-\alpha}{2\alpha}}\rho \sigma^{\frac{1-\alpha}{2\alpha}} \right)^\alpha\right], \,\, \alpha>1
     \end{cases}
\end{equation}
 is the R\'enyi relative entropy and $S_1(\rho||\sigma)=\Tr[\rho(\log \rho-\log\sigma)]$ is found by taking the limit $\alpha\rightarrow 1$. Also note that for $\lim_{\alpha \rightarrow 1 } F_{\alpha}(\rho, \gamma_\beta)$ we obtain the ordinary Helmholtz free energy, see also Eq.~(\ref{eq:HelmholtzFreeEnergy}). The family of inequalities in Eq.~(\ref{secondlaw}) is known as the \emph{second laws of quantum thermodynamics}~\cite{Brandao3275}. In contrast, note that at the macroscopic scale or in the standard thermodynamic limit, transitions are governed by free energy only, and its monotonicity provides the standard second law. Furthermore, \citet{Brandao3275} discussed a generalisation of the above result by considering transformations of arbitrary states and provided some necessary conditions. A state $\rho$ can be transformed into $\rho'$ with arbitrary accuracy only if \cite{Brandao3275}
 \begin{subequations} \label{eq:TherArbSta}
 \begin{equation}\label{eq:TherArbSta1}
     F'_{\alpha}(\rho,\gamma_\beta)\geq F'_{\alpha}(\rho,\gamma_\beta)\,\, \forall \alpha \geq \frac{1}{2},
 \end{equation}
 \begin{equation}\label{eq:TherArbSta2}
     F'_{\alpha}(\gamma_\beta,\rho)\geq F'_{\alpha}(\gamma_\beta,\rho')\,\, \mathrm{for}\,\, \frac{1}{2} \leq \alpha  \leq 1,\,\, \mbox{and},
 \end{equation}
 \begin{equation}\label{eq:TherArbSta3}
     F''_{\alpha}(\rho,\gamma_\beta)\geq F''_{\alpha}(\rho',\gamma_\beta)\,\, \mathrm{for}\,\, 0 \leq \alpha  \leq 2,
 \end{equation}
 \end{subequations}
where $F'$ and $F''$ are defined as
\begin{align}
F'(\rho,\gamma_{\beta}) & =\frac{kT}{\alpha-1}\log\Tr\left[\left(\gamma_{\beta}^{\frac{1-\alpha}{2\alpha}}\rho\gamma_{\beta}^{\frac{1-\alpha}{2\alpha}}\right)^{\alpha}\right]-kT\log Z,\\
F''(\rho,\gamma_{\beta}) & =\frac{kT\mathrm{sgn}(\alpha)}{\alpha-1}\log\Tr\left(\rho^{\alpha}\gamma_{\beta}^{1-\alpha}\right)-kT\log Z.
\end{align}

\citet{Wehner_free_energy} studied how the above mentioned free energies behave in the asymptotic limit for block-diagonal states. As a result of their study, they found out that if $F(\rho)> F(\rho')$, then for any $\varepsilon > 0$ there exists a sufficiently large $n$ and a \textit{catalytic} thermal operation $\Lambda$, taking $\rho^{\otimes n}$ arbitrarily close to $\rho'^{\otimes n}$, i.e., $||\Lambda(\rho^{\otimes n})-\rho'^{\otimes n}||_1\leq \varepsilon$. 

In contrast to the previous result by \citet{Brandao3275}, \citet{Lostaglio2015} showed that only the monotonicity of free energy is required for state transformations if one allows correlations among the particles of the catalyst and an arbitrarily small error in the final state $\rho'$. In their definition of catalytic state transformation, a block-diagonal state $\rho$ is transformed to another block-diagonal state $\rho'$ when for all $\varepsilon>0$, there exists a thermal operation $\Lambda$, a state $\rho_\varepsilon '$ with $||\rho' -\rho_\varepsilon '||<\varepsilon$ and a correlated state $\sigma_{C_1\cdots C_N}$ with marginals $\sigma_{C_1}, \cdots, \sigma_{C_N}$ such that~\cite{Lostaglio2015}
\begin{eqnarray}\label{MarCat}
 \Lambda(\rho\otimes\sigma_{C_1}\otimes\cdots\otimes\sigma_{C_N})=\rho_{\varepsilon}'\otimes \sigma_{C_1\cdots C_N}.
\end{eqnarray}
Such a transformation is possible, if and only if $F(\rho)\geq F(\rho')$ \cite{Lostaglio2015}. However, note that although the local states of the particles remain the same, the global state of the catalyst changes in this procedure. This may lead to embezzling, a phenomenon discussed in more detail in Section \ref{sec:CatalyticEmbezzling}. 

The above result emphasises that if a transformation is possible in the thermodynamic limit then it is also possible in the single copy settings. Hence, it provides an operational meaning to free energy. Motivated by this result,~\citet{Muller_2016IEEE} derived a necessary and sufficient condition for the transformations of probability distributions. For this, they studied transformations of probability distributions under bistochastic maps, which are represented by bistochastic matrices that are composed of non-negative real numbers with every individual row and column summing up to one. In particular, \citet{Muller_2016IEEE} considered transformations of probability distributions of the form
\begin{equation}
    p\otimes r_1\otimes\cdots\otimes r_N\rightarrow q\otimes r_{1,\cdots, N},
\end{equation}
where the probability distribution $p=(p_1,\cdots,p_m)$ is mapped to another probability distribution $q=(q_1,\cdots,q_m)$ with the help of auxiliary systems that get correlated in the procedure, but at the same time, the marginals remain unchanged. In such a scenario, if $p$ and $q$ are not identical up to permutation and both contain non-zero elements, then the transformation from $p$ to $q$ is possible if and only if $H(p)<H(q)$ \cite{Muller_2016IEEE}, where $H(p) = -\sum_i p_i \log p_i$ represents the Shannon entropy. 

In most of the previous studies \cite{Brandao3275,Wehner_free_energy,Lostaglio2015}, only block-diagonal states are considered, which contain no coherence in the eigenbasis of the Hamiltonian. Although the conditions in Eqs. (\ref{eq:TherArbSta}) hold for arbitrary states, the role of coherence in state transformations is not apparent from them.  On the contrary, in \cite{Lostaglio_ncomms7383} the authors considered arbitrary states and derived the necessary conditions for state transformations under catalytic thermal operations. They considered the exact catalytic thermal operation presented in Eq. (\ref{CatTheOp}). If a state $\rho$ can be converted to another state $\rho'$ via a catalytic thermal operation where the catalyst state is in the block-diagonal form, then we have \cite{Lostaglio_ncomms7383}
\begin{equation}
  A_{\alpha}(\rho)\geq A_{\alpha}(\rho')\,\, \forall \,\alpha\geq 0,
\end{equation}
where $A_{\alpha}(\rho)=S_{\alpha}(\rho||D[\rho])$ represents free coherence of the state $\rho$ \cite{Lostaglio_ncomms7383} and $D$ corresponds to the dephasing operation in the given energy eigenbasis. 

In \cite{Wilminge19060241}, the authors considered marginal-catalytic transformation and correlated-catalytic transformation under Gibbs preserving operations and showed that free energy is the only monotonic function under such free operations. In their definition, the marginal-catalytic transformation has a similar structure as presented in Eq. (\ref{MarCat}), with the only difference that the final state is returned exactly and the maps are Gibbs preserving \cite{Wilminge19060241}. On the other hand, in the case of correlated-catalytic transformation of $\rho^S$ into $\mu^S$, we have $\rho^S \otimes \sigma^C \rightarrow \mu^{SC}$, where the final state $\mu^{SC}$ is correlated with marginals $\mu^S$ and $\sigma^C$ \cite{Wilminge19060241}. Then it has been shown that any monotone under these two kinds of catalytic transformations, which is additive on the tensor product and continuous, is proportional to the free energy difference  \cite{Wilminge19060241}. Hence, free energy is the only monotone for such kind of catalytic transformations.

Recently, the role of exact uncorrelated catalysis for the Markovian thermal process has been investigated in \cite{Korzekwa_PhysRevLett.129.040602}. In fact, catalysis has been demonstrated to be advantageous in the cooling of systems. More specifically, the authors demonstrated that a qubit catalyst can cool a system initially attached to a bath to temperatures lower than the bath temperature, whereas without a catalyst, the system can only be cooled to the bath temperature \cite{Korzekwa_PhysRevLett.129.040602}.

In a recent paper \cite{Son_2209.15213}, the role of exact uncorrelated catalysis in elementary thermal operations has been discussed. Elementary thermal operations are a subset of thermal operations, and their practical realizations are apparent in comparison to thermal operations \cite{Lostaglio2018elementarythermal}. In \cite{Son_2209.15213}, the authors provided a few examples to illustrate that for a given initial qutrit state, the set of final states under thermal operations can almost be achieved with elementary thermal operations helped by qubit catalysts. In addition, \citet{Son_2209.15213} demonstrated an example in which the set of final states under thermal operations is strictly smaller compared to the set of final states that can be achieved under elementary thermal operations aided by qubit catalysts. Therefore, as elementary thermal operations can be experimentally realized, this method could provide an experimentally feasible way of achieving the set of final states under thermal operations using only elementary thermal operations and catalysts.

\subsection{Purity} \label{subsec:ExactCatalysis:Purity}

In the resource theory of purity, a key question is to find the minimum amount of noise required to perform a given physical process on a quantum system \cite{GOUR20151,Boes_Cat_rand}. In this theory, the free operations are noisy operations. In general, a noisy operation can be implemented in the following steps: i) first, a given system is attached to an ancillary system prepared in a completely mixed state; ii) then, a unitary operation is performed to couple them; and iii) finally, we discard the ancillary system by taking partial trace over the ancillary system. The overall transformation on a state $\rho^S$ in the Hilbert space $\mathcal{H}_{S}$ can be written as 
\begin{equation}
    \Lambda[\rho^S]=\Tr_{A}\left[U\left(\rho^S\otimes\frac{\openone_A}{d_A}\right)U^\dagger\right],
\end{equation}
where $U$ is in $\mathcal{H}_S\otimes\mathcal{H}_A$, $\mathcal{H}_A$ represents the Hilbert space of the ancillary system, and $d_A$ corresponds to the dimension of the ancillary system. Note that this kind of operation was first reported in \cite{PhysRevLett.90.100402,PhysRevA.67.062104}. All the states except the maximally mixed state are considered resources in this theory. Furthermore, note that without loss of generality, we can assume the states to be diagonal matrices as unitary operations are free in this resource theory. The necessary and sufficient conditions for exact state transformations are given by the majorization relation. More precisely, a transformation from $\rho$ to $\sigma$ is possible under a noisy operation if and only if $\sigma\prec\rho$ \cite{GOUR20151}. In contrast to the entanglement theory, the majorization relation is in the opposite direction. 

An immediate consequence is that there exist transformations that are not achievable only via noisy operations but could be accomplished with the help of a catalyst \cite{GOUR20151}. To analyse state transformations under catalytic noisy operations, one needs to consider a trumping relation $\sigma\prec_T\rho$, which states that there exists a catalyst $\eta$ that helps in converting $\rho$ into $\sigma$ via noisy operations, i.e., $\sigma\otimes\eta\prec\rho\otimes\eta$ \cite{GOUR20151}. The noisy trumping relation is equivalent to \cite{Klimesh0709.3680,GOUR20151} 
\begin{equation}
    f_\alpha(\rho)>f_\alpha(\sigma) \,\,\forall\,\, \alpha \in (-\infty,\infty), 
\end{equation} 
where the functions $f_\alpha$ are the extensions of the functions in Eq.~(\ref{eq:falpha}) to density matrices, i.e.,
\begin{eqnarray}\label{eq:falpha2}
 f_{\alpha}(\rho)=\begin{cases}
 \ln \Tr [\rho^\alpha] & (\alpha >1);\\
\Tr [\rho \ln \rho] &  (\alpha =1);\\
 -\ln \Tr [\rho^\alpha] & (0<\alpha<1);\\
 -\Tr [\ln \rho] &  (\alpha =0);\\
\ln [\Tr \rho^\alpha] & (\alpha <0). 
 \end{cases}
\end{eqnarray}
Similar to the case of entanglement and coherence, an alternative definition of noisy trumping is introduced \cite{GOUR20151}, where an infinitesimally small error is allowed in the initial or final state. More formally a state $\rho$ trumps noisily $\sigma$, if there exists a state $\eta$ such that $\rho\otimes \nu_{\delta}\otimes\eta\rightarrow\sigma\otimes\eta$ for all $\delta>0$. Here $\nu_{\delta}=\left(1/m,\cdots,1/m,0,\cdots,0\right)$ is a sharp state of dimension $n$ with $m\leq n$ and satisfies $||\nu_{\delta}-\openone_n/n||_1\leq \delta$. With this, \citet{GOUR20151} provided necessary and sufficient conditions for noisy trumping as follows: A state $\rho$ noisy trumps $\sigma$, if and only if
\begin{equation}
    H_\alpha(\rho)\leq H_\alpha(\sigma)\,\, \forall \,\, \alpha\geq0,
\end{equation}
where $H_\alpha$ represent the R\'{e}nyi entropies defined earlier in Eq.~(\ref{eq:QuRenEnt}). Furthermore, they also explored approximate catalysis where a state $\rho$ is transformed into $\sigma$ approximately via noisy operations and a catalyst $\eta$. However, no further restriction is imposed on the catalyst except that the final state is $\varepsilon$ close to $\sigma\otimes\eta$. Therefore, the catalyst might change infinitesimally in the procedure. The slight change in the catalyst leads to embezzling, which is discussed in more detail in Section \ref{sec:CatalyticEmbezzling}. In such a scenario, for a given $\varepsilon>0$ the noisy operations are trivial, i.e., every state can be converted to every other state \cite{GOUR20151,Brandao3275}. We also refer to Section~\ref{sec:ApproxCatalysis} for a more detailed analysis of approximate catalysis.

Previously, we discussed the necessary and sufficient conditions for state transformations under catalytic noisy operations. Now, one may wonder what is the minimum amount of randomness required to implement state transitions via noisy operations, and if there is any difference between classical and quantum randomness. Before answering these questions, an understanding of classical and quantum randomness is important. The nature of the source of randomness defines whether it is quantum or classical. A transformation from $\rho$ to $\sigma$ using $m$ units of \emph{classical randomness} is possible if there exists a set of unitaries $\{U_i\}$ such that \cite{Boes_Cat_rand}
\begin{equation}
   \sigma=\frac{1}{m}\sum_{i=1}^{m}U_i\rho U^{\dagger}_{i}
\end{equation} 
and the transformation is denoted by $\rho\xrightarrow[m]{C}\sigma$. On the other hand, a transformation from $\rho$ to $\sigma$ using $m$ units of \emph{quantum randomness} is possible if there exists a unitary $U$ and an ancillary system $A$ of dimension $m$ in the maximally mixed state $\openone/m$ such that \cite{Boes_Cat_rand}
\begin{equation}
     \sigma=\Tr_{A}U\left(\rho\otimes \frac{\openone}{m}\right) U^{\dagger}, \label{eq:QuantumRandomness}
\end{equation}
and the transformation is denoted by $\rho\xrightarrow[m]{Q}\sigma$. When there is no bound on the amount of available randomness (quantified by $m$), the set of transitions that can be achieved with quantum randomness sources can also be achieved with classical randomness sources \cite{Marshall2011,Boes_Cat_rand}. Equivalently we have,
\begin{equation}
    \rho\xrightarrow[\infty]{C}\sigma\iff \rho\xrightarrow[\infty]{Q}\sigma\iff\rho \succ \sigma.
\end{equation}

The transformations of the form~(\ref{eq:QuantumRandomness}) in the limit $m \rightarrow \infty$ have been previously studied in \cite{GOUR20151}, where noise is treated as free or in other words we can use a maximally mixed state of arbitrary dimension to implement a noisy operation. On the contrary, \citet{Boes_Cat_rand} treated noise as a resource and investigated the minimum amount of noise required to accomplish a given state transformation. For a given Hilbert space $\mathcal{H}_S$ of dimension $d$, the minimum amount of noise or randomness can be quantified as 
\begin{equation}\label{exp_rand}
    m_{X}^{*}(d)=\mbox{arg\,min}_{m}\left[\rho\xrightarrow[m]{X}\sigma \,\, \forall\,\, \rho, \sigma \in \mathcal{H_S}\,\, \mbox{with}\,\, \rho \succ \sigma\right],
\end{equation}
where $X$ corresponds to either $C$ or $Q$. It is important to note that for a specific transition, the required randomness could be smaller than the above-specified value in Eq. (\ref{exp_rand}). \citet{Boes_Cat_rand} provided an analytical expression for these quantities as following
\begin{align}
     m_{C}^{*}(d) &= d, \\
     m_{Q}^{*}(d) &= \lceil\sqrt{d}\rceil.
\end{align}
Here, we will focus mainly on quantum randomness. In proving the optimality of $m_{Q}^{*}$, the authors of \cite{Boes_Cat_rand}, used the Schur-Horn lemma~\cite{Shur-horn} with the majorisation relation $\rho \succ \sigma$, which states that for every state transition under noisy operation there exists a map of the form
\begin{equation}\label{noisy_dephase}
     \sigma = \Lambda(\rho) = \mathcal{U}'\circ\Delta_{B}\circ \mathcal{U}  (\rho)
\end{equation}
where, $\mathcal{U}$, $\mathcal{U'}$ are unitary channels that are specific to the states $\rho$ and $\sigma$. Here $\Delta_{B}$ represents a dephasing map in a fixed basis $B=\{\ket{i}\}_{i=1}^d$ of dimension $d$ and defined as $\bra{i}\Delta_B (\rho) \ket{j}=\bra{i}\rho\ket{j}\delta_{ij}$. As the implementations of unitary channels do not require any noise, all the noise needed for a state transition is required to implement the dephasing map $\Delta_B$ in Eq. (\ref{noisy_dephase}). Next, \citet{Boes_Cat_rand} provided an explicit protocol implementing dephasing map $\Delta_{B}$ in any basis $B$, using a noise model and requiring $ m_{Q}^{*}(d)=\lceil\sqrt{d}\rceil$. Specifically, for any dimension $d$ and basis $B$, \citet{Boes_Cat_rand} showed that there exists a unitary $U$, such that 
\begin{align}
     \Tr_{A}\left[U\left(\rho \otimes \frac{\openone}{\lceil\sqrt{d}\rceil}\right) U^{\dagger}\right]=\Delta_{B}(\rho),\label{dephase1}\\
     \Tr_{S}\left[U\left(\rho\otimes \frac{\openone}{\lceil\sqrt{d}\rceil}\right) U^{\dagger}\right]= \frac{\openone}{\lceil\sqrt{d}\rceil}\,\,\forall\,\,\rho.\label{dephase2}
\end{align}
Although the final state may have correlations between the system and the ancilla, the second condition implies that the ancillary system $A$ is returned exactly in the same state as before and thus acts as a catalyst in the process. As mentioned previously, \citet{Boes_Cat_rand} considered noise as a resource, and interestingly, here the noise is not consumed in the process and can be reused again. Furthermore, note that Eq.~(\ref{noisy_dephase}) emphasises that every possible state transition utilising randomness can be accomplished if a dephasing map can be constructed and \citet{Boes_Cat_rand} provided an explicit protocol for that. 

In \cite{Boes_Cat_rand}, only maximally mixed states have been used as a source of randomness. In contrast, \citet{LiePhysRevA.103.042421} considered any arbitrary randomness source $\eta$ and derived a necessary and sufficient condition such that the randomness source $\eta$ can be used to dephase an arbitrary input state $\rho$. In fact, it has been shown that any randomness source $\eta$ can dephase a $d^2$ dimensional input states catalytically iff the following holds \cite{LiePhysRevA.103.042421}
\begin{equation}
    S_{\min}(\eta)\geq \log_2 d,
\end{equation}
where $S_{\min}(\eta)=-\log_2||\eta||$ represents min-entropy and the operator norm $||\eta||$ corresponds to the largest singular value of $\eta$. However, note that in their formulation, the state of the catalyst $\eta$ can change, but the same catalyst can be used repeatedly to dephase arbitrary input states of the same dimension. In validation of this, the authors of \cite{LiePhysRevA.103.042421} showed that randomness described by min-entropy is not consumed in the process and can be reused.  

Based on the result in \cite{Boes_Cat_rand}, \citet{BoesPhysRevLett.122.210402} proved that von Neumann entropy completely characterises single-shot state transitions under unitary operations. To achieve this, the authors in \cite{BoesPhysRevLett.122.210402} used an ancillary system as a catalyst and an environment which dephases a system in a preferred basis. In particular, the following has been shown \cite{BoesPhysRevLett.122.210402}.
\begin{theorem}[See Theorem 1 in \cite{BoesPhysRevLett.122.210402}]
For any two states $\rho$ and $\sigma$ of the same finite dimension and different spectra, the following statements are equivalent\newline
(i) $S(\sigma) >S(\rho)$ and $rank(\sigma) \geq rank(\rho)$\newline
(ii) There exists a unitary $U$ and a finite dimensional catalyst in the state $\eta$, which can be chosen to be diagonal in any basis $B$ such that
\begin{align}
     \Tr_{A}U[\rho\otimes\eta] U^{\dagger}=\sigma,\label{dephase3}\\
     \Delta_{B}(\Tr_{S}U[\rho\otimes\eta] U^{\dagger})=\eta.\label{dephase4}
\end{align}
\end{theorem}
\noindent Note that the result corresponds to Eqs. (\ref{dephase1}) and (\ref{dephase2}) is a special case of the above theorem. Naturally, one may wonder whether the dephasing map given in Eq. (\ref{dephase4}) is at all necessary to implement all state transitions such that the von Neumann entropy does not decrease. In support of this, \citet{BoesPhysRevLett.122.210402} provided the following \emph{catalytic entropy conjecture}.

\begin{conjecture}[See Conjecture 1 in \cite{BoesPhysRevLett.122.210402}] \label{Entropy conjecture}
For any two states $\rho$ and $\sigma$ of the same finite dimension and different spectra, the following statements are equivalent: 
\newline(i) $S(\sigma) >S(\rho)$ and $\mathrm{rank}(\sigma) \geq \mathrm{rank}(\rho)$, 
\newline(ii) There exists a finite dimensional state $\eta$ and a  unitary $U$ such that,
\begin{align}
     \Tr_{A}U[\rho\otimes\eta] U^{\dagger}=\sigma,\\
     \Tr_{S}U[\rho\otimes\eta] U^{\dagger}=\eta.
\end{align}
\end{conjecture}

This conjecture has been proven very recently by \citet{Wilming_2205}. In fact, \citet{Wilming_2205} showed the following: 
Let $\rho$ and $\rho'$ be two states of dimension $d$ in $\mathcal{H}_d$ that are not unitarily equivalent. Then there exist a natural number $n$ and a state $\rho_n'$ on $\mathcal{H}_d^{\otimes n}$ such that 
\begin{eqnarray}
 \rho^{\otimes n}\succ \rho_n'\,\, \mbox{and}\,\, \Tr_{\{1,\cdots,n\}\setminus i} \rho_n'=\rho' \,\, \forall i \in \{1,\cdots, n\}
\end{eqnarray}
holds true if and only if $S(\rho)<S(\rho')$ and $\mbox{rank}(\rho)\leq\mbox{rank}(\rho')$. The validity of the catalytic entropy conjecture is a consequence of this result~\cite{Wilming_2205} 

Recently, in \cite{Lie_PhysRevResearch,PhysRevResearch.3.043089}, randomness to implement quantum channels has been explored. In particular, \citet{Lie_PhysRevResearch} used an ancillary system as a source of randomness such that when this ancillary system is used to implement a quantum channel no information is leaked into it. They named this a \emph{randomness-utilising process}, as no information is leaked into the ancillary system while implementing a quantum channel. As a consequence, the ancillary system after the implementation of the process should not depend on the input state of the channel. A randomness-utilising channel on $\mathcal{B}(\mathcal{H}_S)$ can be expressed as \cite{Lie_PhysRevResearch}
\begin{equation}
    \Lambda(\rho)=\Tr_A\left[U\left(\rho\otimes\sigma\right)U^\dagger\right],
\end{equation}
with a randomness source $\sigma$ on $\mathcal{H}_A$, some unitary $U$ on $\mathcal{H}_S\otimes\mathcal{H}_A$, and the state
$\Tr_{S}[U(\rho\otimes\sigma)U^{\dagger}]$ is independent of the input state. \citet{Lie_PhysRevResearch} called this state of the ancillary system after the implementation of the randomness-utilising process \emph{residue randomness}. If the residue randomness has the same spectrum as $\sigma$, the process is called catalytic. Hence, the randomness source can be reused to implement the same process on another input state. Equipped with this \citet{Lie_PhysRevResearch} showed that any dimension-preserving randomness-utilising process is catalytic. By ``dimension-preserving'' the authors of \cite{Lie_PhysRevResearch} meant that the input and output systems are of the same dimension. Naturally one may wonder whether all dimension-preserving channels can be implemented by randomness-utilising process. It turns out that only unital quantum channels can be implemented using randomness-utilising process \cite{Lie_PhysRevResearch}. A unital quantum channel $\Lambda_U$ keeps the identity operator invariant, i.e., $\Lambda_U(\openone)=\openone$. However, whether all unital quantum channels are catalytic, or in other words, all unital quantum channels can be implemented using randomness-utilising processes, remained open in \cite{Lie_PhysRevResearch}. A negative answer to this problem is given in \cite{PhysRevResearch.3.043089} by showing that not all unital quantum channels are catalytic. Furthermore, \citet{PhysRevResearch.3.043089} defined catalytic entropy to quantify the maximum amount of randomness that can be extracted catalytically from a randomness source $\sigma$. The catalytic entropy of a randomness source $\sigma$ with eigenspace decomposition $\sigma=\sum_i\lambda_iP_i$ is given in \cite{PhysRevResearch.3.043089} by
\begin{equation}
    S_C(\sigma):= -\sum_{i}\lambda_i r_i \log_2 \frac{\lambda_i}{r_i}.
\end{equation}
Here, $P_i$ are the orthogonal projectors associated with the eigenvalue $\lambda_i$ and satisfy $P_i P_j=0$ if $\lambda_i\neq \lambda_j$, and $r_i=\Tr P_i$ represents degeneracy of $\lambda_i$. $S_C(\sigma)$ is the maximum amount of entropy or randomness that can be produced from $\sigma$ \cite{PhysRevResearch.3.043089}. In another recent work, the authors extended the role of catalytic quantum randomness for dynamical maps~\cite{Liearxiv.2206.11469}.

\citet{henao2022catalytic} explored the role of catalysts in reducing the impact of information erasure on the environment. In any information erasing process, an initial state $\rho_s$ of a system $s$ with von Neumann entropy $S(\rho_s)$ is taken to another state $\rho_s'$ that has less entropy than the initial state $\rho_s$, i.e., $\Delta S_s=S(\rho_s)-S(\rho_s')>0$, and the change in entropy $-\Delta S_s$ is dumped as heat in the environment. For a finite-dimensional environment, the authors in \cite{henao2022catalytic} showed that the correlation between the system and the environment can be maneuvered catalytically to reduce the entropy or heat waste in the environment. To set the scene, we introduce an environment $e$ in state $\rho_e$ that is attached to the state $\rho_s$ to reduce the entropy of the system $s$. By applying a global unitary transformation $U_{se}$ on the system and the environment, the uncorrelated state $\rho_s\otimes\rho_e$ is transformed to a final state $\sigma_{se}=U_{se}(\rho_s\otimes\rho_e)U_{se}^{\dagger}$, such that the entropy of the system decreases, i.e., $S(\sigma_s)-S(\rho_s)<0$, where $\sigma_s=\Tr_e \sigma_{se}$. As a result, the heat dumped in the environment is $Q_e=\Tr[H_e(\sigma_e-\rho_e)]$, where $H_e$ represents the Hamiltonian associated with the environment and $\sigma_e=\Tr_s \sigma_{se}$. Now with the help of a catalyst $c$ in the state $\rho_c$, \citet{henao2022catalytic} aimed to reduce the heat dissipation in the environment to a value $Q_e'$ such that $Q_e'<Q_e$. This leads to a two-step process where $\rho_s\otimes\rho_e$ is transformed to $\sigma_{se}$ and then $\sigma_{se}$ is transformed to $\rho_{sec}'$ with the help of a suitable catalyst state $\rho_c$ and a global unitary $U_{sec}$ such that $Q_e'=\Tr[H_e(\rho_e'-\rho_e)]<Q_e$. Here, $\rho_{sec}'=U_{sec}(\sigma_{se}\otimes\rho_c)U_{sec}^\dagger$ and $\rho_e'=\Tr_{sc}\rho_{sec}'$. Apart from ensuring $\rho_s'=\Tr_{ec}\rho_{sec}'=\sigma_s$, we need to make sure that the transformation is catalytic, i.e., $\Tr_{se}\rho_{sec}'=\rho_c$. With this, \citet{henao2022catalytic} proved the following: There exists a unitary transformation $U_{sev}$ and a catalyst state $\rho_c$ such that $\rho_e'\succ\sigma_s$, if and only if $\sigma_{se}$ is a correlated state of the system $s$ and the environment $e$. Furthermore, they showed that if $\rho_e'$ majorises $\sigma_s$, then $Q_e\geq Q_e'$ \cite{henao2022catalytic}. Note that as the unitary is free in this process, the catalyst is the only party responsible for the reduction in heat dissipation to the environment. Furthermore, they showed $S(\rho_e')\leq S(\sigma_e)$ which reflects that the entropy change of the environment can also be reduced. Therefore, in any information erasure process involving a correlated state $\sigma_{se}$, it is possible to reduce the heat and entropy dissipation to the environment by using a suitable catalyst state.

\subsection{Other quantum resource theories}
\label{subsec:ExactCatalysis:OtherResource}

Symmetry arguments play a crucial part in understanding many important results in physics. For example, in many cases when the evolution of a system under some dynamics gets complicated to analyse, one can look for symmetry arguments to derive some useful results. In quantum theory, \citet{Marvian_2013} explored the role that symmetry plays in shaping the dynamics of a quantum system. More precisely, they explored state transformations in the context of time evolution with a given symmetry. To analyse this feature systematically, they introduced the resource theory of asymmetry in \cite{Marvian_2013}. Suppose a group $G$ with elements $g$ represents a set of symmetry transformations. The action of each group element $g$ is described by a unitary transformation $U_g$. In this resource theory, the free states are those states that remain invariant under the action of all $g$, i.e., $\mathcal{U}_g (\rho)=\rho$ $\forall g\in G$, where $\mathcal{U}_g(\rho)=U_g(\rho)U_g^\dagger$. In addition, a quantum operation $\Lambda$ is free if it respects the symmetry, i.e., 
\begin{equation}
\Lambda[U_{g}^{i}\rho(U_{g}^{i})^{\dagger}]=U_{g}^{o}\Lambda[\rho](U_{g}^{o})^{\dagger}\,\,\,\,\,\,\,\forall g\in G,
\end{equation}
where $U_g^i$ and $U_g^o$ represent the input and output representation of $G$, respectively. These types of operations are named as $G$-covariant operations in the literature \cite{Marvian_2013}. 

In such a resource theory, \citet{Marvian_2013} explored the role of catalysis for pure state transformations. If a state $\ket{\psi}$ cannot be transformed into $\ket{\phi}$ under any deterministic $G$-covariant operation but the same transformation can be achieved deterministically with a $G$-covariant operation and a catalyst in a state $\ket{\eta}$ such that $\ket{\phi}\otimes\ket{\eta}\rightarrow\ket{\psi}\otimes\ket{\eta}$, then such a transformation from $\ket{\psi}$ to $\ket{\phi}$ has been called a nontrivial example of catalysis \cite{Marvian_2013}. With this \citet{Marvian_2013} showed that no nontrivial example of catalysis is possible using a finite-dimensional catalyst when the symmetries are associated with compact connected Lie groups. However, this does not work for finite groups and a nontrivial example of catalysis has been shown for finite groups \cite{Marvian_2013}. 

Recently \citet{Wilde_classical} studied the role of catalysis in relative majorization for classical probability distributions. A pair $(p', q')$ of probability distributions $p'$ and $q'$ is relatively majorised by another pair $(p, q)$ of probability distributions $p$ and $q$, i.e., $(p,q)\succ(p', q')$ if and only if there exists a classical channel or a doubly stochastic matrix that transforms $p$ and $q$ to $p'$ and $q'$, respectively \cite{DAHL199953,Renes_2016, PhysRevA.95.012110}. In \cite{Wilde_classical}, the authors reformulated the above-mentioned transformation protocol in the presence of a catalyst pair $(r,s)$ of probability distributions $r$ and $s$. Now the input pair $(p,q)$ is used jointly with the catalyst pair $(r,s)$ to achieve the desired output pair $(p', q')$ by applying a classical channel $\Lambda$ on the joint pair $(p\otimes r, q\otimes r)$. Although, they allowed correlation on $\Lambda(p\otimes r)=t'$, but restrict that the first and second marginals of $t'$ should be $p$ and $r$, respectively. Furthermore, for the second one, they impose no correlation, i.e., $\Lambda(q\otimes s)=q' \otimes s$. As the catalyst is returned exactly it can be reused to transform another independent set of $(p,q)$. With this, the authors in \cite{Wilde_classical} showed that relative entropy completely determines exact catalytic pair transformations. Suppose $p, q, p', q'$ are probability distributions on the same alphabet with $(p,q)$ and $(p',q')$ having different relative spectra, and $q, q'$ have only rational entries and full support. Then the following two statements are equivalent: i) $D(p||q)>D(p'||q')$ and $D_0(p||q)\geq D_0(p'||q')$. ii) For all $\delta>0$, there exists a probability distribution $r$, and a joint probability distribution $t'$ with marginals $p'$ and $r$ such that  $(p\otimes r,q\otimes\nu)\succ(t',q'\otimes\nu)$, and $D(t'||p' \otimes r)<\delta$ \cite{Wilde_classical}. Here $D(p||q)=\sum_{x}p(x)\ln(p(x)/q(x))$, $D_0(p||q)=-\ln\sum_{x:p(x)\neq 0} q(x)$, and $\nu$ is a uniform distribution that is used in place of $s$. The relative spectrum of a pair $(p,q)$ is defined as $\{p(x)/q(x)\}_x$ \cite{Wilde_classical}.
Therefore, relative entropy completely characterises exact transformations under correlated catalysis. 

Note that the correlation can be made arbitrarily small such that $t'$ is almost the same as $p'\otimes r$. The constraints on different relative spectra, $q, q'$ are rational, and $D_0(p||q)\geq D_0(p'||q')$ can be lifted by allowing an infinitesimally small error on the target probability distribution $p'$ \cite{Wilde_classical}. Suppose $p, q, p', q'$ are probability distributions on the same alphabet with $q, q'$ have full support. Then the following two statements are equivalent: i) $D(p||q)\geq D(p'||q')$. ii) For all $\delta>0$ and $\varepsilon > 0$, there exists a probability distribution $r$, and a joint probability distribution $t_\varepsilon'$ with marginals $p_\varepsilon'$ and $r$, such that  $(p\otimes r, q\otimes r)\succ (t_\varepsilon',q'\otimes \nu)$, $||p_\varepsilon'-p'||_1\leq 2\varepsilon$, and $D(t_\varepsilon'||p_\varepsilon'\otimes r)<\delta$. Here, $||p-q||_1=\sum_{i}|p_i-q_i|$.  Considering $q$ and $q'$ to be the uniform distributions, a special case arises. Suppose $p, p'$ are probability distributions on the same alphabet with $p\neq\mathrm{Per}(p')$ for all permutation matrices $\mathrm{Per}$. Then the following two statements are equivalent: i) $H(p)\geq H(p')$. ii) For all $\delta>0$ and $\varepsilon > 0$, there exists a probability distribution $r$, and a joint probability distribution $t_\varepsilon'$ with marginals $p_\varepsilon'$ and $r$, such that  $p\otimes r\succ t_\varepsilon'$, $||p_\varepsilon'-p'||\leq 2\varepsilon$, and $D(t_\varepsilon'||p_\varepsilon'\otimes r)<\delta$ \cite{Wilde_classical}. Here $H$ represents the Shannon entropy of a probability distribution. Therefore, a single entropic condition completely dictates the approximate catalytic relative majorisation of classical probabilities. Note that this result can be applied to other quantum resource theories, for example, purity, quasiclassical thermodynamics, and asymmetric distinguishability \cite{Wilde_classical}. 

The role of catalysis has also been explored in the framework of Gaussian thermal resource theory \cite{CatGauss_2021}. Let us consider an $n$-mode Bosonic continuous variable quantum system on the Hilbert space $\mathcal{H}=\otimes_{k=1}^n\mathcal{H}_k$, with corresponding Bosonic creation and annihilation field operators $\hat{a}_k$ and $\hat{a}_k^\dagger$, respectively. From the field operators, the quadrature operators (or position and momentum-like operators) are constructed as $\hat{x}_k=(\hat{a}_k+\hat{a}_k^\dagger)/\sqrt{2}$ and  $\hat{p}_k=i(\hat{a}_k^\dagger-\hat{a}_k)/\sqrt{2}$. For convenience, all the quadrature operators are grouped together to make a column vector $\hat{\bold{r}}=(\hat{x}_1,\hat{p}_1,\cdots,\hat{x}_n,\hat{p}_n)^T$. For a given quantum state $\rho$, the corresponding first and second moments are defined as \cite{RevModPhys.84.621}
\begin{eqnarray}
 &&\bold{r}=(\langle\hat{x}_1\rangle_\rho,\langle\hat{p}_1\rangle_\rho,\cdots,\langle\hat{x}_n\rangle_\rho,\langle\hat{p}_n\rangle_\rho)^T\,\,\mbox{and},\\
 && \sigma_{ij}=\frac{\langle\{\hat{\bold{r}}_i,\hat{\bold{r}}_j\}\rangle_\rho}{2}-\langle \hat{\bold{r}}_i \rangle_\rho \langle \hat{\bold{r}}_j \rangle_\rho 
\end{eqnarray}
respectively, where the second moments $\sigma_{ij}$ are the elements of the covariance matrix $\sigma$. Gaussian states are completely characterized by their first moments and covariance matrix \cite{RevModPhys.84.621}. 

After defining Gaussian states, we introduce Gaussian operations that preserve Gaussianity, or, in other words, take Gaussian states to Gaussian states. Generally, a Gaussian operation can be realised by allowing an interaction between an environment in a Gaussian state and a given system through a global Gaussian unitary and then performing a projective measurement on the environment \cite{RevModPhys.84.621}. In the case of Gaussian thermal operations, the global unitary and the state of the environment should be energy-preserving and in a thermal state respectively. As Gaussian states are completely characterised by $\bold{r}$ and $\sigma$, it is enough to study the transformations of these two under Gaussian thermal operations. In \cite{CatGauss_2021}, the authors considered Gaussian states with $\bold{r}=0$ and studied the transformation of $\sigma$ under strong and weak catalytic Gaussian thermal operations. Note that the covariance matrix can be characterised by two more matrices $M$ and $A$ with elements $M_{ij}=\langle a_j^\dagger a_i \rangle-\delta_{ij}\nu$ and $A_{ij}=\langle a_j a_i \rangle$, respectively \cite{PhysRevA.49.1567} and this has been used to study state transformations under Gaussian thermal operations \cite{CatGauss_2021,PhysRevLett.124.010602}. Here $\nu=(\mathrm{e}^{\beta\omega}+1)/(\mathrm{e}^{\beta\omega}-1)$, $\omega$ is the eigenfrequency of the modes and $\beta=1/k T$ is the inverse temperature. A strong catalytic Gaussian transformation of $\rho$ to $\rho'$ is possible when there exists a catalyst state $\sigma$ such that $\rho \otimes \sigma\rightarrow\rho'\otimes\sigma$ is achieved under Gaussian thermal operation. On the other hand, a weak catalytic Gaussian transformation of $\rho$ to $\rho'$ is possible when there exists a catalyst state $\sigma$ and a joint final state $\nu^{SC}$ such that $\rho \otimes \sigma \rightarrow \nu^{SC}$ with $\Tr_S\nu = \sigma$ and $\Tr_C\nu = \rho'$ is accomplished under Gaussian thermal operation. Equipped with these tools, the authors in \cite{CatGauss_2021} studied transformations of single-mode Bosonic Gaussian states. The transition from $\sigma$ to $\sigma'$ under strong catalytic Gaussian thermal operation is possible if and only if there exists a $p\in[0,1]$ such that the conditions 
\begin{eqnarray}\label{gaussian1}
 \mu'=p \mu \,\,\, \mbox{and} \,\,\, \alpha'= p \alpha
\end{eqnarray}
hold \cite{CatGauss_2021}. Here $\sigma$ and $\sigma'$ are the covariance matrices of a single-mode Bosonic Gaussian state. For an $n$-mode Bosonic state, $\mu$ and $\alpha$ correspond to the eigenvalues and singular values of the matrix $M$ and $A$, respectively. Note that the conditions in Eq. (\ref{gaussian1}) are exactly the same as the non-catalytic transition presented in \cite{PhysRevLett.124.010602}. Hence, strong catalysis does not provide any advantage in this scenario. Now, if we allow correlations between the system and the catalyst then in the weak catalytic scenario, the above transition is possible if and only if there exists $p, q\in[0,1]$ with $p\geq q$ such that the conditions 
\begin{eqnarray}\label{gaussian2}
 \mu'=p \mu\ \,\, \mbox{and} \,\, \alpha'= q \alpha
\end{eqnarray}
are satisfied \cite{CatGauss_2021}. Furthermore, in \cite{CatGauss_2021} multi-mode state transformations have been discussed as well.

The role of quantum catalysts in the distillation of magic states has also been investigated recently \cite{PhysRevA.83.032317}. Magic states have been considered to be one of the key ingredients for achieving fault-tolerant quantum computation \cite{Knill_2004,Bravyi_PhysRevA.71.022316,Campbell_2017}. Generally, prepared states are noisy, and one needs to distill pure magic states from those to achieve fault-tolerant quantum computation. In magic state distillation a smaller number of more useful pure magic states have been distilled from multiple noisy quantum states \cite{Knill_2004,Bravyi_PhysRevA.71.022316,Campbell_2017,Veitch_2014}. Quantum catalysis has been shown to be useful in magic state distillation as well \cite{PhysRevA.83.032317}. Consider a state $\ket{\psi}\propto \ket{H_{0,0,0}}+\ket{H_{1,1,1}}$, where $\ket{H_{0,0,0}}=\ket{H_0}\otimes\ket{H_0}\otimes\ket{H_0}$, $\ket{H_{1,1,1}}=\ket{H_1}\otimes\ket{H_1}\otimes\ket{H_1}$. $\ket{H_0}$ and $\ket{H_1}$ are qubit magic states and also eigenvectors of the Hadamard matrix. Note that $\ket{\psi}\rightarrow\ket{H_0}$ cannot be achieved deterministically under Clifford unitary operations \cite{PhysRevA.83.032317}. However, if we have an additional state $\ket{H_0}$ in possession, the following transformation  $\ket{\psi}\otimes\ket{H_0}\rightarrow\ket{H_0}\otimes\ket{H_0}$ can be achieved deterministically with Clifford unitary operations \cite{PhysRevA.83.032317}.

Another important aspect in the field of quantum computation is quantum error correction. Recently, the role of catalysis in it has been discussed by \citet{IEEE_Brun}. Consider the following time evolution channel $\mathcal{N}:\mathcal{L}^{\otimes n}\rightarrow \mathcal{L}^{\otimes n}$ between Alice and Bob, which takes quantum states on Alice's system as inputs and provides quantum states on Bob's system as outputs. Here $\mathcal{L}$ represents the space of linear operators acting on the qubit Hilbert space $\mathcal{H}$. Note that here Alice and Bob are separated in time. Suppose Alice and Bob also have access to a noiseless channel $\mbox{id}^{\otimes c}$, and Alice has an initial quantum state of $k$ qbits. Equipped with this a catalytic quantum error correcting code can be defined as the following: i) Alice performs an encoding operation $\mathcal{E}: \mathcal{L}^{\otimes k}\otimes \mathcal{L}^{\otimes c}\rightarrow \mathcal{L}^{\otimes n}$, ii) After the time evolution Bob performs a decoding operation $\mathcal{D}:\mathcal{L}^{\otimes n}\otimes \mathcal{L}^{\otimes c}\rightarrow \mathcal{L}^{\otimes k} $ such that
\begin{equation}\label{cat_error_correction}
   \mathcal{D}\circ(\mathcal{N}\otimes \mbox{id}^{\otimes c})\circ \mathcal{E}=\mbox{id}^{\otimes k}=\mbox{id}^{\otimes k-c}\otimes \mbox{id}^{\otimes c}. 
\end{equation}
The net rate of this error correcting code is $(k-c)/n$ and the noiseless channel $\mbox{id}^{\otimes c}$ served as a catalyst in the process. If we additionally have access to $m$ copies of the channel $\mathcal{N}$, then the following result has been proved \cite{IEEE_Brun}: If Eq. (\ref{cat_error_correction}) holds, then for any arbitrary integer $m\geq0$ there exists an encoding operation $\mathcal{E}_m$ and a decoding operation $\mathcal{D}_m$ for the channel $\mathcal{N}^{\otimes m}$ such that
\begin{equation}
    \mathcal{D}_m\circ(\mathcal{N}^{\otimes m}\otimes \mbox{id}^{\otimes c})\circ \mathcal{E}_m=\mbox{id}^{\otimes m(k-c)}\otimes \mbox{id}^{\otimes c}.
\end{equation}
The above result shows that the size of the catalyst can be reduced from $mc$ to $c$.

In a recent study \cite{KarvonenPhysRevLett.127.160402}, the role of catalysis has been explored in the resource theory of contextuality \cite{Kochen1967,Budroni_2021} and non-locality \cite{BellRevModPhys.38.447,BrunnerRevModPhys.86.419}. \citet{KarvonenPhysRevLett.127.160402} considered the following kind of catalytic transformation: $p\otimes c \rightsquigarrow q\otimes c$, where $p$, $q$, and $c$ represent some correlations, and $q$ can be simulated ($\rightsquigarrow$) from $p$ with the help of a catalyst $c$ that remains intact in the process. Based on this, \citet{KarvonenPhysRevLett.127.160402} proved that catalysis does not help in the resource theory of contextuality and non-locality, or, in other words, there exists no correlation that can be used to catalyse an otherwise impossible transformation. More precisely, if $p\otimes c \rightsquigarrow q\otimes c$ is possible in the resource theory of contextuality and non-locality, then $p\rightsquigarrow q$ is also possible \cite{KarvonenPhysRevLett.127.160402}. 

\section{Approximate catalysis}
\label{sec:ApproxCatalysis}

The definition of exact catalysis, initially introduced for the resource theory of entanglement~\cite{JonathanPhysRevLett.83.3566}, does not allow for any errors in the transformation. In recent years, this notion has been extended, allowing for an error in the initial or final state~\cite{Brandao3275,Lostaglio2015} while at the same time requiring that the catalyst be completely decoupled from the primary system at the end of the process. In parallel, there has been a growing interest in correlated catalysis, allowing the catalyst to build up correlations with the system, which may also be preserved at the end of the procedure~\cite{AbergPhysRevLett.113.150402,Wilminge19060241,PhysRevX.8.041051,shiraishi2020quantum,ShiraishiPhysRevLett.128.089901,BoesPhysRevLett.122.210402,wilming2020entropy,Kondra2102.11136}. 

Recently, the framework of \emph{approximate catalysis} has been proposed in~\cite{Kondra2102.11136,Datta2022entanglement}. While this framework allows for correlations between the main system and the catalyst, it additionally requires that these correlations can be made arbitrarily small. In particular, an approximate catalytic transformation between two states $\rho^S$ and $\sigma^S$ of a system $S$ is possible if and only if for any $\varepsilon > 0$ there exists a catalyst state $\tau^{C}$ and a free operation $\Lambda_f$ acting on the system and the catalyst such that~\cite{Kondra2102.11136,Datta2022entanglement}
\begin{align}
\left\Vert \Lambda_{f}\left(\rho^{S}\otimes\tau^{C}\right)-\sigma^{S}\otimes\tau^{C}\right\Vert _{1} & \leq\varepsilon, \label{eq:ApproximateCatalysis-1}\\
\Tr_{S}\left[\Lambda_{f}\left(\rho^{S}\otimes\tau^{C}\right)\right] & =\tau^{C}. \label{eq:ApproximateCatalysis-2}
\end{align}
In contrast to the exact catalysis discussed in Section~\ref{sec:ExactCatalysis}, the above definition allows for correlations between the system and the catalyst in the final state $\mu^{SC} = \Lambda_{f}(\rho^{S}\otimes\tau^{C})$. However, Eq.~\eqref{eq:ApproximateCatalysis-1} ensures that these correlations can be made arbitrarily small by choosing an appropriate state of the catalyst $\tau^C$ and a free operation $\Lambda_f$. The condition in Eq.~\eqref{eq:ApproximateCatalysis-2} guarantees that the state of the catalyst does not change in the procedure. This ensures that the catalyst can be reused for the same transformation in the future. Moreover, this requirement makes it impossible to ``embezzle'' quantum resources by using small changes of the catalyst state, see also Section~\ref{sec:CatalyticEmbezzling}. 

An alternative framework for approximate catalysis has been proposed in  \cite{PhysRevX.8.041051,ShiraishiPhysRevLett.128.089901,Wilde_classical}.  While Eq.~(\ref{eq:ApproximateCatalysis-1}) represents that the correlations between the system and the catalyst can be made arbitrarily small as quantified by the trace norm, it is possible to formulate a similar requirement in terms of mutual information $I^{A:B}(\rho^{AB})=S(\rho^{A})+S(\rho^{B})-S(\rho^{AB})$. Then, an approximate catalytic transformation from $\rho^S$ into $\sigma^S$ is possible if and only if for any $\varepsilon > 0$ and any $\delta > 0$ there exists a free operation $\Lambda_f$ and a catalyst state $\tau^C$ such that~\cite{PhysRevX.8.041051,ShiraishiPhysRevLett.128.089901,Wilde_classical}
\begin{align}
\left\Vert \Tr_{C}\left[\Lambda_{f}\left(\rho^{S}\otimes\tau^{C}\right)\right]-\sigma^{S}\right\Vert _{1} & \leq\varepsilon, \label{eq:ApproximateCatalysisAlt1}\\
\Tr_{S}\left[\Lambda_{f}\left(\rho^{S}\otimes\tau^{C}\right)\right] & =\tau^{C},\label{eq:ApproximateCatalysisAlt2}\\
I^{S:C}\left[\Lambda_{f}\left(\rho^{S}\otimes\tau^{C}\right)\right] & \leq\delta.\label{eq:ApproximateCatalysisAlt3}
\end{align}
Recently, \citet{Rubboli2111.13356} demonstrated equivalence between this definition and the definition in Eqs.~(\ref{eq:ApproximateCatalysis-1}) and (\ref{eq:ApproximateCatalysis-2}).

Note that the term ``approximate catalysis'' has been previously used by~\citet{GOUR20151} in the resource theory of purity, referring to transformations which fulfill Eq.~(\ref{eq:ApproximateCatalysis-1}) for any $\varepsilon > 0$. Since the condition in  Eq.~\eqref{eq:ApproximateCatalysis-2} is not imposed, this type of transformations leads to embezzling phenomena~\cite{GOUR20151}, see also Section~\ref{sec:CatalyticEmbezzling}. Throughout this article, whenever we refer to approximate catalysis we mean transformations which obey Eqs.~(\ref{eq:ApproximateCatalysis-1}) and (\ref{eq:ApproximateCatalysis-2}). 

In the following, we will review recent results on approximate catalysis for concrete resource theories, including quantum thermodynamics, entanglement, and quantum coherence.

\subsection{Quantum thermodynamics} \label{sec:QuantumThermodynamicsApproximate}

Approximate catalysis has been first investigated in quantum thermodynamics by~\citet{PhysRevX.8.041051}. For the case that the initial and the final state are both block diagonal in the energy eigenbasis, it has been shown that state transformations are completely characterized by the Helmholtz free energy. In particular, a block diagonal state $\rho$ can be converted into another block diagonal state $\sigma$ via thermal operations with approximate catalysis if and only if~\cite{PhysRevX.8.041051}
\begin{equation}
    F(\rho) \geq F(\sigma). \label{eq:QuantumThermodynamicsApproximate}
\end{equation}

\citet{shiraishi2020quantum,ShiraishiPhysRevLett.128.089901} have investigated a closely related setting, replacing thermal operations with the more general class of Gibbs-preserving operations. In this case, Eq. (\ref{eq:QuantumThermodynamicsApproximate}) completely characterizes transformations between all quantum states, leading to the following theorem.

\begin{theorem}[See Theorem 1 in \cite{shiraishi2020quantum}] \label{thm:GibbsPreservingApproximate}
A state $\rho$ can be converted into another state $\sigma$ via Gibbs-preserving operations with approximate catalysis if and only if $F(\rho)\geq F(\sigma)$.
\end{theorem}

It is possible to express the Helmholtz free energy in terms of the quantum relative entropy as follows:
\begin{equation}
   F(\rho) = kT [ S(\rho||\gamma) -\log Z ], \label{RelaFS}
\end{equation}
where $Z$ is the partition function and $\gamma$ is the Gibbs state. Using Eq.~(\ref{RelaFS}), we see that Eq.~(\ref{eq:QuantumThermodynamicsApproximate}) is equivalent to
\begin{equation}
S(\rho||\gamma)\geq S(\sigma||\gamma). \label{eq:QuantumThermodynamicsApproximate2}
\end{equation}

The techniques developed by \citet{shiraishi2020quantum,ShiraishiPhysRevLett.128.089901} to prove the above theorem turned out to be applicable to other resource theories, and we will provide a sketch of the proof in the following. The major insight behind the proof is a connection between asymptotic transformations and approximate catalysis, which has been investigated earlier in \cite{DuanPhysRevA.71.042319}. Let us assume that $\rho$ can be converted into $\sigma$ via Gibbs-preserving operations in the asymptotic limit, i.e., for any $\delta > 0$ there exists an integer $n$ and a Gibbs-preserving operation $\Lambda$ acting on $n$ copies of $\rho$ such that 
\begin{equation}
\left\Vert \Lambda\left(\rho^{\otimes n}\right)-\sigma^{\otimes n}\right\Vert _{1} \leq \delta. \label{eq:GibbsPreservingAsymptotic}
\end{equation}
In the following $S_i$ will denote copies of the system $S$, and we further define $\Gamma = \Lambda(\rho^{\otimes n})$, which is a quantum state on $S_{1}\otimes S_{2}\otimes \cdots\otimes S_{n}$.

\citet{shiraishi2020quantum,ShiraishiPhysRevLett.128.089901} showed that Eq.~(\ref{eq:GibbsPreservingAsymptotic}) implies an approximate catalytic transformation from $\rho$ into $\sigma$. The catalyst achieving this transformation is given as
\begin{equation}
    \tau^C = \frac{1}{n}\sum_{k=1}^{n}\rho^{\otimes k-1} \otimes \Gamma_{n-k} \otimes \ket{k}\!\bra{k}. \label{eq:tau}
\end{equation}
Here, $\Gamma_0 = 1$, and for $i > 0$ the state $\Gamma_{i}$ is the reduced state of $\Gamma$ on $S_{1}\otimes S_{2}\otimes \cdots\otimes S_{i}$. The Hilbert space of the catalyst is in $S^{ \otimes{n-1}}\otimes K$, where $K$ represents an ancillary system of dimension $n$ with a fully degenerated Hamiltonian and basis states $\{\ket{k}\}_{k=1}^n$. 

The Gibbs-preserving operation achieving the desired conversion consists of the following three steps applied onto the total state $\rho^S \otimes \tau^C$~\cite{shiraishi2020quantum,ShiraishiPhysRevLett.128.089901}:

\begin{enumerate}
    \item A controlled Gibbs-preserving operation is performed, with the ancillary system $K$ as the control, and all other particles as the target. Conditioned on the ancilla being in the state $\ket{n}$, the Gibbs-preserving operation $\Lambda$ is applied onto $S_1 \otimes S_2 \otimes \cdots \otimes S_n$ such that Eq.~(\ref{eq:GibbsPreservingAsymptotic}) is fulfilled.
    
    \item A unitary is applied onto the ancillary system that converts $\ket{n}\xrightarrow{}\ket{1}$ and $\ket{i}\xrightarrow{}\ket{i+1}$.
    
    \item The systems are shifted as $S_{i}\xrightarrow{} S_{i+1}$ and $S_{n}\xrightarrow{} S_{1}$.
\end{enumerate}

The initial state of the system and the catalyst has the form
\begin{equation}
    \rho^S \otimes \tau^C = \frac{1}{n}\sum_{k=1}^{n}\rho^{\otimes k} \otimes \Gamma_{n-k} \otimes \ket{k}\!\bra{k}.
\end{equation}
After applying 1. step of the procedure, we obtain the following state:
\begin{equation}
    \mu_1 = \frac{1}{n}\sum_{k=1}^{n-1}\rho^{\otimes k} \otimes \Gamma_{n-k} \otimes \ket{k}\!\bra{k} + \frac{1}{n}\Gamma\otimes \ket{n}\!\bra{n}.
\end{equation}
In 2. step, $\mu_1$ is transformed into the state
\begin{equation}
    \mu_2 = \frac{1}{n}\sum_{k=1}^{n}\rho^{\otimes k-1} \otimes \Gamma_{n+1-k} \otimes \ket{k}\!\bra{k}. \label{eq:Muii}
\end{equation}
Tracing out $S_{n}$ from $\mu_2$ gives the initial state of the catalyst $\tau$, see Eq.~(\ref{eq:tau}). Therefore, the shift in 3. step transforms $\mu_2$ to the final state $\mu^{SC}$ having the property $\Tr_{S}[\mu^{SC}]$ = $\tau^C$, which implies that the condition~(\ref{eq:ApproximateCatalysis-2}) for approximate catalysis is fulfilled.

Using the protocol presented above, \citet{shiraishi2020quantum} further showed that the final system state $\mu^S$ can be made arbitrarily close to the desired state $\sigma^S$, and also the mutual information between the system and the catalyst in the total final state $\mu^{SC}$ can be made arbitrarily small whenever $\rho^S$ can be converted into $\sigma^S$ via Gibbs-preserving operations in the asymptotic limit. Thus, the procedure fulfills the criteria for approximate catalysis in Eqs.~(\ref{eq:ApproximateCatalysisAlt1})-(\ref{eq:ApproximateCatalysisAlt3}). Recalling that these criteria are equivalent to Eqs.~(\ref{eq:ApproximateCatalysis-1}) and (\ref{eq:ApproximateCatalysis-2}), we will now prove that the transformation presented above fulfills Eq.~(\ref{eq:ApproximateCatalysis-1}). For this, we will show that for any $\varepsilon >0$ the protocol described above can be performed such that
\begin{equation}
\left\Vert \mu^{SC}-\sigma^{S}\otimes\tau^{C}\right\Vert _{1}\leq\varepsilon. \label{eq:ApproximateCatalysisProof}
\end{equation}
In this part of the proof we will use the techniques presented for entanglement catalysis in~\cite{Kondra2102.11136,Lipka-Bartosik2102.11846}, adjusting them to the problem under study. Note that $\mu^{SC}$ is equivalent to the state $\mu_2$ in Eq.~(\ref{eq:Muii}) up to a cyclic shift. Thus, our figure of merit $||\mu^{SC} - \sigma^S \otimes \tau^C||_1$ is equal to $||\mu_2 - \nu||_1$, where the state $\nu$ has the form
\begin{equation}
    \nu = \frac{1}{n}\sum_{k=1}^{n}\rho^{\otimes k-1} \otimes \Gamma_{n-k} \otimes \sigma \otimes \ket{k}\!\bra{k}.
\end{equation}
Using the triangle inequality, Eq.~(\ref{eq:GibbsPreservingAsymptotic}), and the monotonicity of trace distance under partial trace we obtain
\begin{align}
\left\Vert \mu^{SC}-\sigma^{S}\otimes\tau^{C}\right\Vert _{1} & =\left\Vert \mu_2-\nu\right\Vert _{1}\\
 & =\frac{1}{n}\sum_{k=1}^{n}\left\Vert \Gamma_{n+1-k}-\Gamma_{n-k}\otimes\sigma\right\Vert _{1}\nonumber \\
 & \leq\frac{1}{n}\sum_{k=1}^{n}\left\Vert \Gamma_{n+1-k}-\sigma^{\otimes(n+1-k)}\right\Vert _{1}\nonumber \\
 & +\frac{1}{n}\sum_{k=1}^{n}\left\Vert \Gamma_{n-k}\otimes\sigma-\sigma^{\otimes(n+1-k)}\right\Vert _{1}\leq2\delta.\nonumber 
\end{align}
Setting $\delta = \varepsilon/2$ we see that Eq.~(\ref{eq:ApproximateCatalysisProof}) is fulfilled. This proves that the Gibbs-preserving operation presented above together with the catalyst state in Eq.~(\ref{eq:tau}) achieves the desired transformation. This holds true whenever $\rho$ can be converted into $\sigma$ via Gibbs-preserving operations in the asymptotic limit. 

\citet{shiraishi2020quantum,ShiraishiPhysRevLett.128.089901} further proved that a conversion from $\rho$ into $\sigma$ via asymptotic Gibbs-preserving operations is possible whenever $F(\rho) > F(\sigma)$ is fulfilled. For proving this statement, \citet{shiraishi2020quantum,ShiraishiPhysRevLett.128.089901} make use of the quantum Stein's lemma~\cite{Hiai1991,Ogawa887855} together with the properties of quantum hypothesis testing relative entropy and relative R\'enyi entropies~\cite{WangPhysRevLett.108.200501,Buscemi2019information,Datta4957651}. Note that this result has previously been proved in a more general setting, specifically for the resource theory of asymmetric distinguishability \cite{WangPRR2019}, and Gibbs-preserving operations can be considered as a special scenario of this resource theory. With the arguments presented above, it follows that for any states $\rho$ and $\sigma$ fulfilling  $F(\rho) > F(\sigma)$, a conversion from $\rho$ into $\sigma$ is possible via Gibbs-preserving operations with approximate catalysis~\cite{shiraishi2020quantum,ShiraishiPhysRevLett.128.089901}. By a simple continuity argument this result can be extended to the case $F(\rho) = F(\sigma)$~\cite{shiraishi2020quantum,ShiraishiPhysRevLett.128.089901}.

It is left to show that a conversion via Gibbs-preserving operations with approximate catalysis is not possible if Eq.~(\ref{eq:QuantumThermodynamicsApproximate}) is violated. For this, consider the following inequalities:
\begin{align}
S(\rho^{S}||\gamma^{S})+S(\tau^{C}||\gamma^{C}) & =S(\rho^{S}\otimes\tau^{C}||\gamma^{S}\otimes\gamma^{C}) \label{eq:GibbsPreservingSuperadditivity}\\
 & \geq S(\Lambda[\rho^{S}\otimes\tau^{C}]||\Lambda[\gamma^{S}\otimes\gamma^{C}])\nonumber \\
 & =S(\Lambda[\rho^{S}\otimes\tau^{C}]||\gamma^{S}\otimes\gamma^{C})\nonumber \\
 & \geq S(\mu^{S}||\gamma^{S})+S(\tau^{C}||\gamma^{C}).\nonumber 
\end{align}
The above result follows from the properties of the quantum relative entropy~\cite{WehrlRevModPhys.50.221,Capel8105839}: the second line follows from the data processing inequality, the third line from the fact that $\Lambda$ is a Gibbs-preserving operation, and the last line follows from superadditivity. Moreover, in the last line of Eq.~(\ref{eq:GibbsPreservingSuperadditivity}) we have used the state $\mu^{SC} = \Lambda[\rho^{S}\otimes\tau^{C}]$, noting that $\mu^C = \tau^C$. Noting that the Gibbs states $\gamma^S$ and $\gamma^C$ have full rank, the quantum relative entropies appearing in Eq.~(\ref{eq:GibbsPreservingSuperadditivity}) are all finite and continuous~\cite{Audenaert2005}. We arrive at $S(\rho^S||\gamma^S) \geq S(\mu^S||\gamma^S)$ for any state $\mu^S$ which is achievable from $\rho^S$ via Gibbs-preserving operations with approximate catalysis. By continuity, it follows that the same inequality also holds for any state $\sigma^S$ which can be achieved with an arbitrary small error as in Eq.~(\ref{eq:ApproximateCatalysis-1}). Recalling that Eq.~(\ref{eq:QuantumThermodynamicsApproximate}) is equivalent to Eq.~(\ref{eq:QuantumThermodynamicsApproximate2}), the proof of Theorem~\ref{thm:GibbsPreservingApproximate} is complete.

The results of~\citet{shiraishi2020quantum,ShiraishiPhysRevLett.128.089901} have also been used by~\citet{wilming2020entropy} to provide a proof of the catalytic entropy conjecture~\cite{BoesPhysRevLett.122.210402} in an approximate form, see also Section~\ref{subsec:ExactCatalysis:Purity}. In particular, for any two system states $\rho$ and $\sigma$ with $S(\rho) \leq S(\sigma)$ and any $\varepsilon > 0$ there exists a catalyst state $\tau$ and a unitary $U$ acting on the system and the catalyst such that~\cite{wilming2020entropy}
\begin{align}
\Tr_{S}\left[U\rho\otimes\tau U^{\dagger}\right] & =\tau,\\
\left\Vert \Tr_{C}\left[U\rho\otimes\tau U^{\dagger}\right]-\sigma\right\Vert _{1} & \leq\varepsilon.
\end{align}

Applications of catalysis in quantum thermodynamics have also been reported by \citet{Bartosik_2022}, who proposed a  notion of temperature of a non-equilibrium state, inspired by the ``zeroth law of thermodynamics''. They showed that for every quantum state there exist two effective temperatures -- hot and cold temperatures -- quantifying the ability of a quantum system to cool down or heat up via thermal operations. Precisely, \citet{Bartosik_2022} showed that, if the temperature of the bath is less than the cold temperature, then the system can never gain energy via any thermal operation (and vice versa). Interestingly, \citet{Bartosik_2022} showed that approximate catalysis leads to much colder or hotter effective temperatures.

\subsection{Entanglement} \label{subsec:ApproximateCatalysisEntanglement}

Approximate catalysis has also been studied within the resource theory of entanglement~\cite{Kondra2102.11136,Datta2022entanglement,Lipka-Bartosik2102.11846}. In this case, \citet{Kondra2102.11136} provided a full characterization of state transformations for all bipartite pure states. 
\begin{theorem}[See Eq. (4) in \cite{Kondra2102.11136}] \label{thm:EntropyAppCata}
Using LOCC and approximate catalysis, Alice and Bob can convert a state $\ket{\psi}^{AB}$ into $\ket{\phi}^{AB}$ if and only if
\begin{equation}\label{EntropyAppCata}
    S(\psi^A) \geq S(\phi^A).
\end{equation}
\end{theorem}
\noindent Similar to the method of \citet{shiraishi2020quantum,ShiraishiPhysRevLett.128.089901} for quantum thermodynamics, the proof of Theorem~\ref{thm:EntropyAppCata} presented by \citet{Kondra2102.11136} uses an analogy between approximate catalysis and asymptotic state transformations via LOCC. For this, recall that the optimal rate for transforming $\ket{\psi}^{AB}$ into $\ket{\phi}^{AB}$ via LOCC in the asymptotic limit is given by $S(\psi^A)/S(\phi^A)$~\cite{BennettPhysRevA.53.2046}. If Eq.~(\ref{EntropyAppCata}) is fulfilled, then asymptotic conversion from $\ket{\psi}^{AB}$ into $\ket{\phi}^{AB}$ is possible with rate at least one. It is then possible to construct a catalyst state and an LOCC protocol achieving the transformation with approximate catalysis, in analogy to the construction proposed by \citet{shiraishi2020quantum,ShiraishiPhysRevLett.128.089901} for quantum thermodynamics, see the previous Section~\ref{sec:QuantumThermodynamicsApproximate} for more details. We also note that the result in Theorem~\ref{thm:EntropyAppCata} does not assume that the state of the catalyst is pure, and in fact a catalyst in a mixed state is sometimes required to achieve the desired transformation~\cite{Kondra2102.11136,Datta2022entanglement}.

\citet{Kondra2102.11136} further used the properties of squashed entanglement~\cite{Christandl_2004} to prove that an approximate catalytic transformation is not possible if Eq.~(\ref{EntropyAppCata}) is violated. Squashed entanglement is defined as~\cite{Christandl_2004}
\begin{equation}
    E_\mathrm{sq}(\rho^{AB}) = \inf \left\{ \frac{1}{2} I(A;B|E):\rho^{ABE} \mathrm{\,\,extension\,\,of\,\,} \rho^{AB} \right\},
\end{equation}
with the quantum conditional mutual information $I(A;B|E)$. As has been shown by \citet{Christandl_2004}, squashed entanglement does not increase under LOCC, and moreover it fulfills 
\begin{equation}
    E_\mathrm{sq}\left(\rho^{AA'BB'}\right) \geq E_\mathrm{sq}\left(\rho^{AB}\right) +E_\mathrm{sq}\left(\rho^{A'B'}\right) \label{eq:SquashedEntanglementSuperadditive}
\end{equation}
with equality if $\rho^{AA'BB'} = \rho^{AB} \otimes \rho^{A'B'}$. For pure states, squashed entanglement coincides with the von Neumann entropy (with logarithm to the base 2) of the reduced state~\cite{Christandl_2004}. Using these properties,~\citet{Kondra2102.11136} proved the following:
\begin{theorem}[See Theorem 2 in \cite{Kondra2102.11136}] \label{thm:SquashedEntanglementMonotonic}
If a bipartite state $\rho^{AB}$ can be transformed into another state $\nu^{AB}$ via LOCC with approximate catalysis, then
\begin{equation}
    E_\mathrm{sq}(\rho^{AB}) \geq E_\mathrm{sq}(\nu^{AB}).
\end{equation}
\end{theorem}
We will present the proof of this theorem following the arguments of~\citet{Kondra2102.11136}. If $\rho^{AB}$ can be transformed into $\nu^{AB}$ with approximate catalysis, then for any $\varepsilon > 0$ there exists a catalyst state $\tau^{A'B'}$ and an LOCC protocol $\Lambda$ such that the final state $\sigma^{AA'BB'}=\Lambda(\rho^{AB}\otimes\tau^{A'B'})$ fulfills 
\begin{align}
\left\Vert \Tr_{A'B'}\left[\sigma^{AA'BB'}\right]-\nu^{AB}\right\Vert _{1} & <\varepsilon, \label{eq:SquashedEntanglementProof}\\
\Tr_{AB}\left[\sigma^{AA'BB'}\right] & =\tau^{A'B'}.
\end{align}
We further obtain
\begin{equation}
    E_\mathrm{sq}\left(\sigma^{AA'BB'}\right)\leq E_\mathrm{sq}\left(\rho^{AB}\otimes\tau^{A'B'}\right) =  E_\mathrm{sq}\left(\rho^{AB}\right)+E_\mathrm{sq}\left(\tau^{A'B'}\right), \label{1ineq}
\end{equation}
where in the inequality we used the fact that squashed entanglement does not increase under LOCC, and the equality follows from the additivity of $E_\mathrm{sq}$ on tensor products. From Eq.~(\ref{eq:SquashedEntanglementSuperadditive}) we further have
\begin{equation}
    E_\mathrm{sq}\left(\sigma^{AA'BB'}\right)\geq E_\mathrm{sq}\left(\Tr_{A'B'}\left[\sigma^{AA'BB'}\right]\right)+E_\mathrm{sq}\left(\tau^{A'B'}\right). \label{2ineq}
\end{equation}
Combining Eqs. (\ref{1ineq}) and (\ref{2ineq}) we obtain
\begin{equation}
    E_\mathrm{sq}\left(\rho^{AB}\right)\geq E_\mathrm{sq}\left(\Tr_{A'B'}\left[\sigma^{AA'BB'}\right]\right).
\end{equation}
The continuity of squashed entanglement~\cite{Alicki_2004} together with Eq.~(\ref{eq:SquashedEntanglementProof}) implies
$E_\mathrm{sq}\left(\rho^{AB}\right) \geq  E_\mathrm{sq}\left(\nu^{AB}\right)$, which completes the proof of Theorem~\ref{thm:SquashedEntanglementMonotonic}. 

Theorem~\ref{thm:SquashedEntanglementMonotonic} together with the faithfulness of squashed entanglement \cite{Li_2014} implies that it is not possible to create entanglement from separable states by using LOCC and approximate catalysis.

The proof of Theorem~\ref{thm:EntropyAppCata} presented in~\cite{Kondra2102.11136} uses the fact that for bipartite pure states $\ket{\psi}^{AB}$ and $\ket{\phi}^{AB}$ fulfilling $S(\psi^A) > S(\phi^A)$ asymptotic conversion from $\ket{\psi}^{AB}$ into $\ket{\phi}^{AB}$ is possible with unit rate~\cite{BennettPhysRevA.53.2046}. States with such properties have also been studied by~\citet{Bennett_PhysRevA.63.012307}, where this feature has been called \emph{asymptotic reducibility}. A general relation between asymptotic reducibility and approximate catalysis has been established by~\citet{Kondra2102.11136} for multipartite systems and multipartite LOCC protocols.
\begin{theorem} [See Theorem 1 in \cite{Kondra2102.11136}] \label{thm:ApproximateCatalysisEntanglement1}
If $\rho$ is asymptotically reducible to $\sigma$, then $\rho$ can be transformed into $\sigma$ via LOCC with approximate catalysis.
\end{theorem}

A stronger form of asymptotic reducibility can be constructed where the parties can additionally use a sublinear amount of entanglement with the standard LOCC protocol to accomplish the same transformation, which is denoted by asymptotic $\mathrm{LOCC}_q$ reducibility \cite{Bennett_PhysRevA.63.012307}. Similarly, catalytic asymptotic reducibility $\mathrm{LOCC}_c$ can be defined as follows: A state $\ket{\phi}$ is asymptotically $\mathrm{LOCC}_c$ reducible to a state $\ket{\phi}$ if there exists a state $\eta$ and a $n$ such that $(\ket{\psi}\otimes\ket{\eta})^{\otimes n}$ can be converted arbitrarily close to $(\ket{\phi}\otimes\ket{\eta})^{\otimes n}$. \citet{Bennett_PhysRevA.63.012307} showed that $\mathrm{LOCC}_c$ is at least as powerful as $\mathrm{LOCC}_{q}$. However, the converse i.e., whether $\mathrm{LOCC}_c$ can be simulated by $\mathrm{LOCC}_q$, has been shown only for some special cases \cite{Bennett_PhysRevA.63.012307}. 

In the context of $\mathrm{LOCC}_c$ reducibility, a recent result on the catalysis of arbitrary entangled states compares single-copy catalysis with asymptotic catalysis \cite{Datta2022entanglement}. A transformation $\rho\rightarrow\sigma$ on system $S$ is possible via asymptotic catalysis if and only if for any $\varepsilon>0$ and any $\delta > 0$ there exist integers $n$ and $m$ with $n>m$, a catalyst state $\tau^C$ and an LOCC protocol $\Lambda$ such that\footnote{The LOCC protocol $\Lambda$ acts on $n$ copies of $S$ and the catalyst $C$, and produces a state on $m$ copies of $S$ and the catalyst.}
\begin{subequations} \label{eq:LambdaCatalyticAsymptotic}
\begin{align}
\left\Vert \Lambda\left[\rho^{\otimes n}\otimes\tau^{C}\right]-\sigma^{\otimes m}\otimes\tau^{C}\right\Vert _{1} & \leq\varepsilon,\label{asymptotic}\\
\mbox{Tr}_{S^{\otimes m}}\left[\Lambda\left(\rho^{\otimes n}\otimes\tau^{C}\right)\right] & =\tau^{C},\\
\frac{m}{n}+\delta &\geq 1. \label{rate}
\end{align}
\end{subequations}
With this \citet{Datta2022entanglement}
showed that state transformations under asymptotic catalysis do not provide any advantage over single-copy catalysis. To be precise, if a state transformation $\rho\rightarrow\sigma$ can be achieved via asymptotic catalysis with unit rate, then the same transformation can be realised with single-copy catalysis too.  

\citet{Datta2022entanglement} also considered asymptotic settings beyond the independent and identically distributed (``iid'') case,
where the quantum state to be transformed asymptotically is no longer assumed to be just the tensor product $\ket{\psi}^{\otimes n}$. In particular, 
\citet{Datta2022entanglement} considered a pure-state distillation scenario,
in which Alice and Bob share a more general total state $\otimes_{i} \ket{\psi_i}^{AB}$,
where each $\ket{\psi_i}^{AB}$ are bipartite states that need not be identical.
As we will see, approximate catalysis provides, in this setting, a significant advantage over non-catalytic LOCC transformations.

By definition, 
a singlet state $\ket{\phi_{2}^{+}}$ can be extracted from a sequence of states $\{\ket{\psi_i}^{AB}\}$ with fidelity $f$ and probability $p$,
if there exists an integer $n$ and a probabilistic LOCC protocol $\Lambda$ such that~\cite{Datta2022entanglement}
\begin{equation}
    \Tr \left( \Lambda \left[ \otimes_{i=1}^{n} \psi_i \right]\right) = p, \quad
    \bra{\phi_{2}^{+}} \Lambda \left[ \otimes_{i=1}^{n} \psi_i \right] \ket{\phi_{2}^{+}} = pf,
\end{equation}
where we set $\psi_i = \ket{\psi_{i}}\!\bra{\psi_{i}}^{AB}$ for brevity.
Compare also Eqs.~(34) and (35) in \cite{Datta2022entanglement}. Moreover, in the context of pure state entanglement as the relevant resource, we can say that the same sequence allows for catalytic extraction of $m$ singlets if the conditions provided in Eqs. \eqref{eq:ApproximateCatalysis-1} and \eqref{eq:ApproximateCatalysis-2} are satisfied, with $\sigma^{S} = \ket{\phi_{2}^{+}}\!\bra{\phi_{2}^{+}}^{\otimes m}$ and $\Lambda$ being an LOCC protocol. With this, \citet{Datta2022entanglement} proved the following:
\begin{theorem}[See Theorem 3 in \cite{Datta2022entanglement}]
\label{thm:CatalysisBeyondIID}
Given fidelity $f>\frac{1}{2}$ and any $\varepsilon>0$, 
there is a sequence of two-qubit states $\{\ket{\psi_i}^{AB}\}$ such that for any $n \geq 1$,
the probability of converting the state $\otimes_{i=1}^n\ket{\psi_i}^{AB}$
into a singlet, via LOCC with fidelity $f$, is smaller than $\varepsilon$.
At the same time, an unbounded number of singlets can be extracted from $\otimes_{i=1}^n\ket{\psi_i}^{AB}$ with the help of catalysis, as $n\rightarrow \infty$.
\end{theorem}
The proof is by explicit construction~\cite{Datta2022entanglement}.
The sequence $\ket{\psi_i}^{AB}$ given in the proof is designed in such a way that although the state $\otimes_{i=1}^n\ket{\psi_i}^{AB}$ stays arbitrarily close to a product state with respect to systems $A$ and $B$,
its total local entropy grows indefinitely as $n \rightarrow \infty$, see \cite{Datta2022entanglement} for more details. The above results demonstrate clearly the advantage of an approximate catalytic transformation over all non-catalytic LOCC protocols.
Whereas for the two-qubit sequence specified in the proof of Theorem~\ref{thm:CatalysisBeyondIID},
the amount of entanglement available via catalytic transformation is unbounded, and not a single maximally entangled state can be extracted with reasonable probability and fidelity without catalysis. It is worth mentioning that entanglement distillation from non-iid sequences of states has been investigated earlier in~\cite{Bowen4567558,Buscemi1.3483717,WaeldchenPhysRevLett.116.020502}.

\citet{EisertPhysRevLett.85.437} have investigated the role of catalysis for approximate transformations between quantum states via LOCC, requiring however that the catalyst is not correlated with the primary system at the end of the procedure. In particular, \citet{EisertPhysRevLett.85.437} demonstrated that catalysis is useful for approximately converting a mixed entangled state into a pure state. For this, \citet{EisertPhysRevLett.85.437} considered the approximate conversion fidelity
\begin{align}
F_{\mathrm{LOCC}}(\rho & \rightarrow\sigma)=\max_{\Lambda\in\mathrm{LOCC}}F(\Lambda[\rho],\sigma),\\
F_{\mathrm{CLOCC}}(\rho & \rightarrow\sigma)=\max_{\Lambda\in\mathrm{CLOCC}}F(\Lambda[\rho],\sigma),
\end{align}
where $\mathrm{CLOCC}$ represents catalytic LOCC. \citet{EisertPhysRevLett.85.437}
proved that the conversion fidelity for CLOCC can exceed the same for LOCC when some initial mixed states $\rho$ are being converted into a pure target state $\ket{\zeta}$: 
\begin{equation}
F_{\mathrm{CLOCC}}\left(\rho\rightarrow\ket{\zeta}\!\bra{\zeta}\right)>F_{\mathrm{LOCC}}\left(\rho\rightarrow\ket{\zeta}\!\bra{\zeta}\right). \label{eq:ApproximateCatalysisEisert}
\end{equation}
To prove the statement, \citet{EisertPhysRevLett.85.437} used the state $\rho$ given in Eq. (\ref{mixed_example1}) with $\ket{\omega}$ given in Eq. (\ref{pure_example3}) and 
\begin{eqnarray}
 \ket{\psi}&=&\varepsilon\left(\sqrt{0.4}\ket{00}+\sqrt{0.4}\ket{11}+\sqrt{0.1}\ket{22}+\sqrt{0.1}\ket{33}\right)+\nonumber\\
 &&\sqrt{1-\varepsilon^2}\ket{44}.
\end{eqnarray}
Setting $\ket{\zeta}=\ket{\phi}$ given in Eq. (\ref{pure_example2}), \citet{EisertPhysRevLett.85.437} proved that the inequality given in Eq. (\ref{eq:ApproximateCatalysisEisert}) holds. The above equation holds for all $\varepsilon\in(\varepsilon',1)$ with some appropriate $\varepsilon'\in(0,1)$ and independent of the value of $\nu\in(0,1)$.

Early results on the role of entanglement catalysis in the presence of noise have also been presented by \citet{VidalPhysRevA.62.012304}. Suppose a catalyst in the state $\ket{\eta}$ improves the performance of the approximate transformation $\ket{\psi}\rightarrow\ket{\phi}$, i.e.,
\begin{equation}
    D(\psi\otimes\eta\rightarrow\phi\otimes\eta)<D(\psi\rightarrow\phi),
\end{equation}
where $D(\rho\rightarrow\sigma)=\min_{\Lambda}D(\Lambda(\rho),\sigma)$, $\Lambda$ represents the set of LOCC operations, and $D(\rho,\sigma)=||\rho-\sigma||_1/2$ is the trace distance. The quantity 
\begin{equation}
    \delta=D(\psi\rightarrow\phi)-D(\psi\otimes\eta\rightarrow\phi\otimes\eta)
\end{equation}
can thus be regarded as a quantifier of the catalytic advantage in this procedure. \citet{VidalPhysRevA.62.012304} have investigated how much noise is allowed in the initial state so that the catalytic advantage remains, i.e., $\delta>0$. Suppose the initial state $\ket{\psi}\otimes\ket{\eta}$ is distorted to $\rho$ such that $D(\rho,\psi\otimes\eta)=\varepsilon$. 
If $\varepsilon<\delta$, it has been shown that the catalytic advantage is preserved as $ D(\rho\rightarrow\phi\otimes\eta)<D(\psi\rightarrow\phi)$ \cite{VidalPhysRevA.62.012304}.

The role of catalysis in the approximate transformation of bipartite quantum channels has been discussed recently in \cite{IEEE_Kim}. In fact, \citet{IEEE_Kim} explored the dynamic resource theory of entanglement using the superchannel theory and also discussed the one-shot catalytic dynamic entanglement cost of bipartie quantum channels. More precisely, the authors provided a lower and an upper bound for the one-shot catalytic dynamic entanglement cost
of an arbitrary bipartite channel. Before presenting the bounds, in the following, we introduce the one-shot catalytic dynamic entanglement cost of a bipartite channel $\mathcal{N}_{AB}$ for a given $\delta>0$ and $\varepsilon\geq 0$ \cite{IEEE_Kim}: 
\begin{eqnarray}
 E_{C,\delta\mathrm{\mbox{-}SEPPSC}}^\varepsilon(\mathcal{N}_{AB})&:=&\min\{\log_2 m^2\,: \\
 &&\!\!\!\!\!\!\!\!\Theta_{A'B'CD\rightarrow ABCD} (\mathcal{F}_{A'B'}^m\otimes\mathcal{F}_{CD}^n)=\mathcal{N}'_{AB}\otimes\mathcal{F}_{CD}^n,\nonumber\\
 &&\!\!\!\!\!\!\!\!\Theta_{A'B'CD\rightarrow ABCD}\in \delta\mathrm{\mbox{-}SEPPSC},\nonumber\\
 &&\!\!\!\!\!\!\!\! ||\mathcal{N}_{AB}-\mathcal{N}'_{AB}||_{\diamond}\leq 2\varepsilon\}\nonumber,
\end{eqnarray}
where $m,n$ represent natural numbers and $\Theta_{A'B'CD\rightarrow ABCD}$ is a $\delta\mbox{-}$separability-preserving superchannel ($\delta\mathrm{\mbox{-}SEPPSC}$) from $A'B'CD$ to $ABCD$. For a given $\delta>0$, a superchannel $\Theta_{AB\rightarrow A'B'}$ is $\delta\mathrm{\mbox{-}SEPPSC}$ when $R(\Theta_{AB\rightarrow A'B'}[\mathcal{S}_{AB}])\leq \delta$ holds for all separable channels $\mathcal{S}_{AB}$, where $R$ is the generalised robustness with respect to the separable channels \cite{IEEE_Kim}.  Here $\mathcal{F}^m_{AB}$ denotes $m$-SWAP channel which can be represented by the application of $m$-SWAP gate $F_{AB}^m=\sum_{i,j=0}^{m-1}\ket{ij}\!\bra{ji}_{A_0B_0\rightarrow A_1B_1}$. With this, we are ready to introduce the following bounds. For $\delta>0$, $\varepsilon\geq 0$, there exists a natural number $n>0$ with $n^2\geq (\delta+1)/\delta$ such that the one-shot catalytic dynamic entanglement cost of an arbitrary bipartite channel $\mathcal{N}_{AB}$ is bounded from below and above as \cite{IEEE_Kim}
\begin{eqnarray}
&& R_L^\varepsilon(\mathcal{N}_{AB}\otimes\mathcal{F}_{CD}^n)-\log_2 n^2-\log_2(1+\delta)\nonumber\\
&&\qquad\qquad \leq E_{C,\delta\mathrm{\mbox{-}SEPPSC}}^\varepsilon(\mathcal{N}_{AB})\leq\\
&& R_L^{\varepsilon'}(\mathcal{N}_{AB}\otimes\mathcal{F}_{CD}^n)-\log_2 n^2-\log_2(1-2\varepsilon')+2,\nonumber
\end{eqnarray}
where $\varepsilon'=\varepsilon^2/(2|A_0|^2|B_0|^2)$, $R_L^\varepsilon(\mathcal{N}_{AB})=\min_{\mathcal{N}'_{AB}}R_L(\mathcal{N}'_{AB})$, with $||\mathcal{N}_{AB}-\mathcal{N}'_{AB}||_{\diamond}\leq 2\varepsilon$, and $R_L(\mathcal{N}_{AB})=\log_2(1+R(\mathcal{N}_{AB}))$.

\subsection{Quantum coherence}

As discussed previously in Section~\ref{subsec:ExactCatalysis:Coherence}, correlated catalysis does not help in coherence broadcasting under covariant operations \cite{LostaglioPhysRevLett.123.020403, Marvian_2019}. Due to these limitations on coherence transformation under covariant operations, \citet{Takagi2106.12592} investigated a different kind of catalyst known as marginal catalysts \cite{Lostaglio2015, Wilminge19060241} that could help in enhancing covariant transformations. Marginally catalytic covariant transformations are defined as follows: Suppose $\rho$ is transformed into $\rho'$ via a marginal catalytic conversion. Then there exists a covariant operation $\Lambda$, a constant $N$ and a finite set of states $\otimes_{j=1}^N\tau_{C_j}$ such that \cite{Takagi2106.12592}
\begin{eqnarray}
    &&\Lambda\left(\rho\otimes_{j=1}^N\tau_{C_j}\right)=\rho'\otimes \tau_{C_1\cdots C_N}\,\,\mbox{and}\\ 
    && \Tr_{\bar{C_j}}\tau_{C_1\cdots C_N}=\tau_{C_j} \,\, \mbox{for all}\,\, j,
\end{eqnarray}
where $\bar{C_j}$ represents all the systems except $C_j$. Although in the final state, there are correlations among different catalyst systems, the marginal state of each catalyst system remains unchanged. Therefore, the change in the global state of the catalyst produces an embezzling-like phenomenon. For more details on embezzling see Section \ref{sec:CatalyticEmbezzling}. With this  \citet{Takagi2106.12592} showed that marginal catalysts can provide enormous power in state transformations under covariant transformations. Indeed for any state $\rho$, $\rho'$ and $\varepsilon>0$, $\rho$ can be transformed into a state $\rho_\varepsilon'$ that is arbitrarily close to $\rho'$ in trace distance, i.e., $||\rho'-\rho_\varepsilon'||_1\leq 2\varepsilon$ by a marginal-catalytic covariant transformation. Hence, marginal catalysts exceptionally simplify coherence transformations in this theory. Similar results for qubit states have been presented earlier by \citet{Ding2021}. 

\citet{Fang_2018} studied one-shot probabilistic coherence distillation with catalytic assistance. The idea is to transform a coherent state $\rho$ close to a maximally coherent state $\ket{\Psi_N}=\sum_{i=0}^{N-1}\ket{i}/\sqrt{N}$ of dimension $N$ with the help of a catalyst $\eta$. In such a scenario, \citet{Fang_2018} showed that the maximum success probability of coherence distillation is given by the maximum value of $p$, such that
\begin{eqnarray}
 &&\Lambda(\rho\otimes\eta)=\left(p\ket{0}\!\bra{0}\otimes\sigma+(1-p)\ket{1}\!\bra{1}\frac{\openone}{m}\right)\otimes\eta,\\
 &&F(\sigma,\Psi_m)\geq 1-\varepsilon.
\end{eqnarray}
The result holds if $\Lambda$ is an incoherent operation, but it can also be applied to other classes of free operations studied in the resource theory of coherence~\cite{Fang_2018,StreltsovRevModPhys.89.041003}. Note that the catalyst state remains unchanged regardless of the outcome. However, this constraint was not enforced in the previous probabilistic transformation protocol \cite{BuPhysRevA.93.042326}. With the following example, a significant improvement has been shown in probabilistic coherence distillation. Consider the initial state
\begin{equation}
\rho=\frac{1}{2}(\ket{u_{1}}\!\bra{u_{1}}+\ket{u_{2}}\!\bra{u_{2}})
\end{equation}
with 
\begin{align}
\ket{u_{1}} & =\frac{1}{2}(\ket{00}-\ket{01}-\ket{10}+\ket{11}),\\
\ket{u_{2}} & =\frac{1}{5\sqrt{2}}(2\ket{00}+6\ket{01}-3\ket{10}+\ket{11}).
\end{align}
The probability of distilling the state $\ket{\Psi_2}$ (with an error $\varepsilon\leq 0.01$) from $\rho$ via dephasing-covariant incoherent operations (DIO) \cite{PhysRevLett.117.030401,ChitambarPhysRevA.94.052336,Marvian_PhysRevA.94.052324} can be increased significantly with the help of a catalyst in the state $\ket{\Psi_2}$ \cite{Fang_2018}. In contrast to the deterministic-exact catalytic transformation where maximally coherent states are useless~\cite{PhysRevA.91.052120}, this result shows that a maximally coherent state can be used as a catalyst in stochastic-approximate transformations \cite{Varun2111.12646}.    

Recently,~\citet{Chen2019} studied the role of catalysis in one-shot coherence distillation under incoherent operations. In a one-shot catalytic coherence distillation protocol, the distillable coherence of $\rho$ can be defined as follows:
\begin{equation}\label{coherence_distillation}
   \mathcal{C}_{\mathcal{O},c}^\varepsilon(\rho)=\max_{\Lambda\in\mathcal{O}}\max_{\dim\eta<\infty}\{\log_2 N\,\,:\,\, P(\Lambda(\rho\otimes\sigma),\Psi_N\otimes\sigma)\leq\varepsilon\}
\end{equation}
where $\rho$ is the initial state, $\mathcal{O}$ corresponds to different classes of incoherent operations, and $P$ is the purified distance given as $P(\rho,\sigma)=\sqrt{1-F(\rho,\sigma)}$. This definition has been adapted from the recently introduced one-shot coherence distillation protocol \cite{Regula_2018,Zhao_2019}. Note that there is no further restriction on the catalyst $\sigma$ except that the dimension is finite. In this setup, one can extract as much coherence as one wishes from any initial state $\rho$ with some arbitrary small error $\varepsilon$. As there is no restriction on the catalyst, in principle the catalyst can change slightly in this procedure keeping the error within $\varepsilon$. Since the dimension of the catalyst could be arbitrarily large, a small change in coherence in it can provide a large value of $C_{\mathcal{O},c}^\varepsilon(\rho)$. This phenomenon is known as embezzling and is discussed in more detail in Section \ref{sec:CatalyticEmbezzling}. As the previous case trivializes the transformation, in the next case \citet{Chen2019} studied coherence distillation when the dimension of the catalyst is not more than $M$. In this case, they provide a lower bound on the amount of coherence that can be distilled. In fact, for an integer $M$ and a positive number $\varepsilon$ with the constraint $\varepsilon^2\log_2(M-1)/4\geq 1$, the distillable coherence of a state $\rho$ can be lower bounded as
\begin{equation}
    C_\varepsilon(\rho)\geq \frac{\varepsilon^2[\log_2(M-1)+1]-2}{2}.
\end{equation}
This result holds for incoherent operations, and can also be extended to other operations investigated in the resource theory of coherence \cite{Chen2019}. However, this bound appears to be loose as it is independent of the coherence of the initial state $\rho$, and a better bound may be possible. To avoid such embezzling they proposed a perfect catalysis protocol with pure catalyst states where the catalyst is returned exactly, i.e., $\Tr_S\Lambda(\rho\otimes\eta)=\eta$ and the error is allowed only in the final state, i.e., $P(\Tr_C\Lambda(\rho\otimes\eta),\Psi_N)\leq \varepsilon$. As shown in \cite{Chen2019}, a catalyst in a pure state does not provide any advantage over standard one-shot coherence distillation where no catalyst is present to help. It is an open question whether the result is also true for a catalyst in a mixed state.  

Recently, \citet{char2021catalytic} explored the approximate catalytic state transformations in coherence theory. Their results provide a connection between asymptotic transformation and single copy transformation with a catalyst. In fact, the following has been shown: If there exists an asymptotic incoherent operation transforming a state $\rho$ into $\sigma$ with unit rate, then $\rho$ can be transformed into $\sigma$ via incoherent operations with approximate catalysis~\cite{char2021catalytic}. An analogous result in entanglement theory has been reported earlier by~\citet{Kondra2102.11136}, see also Theorem~\ref{thm:ApproximateCatalysisEntanglement1}.
Furthermore, \citet{char2021catalytic} showed that the relative entropy of coherence is a monotone under incoherent operations with approximate catalysis. A direct consequence of these results gives a necessary and sufficient condition for pure state transformations under incoherent operations with approximate catalysis. Particularly, the transformation from $\ket{\psi}$ to $\ket{\phi}$ is possible if and only if
\begin{equation}
    S(\Delta[\psi])\geq S(\Delta[\phi]). \label{eq:CoherenceCatalysis}
\end{equation}
The above result is analogous to the result in entanglement theory describing catalytic transformations for pure bipartite states~\cite{Kondra2102.11136}, see also Theorem~\ref{thm:EntropyAppCata}. \citet{char2021catalytic} prove Eq.~(\ref{eq:CoherenceCatalysis}) making use of the methods proposed by \citet{shiraishi2020quantum} in quantum thermodynamics, see Section \ref{sec:QuantumThermodynamicsApproximate}.

\subsection{Other quantum resource theories}

One of the ways to quantify the amount of resource present in a state is the minimum amount of noise or randomness required to make it a free state \cite{PhysRevA.72.032317}. Recently, this idea has been put forward for general resource theories with the additional freedom of allowing free states to help in the process as a catalyst \cite{Anshu_2018}. In such a scenario, we want to quantify the resource content of a state $\rho$, or in other words, the minimum amount of noise needed to make it a free state. To achieve this, we add an extra system in a free state $\rho_f$ to it and apply free operations on the joint state $\rho\otimes\rho_f$, such that the final state $\Lambda_f(\rho \otimes\rho_f)$ becomes arbitrarily close to a free state of the form $\sigma_f\otimes\rho_f$, where $\sigma_f$ is a free state. The last point ensures that the additional state $\rho_f$ remains almost the same in the procedure and, thus, serves as a catalyst. However, note that there is no additional constraint on the catalyst, like Eq. (\ref{eq:ApproximateCatalysis-2}), which ensures that the catalyst remains exactly the same. In any case, as free states (which are not expensive) are being used as catalysts, slight changes in them do not prove to be costly. Equipped with this, \citet{Anshu_2018} showed that the asymptotic randomness rate of catalytic transformation, or the resource content of a state $\rho$, is given by the regularised relative entropy of the state 
\begin{equation}
    E^{\infty}(\rho)=\lim_{n\rightarrow\infty}\frac{E(\rho^{\otimes n})}{n},
\end{equation}
where $E(\rho)=\inf_{\sigma\in\mathcal{F}}S(\rho||\sigma)$ and $\mathcal{F}$ represents the set of free states. Furthermore, the resource content of a state in a single copy regime has also been discussed in \cite{Anshu_2018}. As the above result has been derived in a general resource theoretic framework, it is applicable to resource theories of entanglement, coherence, purity, and asymmetry \cite{Anshu_2018}. 

While the preceding discussion shows when an approximate catalytic transformation is possible, it does not provide any information about the catalyst's resource content. \citet{Rubboli2111.13356} recently provided some positive answers in this direction for any general resource theory where certain monotones are additive on the tensor product. Assume that a quantity $\mathcal{D}$ completely characterizes approximate catalytic state transformations in a resource theory, i.e., a transformation from $\rho$ to $\sigma$ is possible under approximate catalysis iff $\mathcal{D}(\rho)\geq \mathcal{D(\sigma)}$, where $\mathcal{D}(\rho)=\min_{\rho'\in \mathcal{F}}S(\rho||\rho')$. Keeping this in mind, we can consider $\mathcal{D}$ as a relevant resource measure to quantify the catalyst's resourcefulness. In fact, the order of the resource content of a catalyst has been provided for a given error in \cite{Rubboli2111.13356}. 

\begin{theorem}[See Theorem 1 in \cite{Rubboli2111.13356}]\label{resource content catalyst}
Consider two quantum states $\rho$ and $\sigma$, and $\alpha \in [1/2,1)$ such that $\mathcal{D}_\alpha(\rho)<\mathcal{D}_\alpha(\sigma)$ and $\mathcal{D}_\alpha$ is additive for the state $\sigma$. Then, for any given error $\varepsilon$ in an approximate catalytic transformation of $\rho$ into $\sigma$ with catalyst $\eta$, we have
\begin{equation}
    \mathcal{D}(\eta)=\Theta\left(\log \frac{1}{\varepsilon}\right).
\end{equation}
\end{theorem}
Here, $\mathcal{D}_\alpha(\rho)=\min_{\rho'\in \mathcal{F}}S_{\alpha}(\rho||\rho')$ and $S_{\alpha}(\cdot||\cdot)$ is defined earlier in Eq.~(\ref{eq:RenyiRelEntropy}). Therefore, the catalyst's resourcefulness grows as the error reduces, and in the limit of zero error, the catalyst's resource content is unbounded. In \cite{Rubboli2111.13356}, the authors also derived a quantitative upper bound on the resource content of a catalyst. Furthermore, in \cite{Rubboli2111.13356}, it has been discussed that the additivity assumption, combined with $\mathcal{D}_\alpha(\rho)<\mathcal{D}_\alpha(\sigma)$ for some $\alpha \in [1/2,1)$, implies that the approximate catalytic transformation of $\rho$ into $\sigma$ is only possible if a correlation between the system and the catalyst is allowed. Note that this kind of result has been discussed independently in the context of the resource theory of entanglement \cite{Datta2022entanglement}.

\begin{figure*}
\includegraphics[width=0.7\paperwidth]{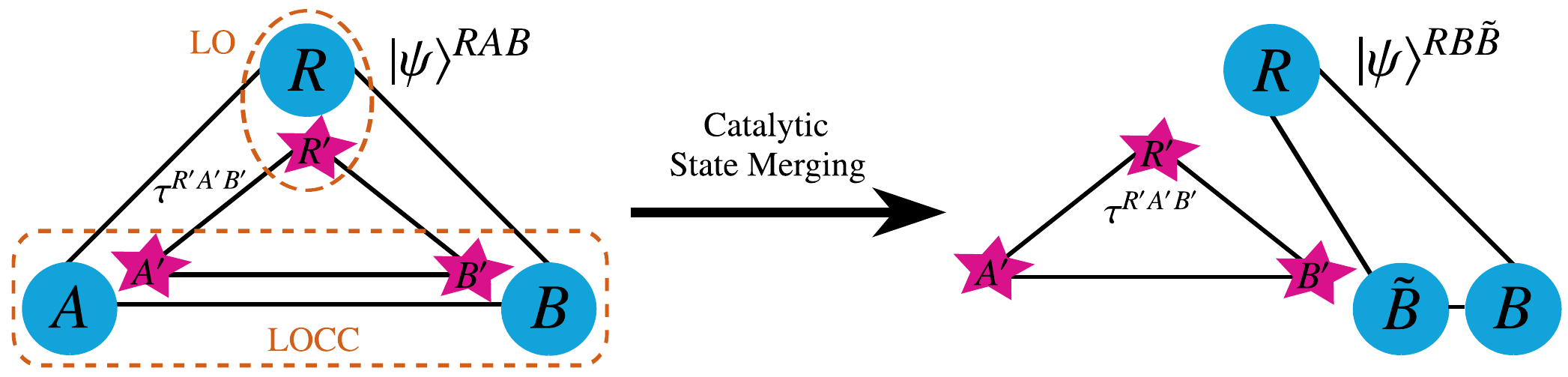}

\caption{\label{fig:CatalyticMerging} Catalytic quantum state merging. Alice, Bob, and a reference system share a single copy of $\ket{\psi}$. Alice aims to send her part of the state
to Bob by using LOCC and approximate catalysis with catalyst states $\tau$. Additionally, local operations (LO) are allowed on the reference system. The process is completely characterized by the
quantum conditional entropy $S(A|B)$. Figure is taken from \cite{Kondra2102.11136}.}

\end{figure*}

\section{Applications of quantum catalysis}
\label{sec:Applications}

\subsection{Quantum state merging}

Catalysis has proven useful for quantum state merging, an important protocol in quantum information theory. Before we review catalytic quantum state merging as discussed by~\citet{Kondra2102.11136}, we consider the original quantum state merging introduced in~\cite{Horodecki_2005,Horodecki_2006}. In the standard quantum state merging protocol, Alice ($A$), Bob ($B$), and a reference system ($R$) have access to many copies of a pure state $\ket{\psi}^{ABR}$. By performing LOCC operations between Alice and Bob, Alice aims to transfer her part of the state to Bob while preserving correlation with the reference system. For this, Alice and Bob can share additional singlets. In the asymptotic limit, where many copies of the overall state $\ket{\psi}^{ABR}$ are available, the minimal singlet rate for this procedure is given by the quantum conditional entropy\footnote{In Section~\ref{sec:Applications}, the von Neumann entropy is defined with a logarithm to the base 2, i.e., $S(\rho) = -\Tr[\rho \log_2 \rho]$.}~\cite{Horodecki_2005,Horodecki_2006}:
\begin{equation}
    S(A|B)=S(\psi^{AB})-S(\psi^B). \label{eq:ConditionalEntropy}
\end{equation}
Depending on the value of $S(A|B)$ two situations can arise. When $S(A|B)>0$, Alice and Bob need to share additional singlets at a rate $S(A|B)$ to accomplish the state merging protocol. However, for $S(A|B)\leq 0$, Alice and Bob can achieve state merging and gain extra singlets at a rate $|S(A|B)|$ to their disposal. 

Catalytic quantum state merging \cite{Kondra2102.11136} allows Alice and Bob to achieve merging for a single copy of the state $\ket{\psi}^{ABR}$ by using approximate catalysis, see Section~\ref{sec:ApproxCatalysis}. In this case, Alice, Bob, and the reference system share only one copy of the state $\ket{\psi}^{ABR}$. Additionally, they share a tripartite catalyst state $\tau^{A'B'R'}$, see also Fig.~\ref{fig:CatalyticMerging}. In contrast to standard quantum state merging, in the catalytic case local unitaries are allowed on the reference system. However, no communication between the reference system and the other parties is required~\cite{Kondra2102.11136}. Similar to standard quantum state merging, the performance of this procedure can be quantified via the conditional entropy given in Eq.~(\ref{eq:ConditionalEntropy}). For $S(A|B)>0$, Alice and Bob can achieve catalytic state merging if they additionally have access to a pure entangled state with entanglement entropy $S(A|B)$. For $S(A|B) \leq 0$ Alice can merge her part of the state with Bob without any additional entanglement. Apart from merging the state, Alice and Bob will gain a pure entangled state having entanglement entropy $|S(A|B)|$. The optimality of both scenarios is discussed in~\cite{Kondra2102.11136}.

The standard quantum state merging protocol has also been studied in the context of incoherent operations, known as incoherent quantum state merging \cite{Streltsov_2016}. In incoherent quantum state merging the main idea remains the same with some subtle changes. In comparison to LOCC in the standard state merging protocol \cite{Horodecki_2005,Horodecki_2006}, here, the only change is that Bob's operations are restricted to local incoherent operations. These kinds of operations between Alice and Bob are denoted by local quantum-incoherent operations and classical communications (LQICC)~\cite{Chitambar_2016,StreltsovPhysRevX.7.011024}. Furthermore, Alice and Bob can have access to additional singlets at a rate $E$ and only Bob has access to maximally coherent states at a rate $C$. Then the question is what are the achievable pairs of $(E,C)$ such that the merging is accomplished in the asymptotic limit? In fact, for a tripartite pure state $\ket{\psi}^{ABR}$, an achievable pair of $(E,C)$ should satisfy $E+C\geq S(A|B)_{\psi_D^{AB}}$, where $\psi_D^{AB}=\Delta(\psi^{AB})$ and $\psi^{AB}=\Tr_R\ket{\psi}\!\bra{\psi}^{ABR}$. As shown in \cite{Streltsov_2016}, the bound can be saturated with $(E_0,0)$, where $E_0=S(\psi_D^{AB})-S(\psi_D^B)$. Recently, this result has been translated to single copy scenarios with the help of approximate catalysis  \cite{char2021catalytic} and named as catalytic incoherent state merging. As a matter of fact, it has been shown that under catalytic LQICC incoherent state merging of a single copy of $\ket{\psi}^{ABR}$ is possible when Alice and Bob have access to an additional pure state of entanglement entropy $E_0$ \cite{char2021catalytic}. Note that no additional coherence is needed on Bob's side.  Furthermore, optimality has been shown by verifying that $E_0$ is the minimum amount of entanglement required to achieve the state merging protocol. Therefore, catalysis helps us translate the asymptotic state merging protocol in the single copy scenario.

\subsection{Quantum teleportation}

The role of approximate entanglement catalysis for quantum teleportation has been studied by~\citet{Lipka-Bartosik2102.11846}. Before presenting the result, we first briefly recapitulate the standard teleportation protocol \cite{BennettPhysRevLett.70.1895,Popescu_1994}. In general, there are two distant parties Alice and Bob sharing a quantum state $\rho^{AB}$. Additionally, Alice holds a particle $\tilde A$ in an unknown state $\ket{\psi}^{\tilde{A}}$. The goal of teleportation is to transfer Alice's state $\ket{\psi}^{\tilde{A}}$ to Bob using only LOCC. After performing an LOCC protocol $\Lambda$, the final state of Bob can be expressed as
\begin{equation}
    \eta^B=\Tr_{A\tilde{A}} \Lambda\left[\rho^{AB}\otimes\psi^{\tilde{A}}\right].
\end{equation}
The merit of the above protocol can be quantified by the teleportation fidelity \cite{Popescu_1994,Horodecki_1999} expressed as
\begin{equation}
    F_s(\rho_{AB})=\max_{\Lambda}\int d\psi \bra{\psi}\Tr_{A\tilde{A}} \Lambda\left[\rho^{AB}\otimes\psi^{\tilde{A}}\right]\ket{\psi},
\end{equation}
where the maximization is performed over all possible LOCC operations and the integral is taken over uniform distribution $d \psi$ with respect to all pure input states $\ket{\psi}$. The fidelity of teleportation can also be expressed as
\begin{equation}
    F_s(\rho^{AB})=\frac{f(\rho^{AB})d_{\tilde{A}}+1}{d_{\tilde{A}}+1},
\end{equation}
where $d_{\tilde{A}}$ is the dimension of the input state $\ket{\psi}^{\tilde{A}}$ and $f(\rho^{AB})=\max_{\Lambda}\bra{\Phi^+}\Lambda(\rho^{AB})\ket{\Phi^+}$ is the singlet fraction maximised over all LOCC protocols \cite{Horodecki_1999}. 

Recently, the standard teleportation protocol has been studied in a catalytic scenario to see whether there is an advantage in teleportation fidelity \cite{Lipka-Bartosik2102.11846}. In contrast to the standard teleportation protocol, here in the catalytic teleportation protocol Alice and Bob share an additional entangled state $\tau^{A'B'}$ such that it remains the same after the process. To be precise, the catalytic teleportation fidelity can be expressed as \cite{Lipka-Bartosik2102.11846} 
\begin{equation}
 F_c(\rho_{AB})=\max_{\Lambda,\tau^{A'B'}}\int d\psi \bra{\psi}\Tr_{A\tilde{A}A'B'} \Lambda \left[\rho^{AB}\otimes\tau^{A'B'}\otimes\psi^{\tilde{A}}\right]\ket{\psi}
\end{equation}
with the constraint $\Tr_{A\tilde{A}B} \Lambda[\rho^{AB}\otimes\tau^{A'B'}\otimes\psi^{\tilde{A}}]=\tau^{A'B'}$ which ensures that the catalyst remains unchanged in the process. Furthermore, in \cite{Lipka-Bartosik2102.11846} an achievable lower bound of $F_c(\rho^{AB})$ has been derived: 
\begin{equation}
    F_c(\rho^{AB})\geq\frac{f_r(\rho^{AB})d_{\tilde{A}}+1}{d_{\tilde{A}}+1}, 
\end{equation}
where $f_r(\rho^{AB})=\lim_{n\rightarrow\infty}f_n(\rho^{\otimes n})/n$ is the regularised version of the singlet fraction and $f_n(\sigma)=\max_\Lambda\sum_{i=1}^n
\bra{\Phi^+}\Tr_{\bar{i}}\Lambda(\sigma)\ket{\Phi^+}$, where $\Tr_{\bar{i}}$ represents partial trace over systems $1$ to $n$ except for $i$. As $f_r(\rho^{AB})\geq f(\rho^{AB})$ for all states $\rho^{AB}$, we have $F_c(\rho^{AB})\geq F_s(\rho^{AB})$ \cite{Lipka-Bartosik2102.11846}. In addition, when $\rho^{AB}$ is pure, \citet{Lipka-Bartosik2102.11846} showed that the catalytic teleportation outperforms the standard teleportation protocol for a large variety of generic quantum states. Furthermore, they studied the advantage when the dimension of the catalyst is small.

\subsection{Assisted distillation of entanglement and coherence}

Catalytic-assisted entanglement distillation has been introduced and studied by~\citet{Kondra2102.11136}. In the standard assisted entanglement distillation scenario, Alice, Bob, and Charlie share many copies of a tripartite state $\ket{\psi}^{ABC}$, and their aim is to extract singlets between Alice and Bob by performing LOCC \cite{DiVincenzo10.1007/3-540-49208-9_21,SmolinPhysRevA.72.052317}. In \cite{SmolinPhysRevA.72.052317}, the singlet rate has been given in the standard asymptotic scenario as $\min\{S(\psi^A),S(\psi^B)\}$. In catalytic assisted entanglement distillation~\cite{Kondra2102.11136} the parties have access to one copy of a state $\ket{\psi}^{ABC}$, and can make use of tripartite entangled catalysts. By using approximate catalysis as discussed in Section~\ref{sec:ApproxCatalysis}, a single copy of $\ket{\psi}^{ABC}$ can be converted into a pure entangled state shared by Alice and Bob with entanglement entropy given by $\min\{S(\psi^A),S(\psi^B)\}$. The optimality of this procedure and more details are discussed in~\cite{Kondra2102.11136}.

The role of catalysis has also been studied for assisted coherence distillation \cite{Chitambar_2016} in \cite{char2021catalytic}. In assisted coherence distillation Alice and Bob share many copies of a state $\rho^{AB}$ and their target is to maximize the coherence on Bob's side by performing LQICC \cite{Chitambar_2016}. The quantifier of the performance $C_d^{A|B}(\rho^{AB})$ represents the optimal rate at which maximally coherent states per copy of $\rho^{AB}$ can be obtained on Bob's side. If the initial state is pure then it has been shown that \cite{Chitambar_2016} 
\begin{equation}\label{distill_coherence}
    C_d^{A|B}(\ket{\psi^{AB}})=S(\Delta [\psi^B]).
\end{equation}
With the help of an additional system as a catalyst, the above protocol can be translated to the single copy scenario where a single copy of state $\ket{\psi}^{AB}$ is used to maximize coherence on Bob's side by performing  catalytic LQICC \cite{char2021catalytic}. Under catalytic LQICC, the optimal coherence achieved on Bob's side is $S(\Delta [\psi^B])$ \cite{char2021catalytic}, which is the same as the optimal rate at which maximally coherent states per copy of $\ket{\psi}^{AB}$ are obtained on Bob's side in the standard assisted coherence distillation protocol \cite{Chitambar_2016}. 

\subsection{Catalysis of noisy quantum channels}

The role of entanglement catalysis for quantum communication via noisy quantum channels has been studied by~\citet{Datta2022entanglement}. For this,~\citet{Datta2022entanglement} introduced and studied catalytic quantum capacity, capturing the ability of a quantum channel to transmit qubits in the presence of entangled catalysts. In the standard asymptotic setting, where the communicating parties have access to many copies of a quantum channel which can be used in parallel, the performance is quantified by quantum capacity. It represents the maximum rate at which qubits can be faithfully transmitted to a distant receiver via a channel $\Lambda$ \cite{LloydPhysRevA.55.1613,SchumacherPhysRevA.54.2629,Horodecki_2000}, and can be expressed as \cite{Shor_2002,Devetak1377491}
\begin{equation}
    Q(\Lambda)=\lim_{n\rightarrow\infty} \frac{1}{n}\max_{\rho_n}I(\rho_n,\Lambda^{\otimes n}).
\end{equation}
Here, $I$ represents coherent information and can be expressed as $I(\rho,\Lambda)=S(\Lambda[\rho])-S(\openone\otimes\Lambda[\ket{\psi_\rho}\!\bra{\psi_\rho}])$ \cite{SchumacherPhysRevA.54.2629,LloydPhysRevA.55.1613} and $\ket{\psi_\rho}$ represents some purification of $\rho$. 

If the communicating parties can use the quantum channel only once, they will typically not be able to transmit qubits faithfully through the channel. However, the situation changes if they have access to entangled catalysts~\cite{Datta2022entanglement}. The figure of merit in this setup is quantified via catalytic quantum capacity~\cite{Datta2022entanglement}
\begin{equation}
    Q_c(\Lambda)=\max\left\{m: \lim_{n\rightarrow\infty} \Vert\mu_n-\ket{\phi_{2^m}^+}\bra{\phi_{2^m}^+}\otimes \tau_n\Vert_1=0\right\},
\end{equation}
where $\{\mu_n\}$ is a sequence of states obtained by a single use of the noisy quantum channel $\Lambda$ and approximate catalysis~\cite{Datta2022entanglement}. Note that the catalytic quantum capacity is an integer, and can be zero even for channels which are not entanglement-breaking. A lower bound on the catalytic quantum capacity can be obtained as~\cite{Datta2022entanglement}
\begin{equation}
Q_{\mathrm{c}}(\Lambda)\geq\left\lfloor \sup_{\psi}E_{\mathrm{d}}\left(\openone\otimes\Lambda\left[\psi\right]\right)\right\rfloor,
\end{equation}
where $E_{\mathrm d}$ is the distillable entanglement \cite{BennettPhysRevA.53.2046,BennettPhysRevLett.76.722} and the supremum is taken over all bipartite pure states $\psi = \ket{\psi}\!\bra{\psi}$. For achieving this bound, Alice can prepare the two-particle state $\ket{\psi}$ locally, and send one half of it to Bob via the channel $\Lambda$. In this way, Alice and Bob end up sharing the state $\openone \otimes \Lambda[\psi]$, which can be converted into any pure state with entanglement entropy $E_{\mathrm d}(\openone \otimes \Lambda [\psi])$~\cite{Datta2022entanglement}. 

As distillable entanglement is lower bounded by $E_{\mathrm d}(\rho^{AB}) \geq S(\rho^A)-S(\rho^{AB})$ \cite{Devetak2005}, we can obtain an easily computable lower bound for the catalytic quantum capacity~\cite{Datta2022entanglement}:
\begin{equation}
Q_{\mathrm{c}}(\Lambda)\geq\left\lfloor \log_{2}d-S\left(\openone\otimes\Lambda\left[\phi_{d}^{+}\right]\right)\right\rfloor,
\end{equation}
where $d$ is the dimension of the Hilbert space on which $\Lambda$ is acting and $\ket{\phi^+_d} = \sum_{i=0}^{d-1}\ket{ii}/\sqrt{d}$. For a general quantum channel with Kraus operators $K_i$~\citet{Datta2022entanglement} further obtained the lower bound
\begin{equation}
    Q_{\mathrm c}(\Lambda)\geq \lfloor \log_2 d - H(p) \rfloor,
\end{equation}
where $H(p)$ is the Shannon entropy of the probability distribution 
\begin{equation}
p_{i}=\Tr\left[(\openone\otimes K_{i})\ket{\phi_{d}^{+}}\!\bra{\phi_{d}^{+}}(\openone\otimes K_{i}^{\dagger})\right].
\end{equation}
As an immediate result, we observe that a channel $\Lambda$ of dimension $d\geq 4$ has a catalytic capacity of at least one if it can be decomposed into at most two Kraus operators \cite{Datta2022entanglement}. An upper bound on the catalytic quantum capacity has also been obtained by \citet{Datta2022entanglement}:
\begin{eqnarray}
 Q_\mathrm{c}(\Lambda)\leq \Delta E_{\mathrm {sq}}(\Lambda), 
\end{eqnarray}
with
\begin{equation}
    \Delta E_\mathrm{sq} (\Lambda) = \sup_{\rho^{ABC}} \left\{ E^{A|BC}_\mathrm{sq}(\Lambda^C[\rho^{ABC}]) - E^{AC|B}_\mathrm{sq}(\rho^{ABC}) \right\}, \label{eq:TransmittedEntanglement}
\end{equation}
where $E_\mathrm{sq}$ is the squashed entanglement~\cite{Christandl_2004}. $\Delta E_\mathrm{sq} (\Lambda)$ can be interpreted as the amount of entanglement that the channel $\Lambda$ can transmit. Generally, computation of $\Delta E_\mathrm{sq}(\Lambda)$ is a formidable task. However, for certain types of channels an upper bound on it can be easily computed, e.g. for multiple copies of a single-qubit Pauli channel $\Lambda_p(\rho)=\sum_i p_i\sigma_i\rho\sigma_i$. In particular, the catalytic quantum capacity of $\Lambda_p^{\otimes n}$ is bounded as follows~\cite{Datta2022entanglement}:
\begin{equation}
\left\lfloor nE_{\mathrm{f}}(\openone\otimes\Lambda_{p}[\phi_{2}^{+}])\right\rfloor \geq Q_{\mathrm{c}}(\Lambda_{p}^{\otimes n})\geq\left\lfloor n-nH(p_{i})\right\rfloor,
\end{equation}
where $E_\mathrm{f}$ represents entanglement of formation \cite{BennettPhysRevA.54.3824}. The upper bound on $Q_\mathrm{c}$ can be easily evaluated noting that entanglement of formation can be computed for all two-qubit states \cite{WoottersPhysRevLett.80.2245,BennettPhysRevA.54.3824}.  

\begin{figure}
\includegraphics[width=0.9\columnwidth]{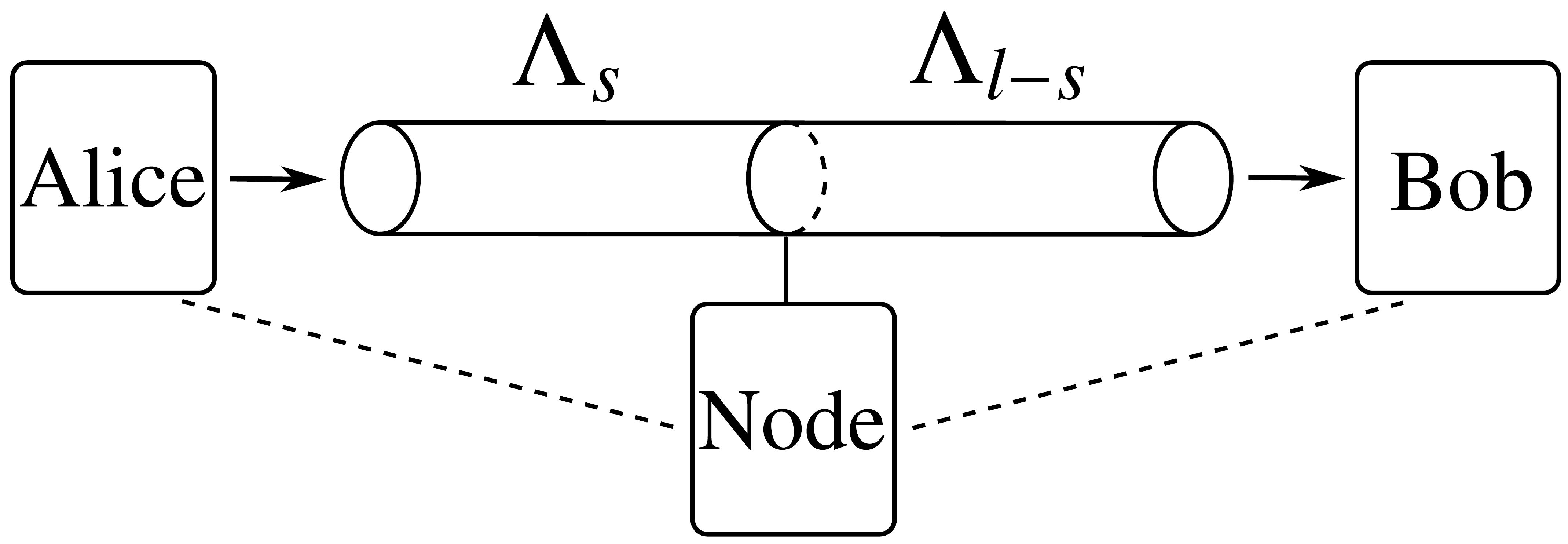}

\caption{A schematic of entanglement distribution between Alice and Bob via a depolarising qubit channel of length $l$. To assist the parties in the procedure an intermediate node is considered at a distance of $s$ from Alice. Note that the node can perform local quantum operations and communicate classically (dashed lines) with the parties. The figure is taken from \cite{Datta2022entanglement}.} \label{fig:EntanglementDistribution}

\end{figure}

\citet{Datta2022entanglement} also discussed the advantage of catalysis for entanglement distribution between two distance parties. Consider a scenario where Alice wants to distribute entanglement with Bob via a single qubit depolarising channel of the form \begin{equation}
    \Lambda_l(\rho)=e^{-\alpha l}\rho+ (1-e^{-\alpha l}) \frac{\openone}{2},
\end{equation}
where $l$ is the length of the channel and $\alpha\geq 0$ represents a damping parameter. The channel is entanglement breaking when $l\geq \ln 3/\alpha$. Suppose there is a node in between Alice and Bob to assist them in the entanglement distribution. More precisely, the node is allowed to perform any local operations and communicate classically with Alice and Bob, see also Fig.~\ref{fig:EntanglementDistribution}. Nevertheless, entanglement distribution is still not possible for $l\geq \ln 3/\alpha$ as shown in \cite{Datta2022entanglement}. However, if we have an entangled catalyst shared between Alice and the node then entanglement distribution between Alice and Bob is possible for $l<2 \ln 3/\alpha$ \cite{Datta2022entanglement}. To see this, let us assume that Alice prepares a maximally entangled state locally and sends one half of it to Bob through the channel. As a result, the state between the node and Alice becomes mixed, and to ensure that the resulting state between them is entangled the position ($s$) of the node should satisfy $s< \ln 3/\alpha$.  As all two-qubit entangled states are distillable \cite{HorodeckiPhysRevLett.78.574}, it is possible to create a pure entangled state between Alice and the node using the results in \cite{Kondra2102.11136}. After that, the particle at the node is sent to Bob via the channel, leading to an entangled state between Alice and Bob. Therefore, the introduction of an intermediate node and a suitable catalyst makes a useless depolarising channel for entanglement distribution useful.

\subsection{Authentication protocols based on catalysis}
 
Catalysis has shown to be useful in authentication protocols as well \cite{barnum1999quantum}. Suppose there are two parties Alice, Bob and Bob needs to authenticate himself to Alice. As we know there exist incomparable states $\ket{\psi}$ and $\ket{\phi}$ which can not be transformed into each other via LOCC. Let us say, Alice prepares locally a bipartite state $\ket{\psi}$ and sends one half of it to Bob. Additionally, Alice and Bob share a catalyst whose presence accomplishes the forbidden transformation $\ket{\psi}\nrightarrow\ket{\phi}$. Now they perform catalytic LOCC on the state $\ket{\psi}$, such that they achieved the desired final state. In the next step, Bob will send his half of the final state to Alice and Alice confirms the presence of Bob by performing a measurement in the basis $\ket{\phi}$. Suppose, there is some third-party intruder Charlie on the other side of Alice's communication channel. However, they will not be able to transform the state $\ket{\psi}$ to the desired final state $\ket{\phi}$ as Charlie does not possess the other half of the catalyst state. Note that the requirement of incomparable states is essential for this protocol. Otherwise, Alice and Charlie would have accomplished the desired transformation without the catalyst. For the error probability and security analysis we refer to \cite{barnum1999quantum}. 

\subsection{Catalytic decoupling}

Catalysis has also been useful in the decoupling of quantum information \cite{Majenz_2017}. In the standard decoupling technique, we want to approximately decouple $S$ and $E$ from a bipartite state $\rho^{SE}$ by erasing a part of $S$ \cite{Horodecki_2005,Horodecki_2006,Dupuis_2014}. Precisely, the bipartite state $\rho^{SE}$ is $\varepsilon$-decoupled, if there exists a unitary operation $U_S$ such that the following holds \cite{Majenz_2017} 
\begin{equation}
    \min_{\omega_{S_1}\otimes \omega_E} P\left(\Tr_{S_2} U_S\rho^{SE}U_S^\dagger, \omega_{S_1}\otimes \omega_E\right)\leq \varepsilon,
\end{equation}
with the purified distance $P(\rho,\sigma)=(1-F(\rho,\sigma))^{1/2}$ and $S=S_1S_2$. Therefore, we successfully decouple $S_1$ from $E$ at the price of losing $S_2$ in the process. The figure of merit is the minimum dimension of $S_2$ or to be precise $\log_2|S_2|$ to achieve the $\varepsilon$-decoupling and it is denoted by $R^\varepsilon(S;E)_{\rho}$. In fact, the minimum system size satisfies the following \cite{Majenz_2017}
\begin{equation}\label{decouple_bound}
    R^\varepsilon(S;E)_{\rho}\geq \frac{1}{2} I_{\max}^\varepsilon(E;S)_{\rho},
\end{equation}
where $I_{\max}^\varepsilon(E;S)_{\rho}$ represents smooth max-mutual information in the initial state $\rho^{SE}$. It is defined as
\begin{equation}
    I_{\max}^\varepsilon(E;S)_{\rho}=\min_{\tilde{\rho}}I_{\max}(E;S)_{\tilde{\rho}},
\end{equation}
where the minimization is taken over all bipartite states $\tilde{\rho}^{SE}$ such that $P(\rho^{SE},\tilde{\rho}^{SE})\leq\varepsilon$ and   
\begin{equation}
    I_{\max}(E;S)_{\tilde{\rho}}=\log_2\min_{\sigma^S}\{\Tr_{\sigma^S}|\sigma^S\otimes\tilde{\rho}^E\geq\tilde{\rho}^{SE}\}.
\end{equation}
However, the bound is not achievable for all states. On the contrary, if we allow catalytic decoupling the above bound can be achieved for any bipartite system \cite{Majenz_2017}. As a formal definition of catalytic decoupling, we say that a bipartite state $\rho^{SE}$ is $\varepsilon$-decoupled with the help of a catalyst in  state $\rho^{S'}$, if there exists a unitary operation $U_{\bar{S}}$ such that
\begin{equation}
    \min_{\omega^{S_1 S'}\otimes \omega^E} P\left(\Tr_{S_2} U_{\bar{S}} \rho^{\bar{S}E} U_{\bar{S}}^\dagger , \omega^{S_1S'}\otimes \omega^E\right)\leq \varepsilon,
\end{equation}
with $\bar{S}=SS'=S_1S_2S'$, and $\rho^{\bar{S}E}=\rho^{SE}\otimes\rho^{S'}$. Here $\omega^{S_1 S'}$ represents uncorrelated state of $S_1$ and $S'$. As mentioned in \cite{Majenz_2017}, the protocol is catalytic in the sense that the share of the ancillary state $\rho^{S'}$, which becomes part of the decoupled system $S_1S'$, remains uncorrelated with $S_1$ after the decoupling. However, note that other than the catalyst remaining decoupled from the system, no further restriction has been imposed on the catalyst state $\rho^{S'}$, such that it remains unaltered after the protocol. Therefore, it may not be catalytic in the sense that the state of the catalyst should remain unchanged. The minimum system size required in this scenario can be quantified as
\begin{equation}
    R^\varepsilon_c(S;E)_{\rho}=\min_{\mathcal{H}_{S'},\sigma^{S'}}R^\varepsilon(SS';E)_{\rho},
\end{equation}
where the minimum is taken over all possible Hilbert spaces $\mathcal{H}_{S'}$ and all states $\sigma^{S'}$ belong to the Hilbert space $\mathcal{H}_{S'}$. It is straightforward to check that $ R^\varepsilon_c(S;E)_{\rho}\geq  R^\varepsilon(S;E)_{\rho}$. With this \citet{Majenz_2017} showed that the bound in Eq. (\ref{decouple_bound}) is achievable as we have
\begin{equation}
    R^\varepsilon_c(S;E)_{\rho} \lesssim \frac{1}{2} I_{\max}^{\varepsilon-\delta}(E;S)_{\rho}, 
\end{equation}
where $1\geq\varepsilon\geq\delta>0$ and ``$\lesssim$'' is up to terms of the order $\mathcal{O}(\log_2(1/\varepsilon))$. Furthermore, they discuss some applications of catalytic decoupling. 

\subsection{Catalytic cooling}

\citet{Henao2021catalytic} explored the advantage of catalysts to cool a system. In \cite{Henao2021catalytic}, the authors considered a passive state $\rho_b$ as an initial state that is to be cooled. Note that a passive state cannot be cooled alone as its average energy $\langle H_b \rangle=\Tr (H_b \rho_b)$ cannot be lowered using any local unitary $U_b$, where $H_b$ is the corresponding Hamiltonian \cite{Allahverdyan_2004}. More precisely a state $\rho_b$ is called passive if and only if the change in average energy under any unitary operation $U_c$ is greater or equal to zero, i.e., $\Delta \langle H_c \rangle = \Tr [H_b(U_b\rho_b U_b^\dagger-\rho_b)]\geq 0$ \cite{Allahverdyan_2004}. As $\rho_b$ cannot be cooled alone, we need to attach some ancillary systems to it and perform a global unitary to cool the system. If we attach a hot object which is not in a passive state then cooling is possible. However, in \cite{Henao2021catalytic}, the authors considered a hot object in a passive state $\rho_r$ and used a catalyst in a non-passive state $\rho_c$ to cool the system in state $\rho_b$. Therefore, under a global unitary $U_{brc}$ the initial state $\rho=\rho_b\otimes\rho_r\otimes\rho_c$ transforms to $\rho'= U_{brc}\rho U_{brc}^\dagger$ and the process is catalytic if we have $\Tr_{br}\rho'=\rho_c$ additionally. To ensure that the system in state $\rho_b$ cools we should have $\Delta \langle H_b \rangle <0$, and in such a scenario, the transformation is called a catalytic cooling transformation \cite{Henao2021catalytic}. Equipped with this, \citet{Henao2021catalytic} showed that catalytic cooling is always possible with a sufficiently large dimensional catalyst state. More precisely, they showed the following: For a passive state $\rho_b\otimes\rho_r$ with $\rho_r\neq\openone/d_r$ ($d_r$ is the dimension of $\rho_r$), there exists a global unitary $U_{brc}$ and a suitable catalyst state $\rho_c$ such that catalytic cooling transformation is possible \cite{Henao2021catalytic}. Furthermore, they also showed that cooling of a qubit system can be maximised with a hot qubit and a three-dimensional catalyst \cite{Henao2021catalytic}.

\section{Universal catalysis}
\label{sec:UniversalCatalysis}

So far, for the results presented above,
the construction of the catalytic state for a non-trivial quantum transformation depends strongly on the choice of input and output states that are to be transformed.
A natural question arises then,
whether some states are more suitable than others to serve as a catalyst for a wider range of desired transformations.  
Perhaps one could even find a single quantum state that would be a catalyst for all allowed transformations within a particular resource theory?
As we will see,
such a \emph{universal catalyst} state can be readily constructed as long as the allowed catalytic transformations induce a continuous partial order on the compact state space.

In this section,
we will review the nascent theory of universal catalysis, 
when the catalyst state does not depend on the states undergoing the transition.
The question about the existence of a universal catalysis state in the context of quantum thermodynamics was raised,
and found an intriguing answer,
by~\citet{Lipka-Bartosik2006.16290}.
Soon afterwards, a similar question was posed by \citet{Kondra2102.11136} for the theory of pure state quantum entanglement,
and an answer was given in the subsequent work \cite{Datta2022entanglement}.
Here, we are going to focus our attention first on the results regarding quantum thermodynamics;
then we will proceed to the case of quantum entanglement,
leaving the general discussion about the implications of the universal catalysis to the final part of the section.

\subsection{Quantum thermodynamics}
\label{subsec:UniversalCatalysis:QauntumThermodynamics}
For a given pair $\rho_S, \sigma_S$, of states that are block-diagonal in the energy eigenbasis of a Hamiltonian $H_S$ of the system $S$,
let vectors $\bm{p} = \mathrm{diag}[\rho_{\rm S}]$ and $\bm{q} = \mathrm{diag}[\sigma_{\rm S}]$ denote their diagonal parts.
Let also $\{E_i\}$ be the eigenvalues of the Hamiltonian $H_S$.
We set
$\gamma_S = \mathrm{diag}[\bm{g}] = (g_1, g_2, \ldots, g_{d_S})$ be the system's thermal state,
diagonal in the same basis,
with $g_i = e^{-\beta E_i}/ Z_S$,
$Z_S = \Tr{ e^{-\beta H_S} }$.
Recall that, for a block-diagonal state in the energy eigenbasis $\rho_S$,
such that  $\bm{p} = \mathrm{diag}[\rho_{\rm S}]$, with Hamiltonian $H$
and in contact with the environment at inverse temperature $\beta$, the generalized free energies are defined as
\begin{equation}
    F_{\alpha}(\bm{p}, \bm{g}) = S_{\alpha}(\bm{p} || \bm{g}) -Z,
\end{equation}
where $S_{\alpha}(\bm{p} \, || \, \bm{g}) = \frac{\mathrm{sgn}(\alpha)}{1-\alpha}  \log \left(\sum_{i} p_{i}^{\alpha} g_{i}^{1-\alpha} \right)$ is the R\'enyi relative entropy \cite{muller2013quantum},
and $Z = \Tr e^{-\beta H}$ is the total partition function, compare also Eqs.~\eqref{eq:generalisedFreeEnergies} and \eqref{eq:RenyiRelEntropy}.

Now, the main result concerning the universal catalytic transformation in quantum thermodynamics can be stated as follows.
\begin{theorem}[See \cite{Lipka-Bartosik2006.16290}, Theorem~2]
\label{thm:lipkaPhysRevX_Thm2}
Let $\rho_{\rm S}$ and $\sigma_{\rm S}$ be two states with corresponding representations $\bm{p} = \emph{diag}[\rho_{\rm S}]$ and $\bm{q} = \emph{diag}[\sigma_{\rm S}]$ which satisfy:
\begin{align}
    \label{eq:seq_laws2}
    F_{\alpha}(\bm{p}, \bm{g}) > F_{\alpha}(\bm{q}, \bm{g}) \qquad \forall\, \alpha \geq 0,
\end{align}
Then, for any catalyst state $\omega_{\rm C}$ with  $\bm{c} = \emph{diag}[\omega_{\rm C}]$ and sufficiently large $n$ there exists a thermal operation $\mathcal{T}_{\rm SC}$ such that:
\begin{align}
\label{eq:AproxCatThermalUniv}
    \mathcal{T}_{\rm SC}\left[\rho_{\rm S} \otimes \omega_{\rm C}^{\otimes n} \right] = \sigma_{\rm SC}', 
\end{align}
and the errors on the system and the catalyst satisfy:
\begin{align}
    \label{eq:err_sys_gcat}
    \epsilon_{C} &:= || \Tr_{\rm S} \left[\sigma_{\rm SC}'\right] - \omega_{\rm C}^{\otimes n} ||_{1} \leq \mathcal{O}\left(e^{- n^{\kappa}}\right), \\
    \epsilon_{S} &:= || \Tr_{\rm C} \left[\sigma_{\rm SC}'\right] - \sigma_{\rm S} ||_{1} = 0,
\end{align}
where $\kappa \in (0, 1)$ can be chosen arbitrarily.
\end{theorem}

A remark regarding the expression ``any catalyst state'' is in order here.
In the proof of the above result, \citet{Lipka-Bartosik2006.16290} assume that the catalyst 
system of finite dimension $d_C$ is prepared in the state $\omega_C$ that is diagonal in the eigenbasis of the system's Hamiltonian $H_C$.
It is known that if a catalytic transformation between diagonal states
$\rho_S$ and $\sigma_S$ is admissible, it is also possible to transform
$\rho_S$ into $\sigma_S$ as in Eq.~\eqref{eq:AproxCatThermalUniv} with the help of a catalyst state that has a fully degenerate spectrum, compare the proof of Theorem~18, Supplementary Information in \cite{Brandao3275}. Hence, without loss of generality, 
we can assume that the Hamiltonian $H_C$ is trivial, $H_C \varpropto \mathbbm{1}$.
This, in turn, means that \emph{any} quantum state $\omega_C$ can be considered a viable catalyst state, since we require only that it be diagonal in the eigenbasis of $H_C$.
In this sense, any quantum state $\omega_C$,
as long as sufficiently many copies are available,
can act as a universal catalyst.

The family of equations given in Eq. \eqref{eq:seq_laws2} are referred to as ``second laws of quantum thermodynamics'' \cite{Brandao3275}, as they induce a partial order on the state space, which describes allowed catalytic transformations, see also Section~\ref{subsec:ExactCatalysis:Thermodynamics}.
It is important to notice here that if one relaxes the strict condition in Eq.~\eqref{eq:err_sys_gcat} on the amount of error accumulated by the transformation,
in a sense that the error is allowed to vanish at the rate slower than exponential in $n$,
the order will ``collapse'' and transformations between essentially all states become possible, as in  \cite{Lipka-Bartosik2006.16290}, Theorem 1.  We will discuss this phenomenon, known in general as \emph{catalytic embezzling} of quantum resources, thoroughly in the subsequent Section \ref{sec:CatalyticEmbezzling}.

An intuitive explanation of the above result is the following.
Because of the law of large numbers, for an arbitrary state $\omega_C$ that is diagonal in the energy eigenbasis,
as the number of copies of the state increases in the asymptotic regime ($n\rightarrow \infty)$,
a subset of eigenvalues of $\omega_{C}^{\otimes n}$
forms a typical set: almost all probability weight is concentrated in a set that is uniformly distributed.
Using a result from
\cite{Chubb_PhysRevA.99.032332}, see Theorem 2 therein,
\citet{Lipka-Bartosik2006.16290} claim that the state $\omega_C^{\otimes n}$
can be converted reversibly into multiple copies of yet another state, useful for catalytic transformation,
at a rate given approximately by relative entropy.
The exact quantification of the typicality property is the source of the error as in Eq.~\eqref{eq:err_sys_gcat}. 
The proof of Theorem~\ref{thm:lipkaPhysRevX_Thm2} utilises this kind of intuition  and focuses on specifying exactly the thermal operation $\mathcal{T}_{SC}$.
The actual catalytic transformation allows for a change of the catalyst.  To avoid embezzling of the quantum resource, 
care needs to be taken in analysing the magnitude of that change, as in Eq.~\eqref{eq:err_sys_gcat}.

\subsection{Pure state quantum entanglement}
\label{subsec:UniversalCatalysis:Entanglement}

As pointed out in Section~\ref{subsec:ApproximateCatalysisEntanglement}, for two bipartite pure states
$\ket{\psi}^{AB}$, $\ket{\phi}^{AB}$, 
an approximate catalytic conversion is possible if and only if
$S(\psi^{A}) \geq S(\phi^{A})$. This result was first proven in \cite{Kondra2102.11136}, where an explicit construction for the catalyst state was provided. The state of the catalyst given in \cite{Kondra2102.11136} depends strongly on the states $\ket{\psi}^{AB}$, $\ket{\phi}^{AB}$, 
and a natural question was raised in  \cite{Kondra2102.11136} 
whether there exists a catalyst state that can catalyze all allowed transformations, even if only to some arbitrary precision. \citet{Datta2022entanglement} give a positive answer to that question.

\begin{theorem}[See \cite{Datta2022entanglement}, Theorem~1]
\label{thm:UniversalCatalysisEntanglement}
Consider a bipartite Hilbert space of arbitrary but finite dimension. For every $\varepsilon>0$
there exists a universal catalyst state $\tau_{\varepsilon}^{A'B'}$ such that for every pair of pure states $\ket{\psi}^{AB}$ and $\ket{\phi}^{AB}$ with $S(\psi^A) \geq S(\phi^A)$ there is an LOCC protocol $\Lambda$ for which 
\begin{subequations} \label{eq:UniversalCatalysis}
\begin{align}
\label{eq:UniversalCatalysisUnchanged}
\mathrm{Tr}_{AB}\left[\Lambda\left(\psi^{AB}\otimes\tau_{\varepsilon}^{A'B'}\right)\right] & =\tau_{\varepsilon}^{A'B'}, \\
\label{eq:UniversalCatalysisEpsilon}
\left\Vert \Lambda\left(\psi^{AB}\otimes\tau_{\varepsilon}^{A'B'}\right)-\phi^{AB}\otimes\tau_{\varepsilon}^{A'B'}\right\Vert _{1} & <\varepsilon.
\end{align}
\end{subequations}
\end{theorem}

As we can see, contrary to Eq.~\eqref{eq:err_sys_gcat} in the previous section \ref{subsec:UniversalCatalysis:QauntumThermodynamics}, Eq.~\eqref{eq:UniversalCatalysisUnchanged} assures that the universal catalyst state $\tau_{\varepsilon}^{A'B'}$ remains unchanged as a result of the LOCC transformation $\Lambda$. It is generally understood that the requirement in Eq. \eqref{eq:UniversalCatalysisUnchanged} is essential if we want to talk meaningfully about universal catalysis for pure state entanglement. 
This is perhaps the most natural assumption to make if one wants to avoid 
``quite implausible and unphysical'' \cite{PhysRevX.8.041051}
effects like embezzling of entanglement (cf. Sec.~\ref{sec:CatalyticEmbezzling}).
The requirement for \textit{approximate} catalysis,
as in Sec.~\ref{sec:ApproxCatalysis} above,
is encoded in  Eq.~\eqref{eq:UniversalCatalysisEpsilon}.
We should notice here that the amount of error allowed during the transformation
does not depend on the choice of the input and output states.  
Hence, an arbitrary precision $\varepsilon>0$ can be achieved for all possible catalytic transformations using only one state $\tau_{\varepsilon}^{A'B'}$.
In this sense, one can speak of a universal catalytic state for pure-state entanglement.

The proof of Theorem \ref{thm:UniversalCatalysisEntanglement},
reported in~\cite{Datta2022entanglement},
makes use of the fact that the set of quantum states is compact. 
By discretizing appropriately the set of states, one can find a catalytic state for every pair of initial and final states from the finite collection of discrete ``points'' that allow a catalytic transformation as in Sec.~\ref{sec:ApproxCatalysis}.
Then the universal catalytic state is obtained by taking the tensor product 
of that finite number of catalysts.
By controlling how fine the discretization should be,
a bound for the error of the universal catalytic transformation is achieved as in Eq.~\eqref{eq:UniversalCatalysisEpsilon}.

\subsection{Comparison of universal catalysis of entanglement and quantum thermodynamics}
\label{subsec:UniversalCatalysis:Discusion}

The two constructions of a universal catalytic state given above are rather different.  
On the one hand, the construction in \cite{Lipka-Bartosik2006.16290} for quantum thermodynamics utilizes the typicality of a sufficiently large tensor power of any quantum state; on the other, the method employed in \cite{Datta2022entanglement} for the resource theory of pure state entanglement relies on the discretization of the compact state space.
Both approaches, however, share a common assumption about what catalytic transformations are in general possible with a particular resource theory.  That is the existence of a continuous function, or a family of functions, inducing a partial order of allowed transformations on the compact state space.
Additionally, both results are believed to be extendable to a broader context beyond their respective resource theories: compare Section III.F in \cite{Lipka-Bartosik2006.16290} and Discussion in  \cite{Datta2022entanglement}.
It is then of merit to ask at this point if both methods can be unified somehow and expanded to a general theory of universal catalytic states.
It seems like in both cases the universality of the construction is achieved by taking a state that in some sense contains all other catalysts, at least within an assumed margin of error.   Hence, the existence of a universal catalyst should be considered as another manifestation of the ``second laws'' governing the allowed catalytic transformation between resourceful states.

\section{Catalytic embezzling phenomena}
\label{sec:CatalyticEmbezzling}

In most parts of the previous discussion, we focused on catalytic transformations with the requirement that the state of the catalyst is left unchanged by the overall procedure. Here we will discuss phenomena which occur if a change of the catalyst state is allowed. 

One of the first results in this direction has been presented by~\citet{vanDamPhysRevA.67.060302}. 
The authors investigated exact entanglement catalysis, allowing the catalysts to change by an arbitrarily small amount, as measured by fidelity between the initial and final state of the catalyst system. In this setting, \citet{vanDamPhysRevA.67.060302} proved that there exists a family of pure state $\ket{\mu_n}$ such that for any bipartite state $\ket{\psi}^{AB}$ there is an LOCC protocol: $\ket{\mu_n} \rightarrow \ket{\psi}^{AB} \otimes \ket{\omega_n}$
with $| \langle \mu_n | \omega_n \rangle | > 1-\varepsilon$ and vanishing $\varepsilon >0$ as $n \rightarrow \infty$. Thus, if we consider approximate entanglement catalysis with Eq.~(\ref{eq:ApproximateCatalysis-1}) being the only constraint, the transformations obtained in this way allow to transform any bipartite pure state into any other bipartite pure state. In particular, for any $\varepsilon > 0$ and any two states $\ket{\psi}^{AB}$ and $\ket{\phi}^{AB}$ there exists a catalyst state $\tau$ and an LOCC protocol $\Lambda$ such that Eq.~(\ref{eq:ApproximateCatalysis-1}) is fulfilled. This applies also to the setting where the initial state $\ket{\psi}^{AB}$ is product, and the final state $\ket{\phi}^{AB}$ corresponds to $n$ singlets, where $n$ is an arbitrary integer. Thus, the procedure allows to extract an unbounded number of singlets from a shared catalyst, inducing only a small change in the quantum state of the catalyst. Due to this behavior, this phenomenon has also been termed ``embezzling entanglement''~\cite{vanDamPhysRevA.67.060302}. 

The result of \citet{vanDamPhysRevA.67.060302} indicates that perfect embezzlement with zero error is not possible in a finite-dimensional state. This result has recently been extended to infinite-dimensional systems, demonstrating that perfect embezzlement is also impossible in this case \cite{Cleve_2017}. Furthermore, in contrast to the tensor product framework, \citet{Cleve_2017} have demonstrated that perfect embezzlement is possible in a commuting operator framework.

\citet{vanDamPhysRevA.67.060302} also proved that universal entanglement embezzling is possible, which means that for any given $\varepsilon >0$ and given dimensions $d_A$ and $d_B$ there exists a catalyst achieving the conversion $\ket{\psi}^{AB} \rightarrow \ket{\phi}^{AB}$ for any initial and final state. Moreover, the conversion can be achieved by using local unitaries, and no communication between the parties is required. Additional families of states for universal entanglement embezzling have been presented by~\citet{LeungPhysRevA.90.042331}, and entanglement embezzling has also been used for an alternative proof for the quantum reverse Shannon theorem~\cite{Berta2011,Bennett6757002}. 

A detailed study of catalytic embezzling phenomena in quantum thermodynamics has been performed by~\citet{Brandao3275}. The authors introduced catalytic thermal operations (see also Sec.~\ref{subsec:ExactCatalysis:Thermodynamics}), and investigated different regimes for the change of the catalyst. For the case that the catalyst is returned unchanged, the transitions are characterized by the second laws as described in Sec.~\ref{subsec:ExactCatalysis:Thermodynamics}. The same holds true if the state of the catalyst changes in such a way, that the change can be corrected by investing a small amount of work~\cite{Brandao3275}. The third regime considered by~\citet{Brandao3275} requires that the state of the catalyst changes at most by $\varepsilon/\log d$ in trace norm, where $d$ is the dimension of the catalyst. In this case, it is shown that the transitions are completely characterized by the standard second law, under the assumption that the initial and the final states are diagonal in the energy eigenbasis. 

For the resource theory of purity, a careful investigation of embezzling phenomena has been performed by~\citet{Ng_2015}. The following condition must hold for the catalyst state, if the catalyst is to enable all state transformations via unital operations~\cite{Ng_2015}:
\begin{equation}
\frac{1}{2}\left\Vert \omega-\omega'\right\Vert _{1}\geq\frac{d_{S}-1}{1+(d_{S}-1)\frac{\log_{2}d_{C}}{\log_{2}d_{S}}}, \label{eq:PurityEmbezzling}
\end{equation}
where $d_S$ and $d_C$ are the dimensions of the system and the catalyst, respectively. In particular, for any system states $\rho$ and $\sigma$ there exist states of the catalyst $\omega$ and $\omega'$ fulfilling Eq.~(\ref{eq:PurityEmbezzling}) such that $\rho \otimes \omega$ can be transformed into $\sigma \otimes \omega'$ via an unital operation. Moreover, this bound is optimal: it is not possible to enable all state transformations with a catalyst violating Eq.~(\ref{eq:PurityEmbezzling}).

For general resource theory, an interesting result regarding the resource content of a catalyst in the case of embezzlement has recently been discussed in \cite{Rubboli2111.13356}. In fact, \citet{Rubboli2111.13356} demonstrated that the Theorem \ref{resource content catalyst} holds true even when the catalyst changes in the procedure, resulting in resource embezzlement. As a consequence, the results indicate that a highly resourceful catalyst is required to achieve small errors in state transformations.

In \cite{Leung2008}, catalytic embezzling phenomena have also been studied in the context of coherent state exchange. Suppose there are $m$ parties, and they want to transform a $m$-partite state $\ket{\psi}$ into another $m$-partite state $\ket{\phi}$ by a coherent process and without any communication, i.e., 
\begin{equation}
    \alpha \ket{0}^{\otimes m}\ket{\gamma}+\beta \ket{1}^{\otimes m} \ket{\psi} \rightarrow \alpha \ket{0}^{\otimes m}\ket{\gamma}+\beta \ket{1}^{\otimes m} \ket{\phi}. 
\end{equation}
Note that in the above process the coherence remains intact. In general, the above task cannot be always accomplished for $m\geq 2$ \cite{Leung2008}. However, with an additional catalyst state shared between $m$ parties and allowed to change slightly in the process, it has been shown that coherent state exchange is always possible for any number of parties \cite{Leung2008}. Furthermore, \citet{Leung2008} showed that universal embezzling is possible for any $m$. In another recent study, universal embezzling has been explored for entangled projection games \cite{Dinur2013}.

\section{Perspectives and open problems}
\label{sec:Perspectives}
While significant progress has been achieved in the investigation of quantum catalysis in recent years, some fundamental problems in this research area are still open. One such open problem concerns exact entanglement catalysis, where the catalyst is left uncorrelated with the primary system at the end. To the best of our knowledge it is not known whether an entangled catalyst in a mixed state can offer an advantage over pure-state catalysts. This question is open even for transformations between bipartite pure states. In fact, the trumping conditions \cite{Turgut_2007,Klimesh0709.3680} as well as the condition based on R\'enyi entropies \cite{Brandao3275} all assume that the catalyst is in a pure state. It is not known whether or not relaxing this condition will enrich the possible state transformations. Thus, a complete characterization of exact catalytic transformations between bipartite pure states remains a pressing open question. It is worth mentioning that the above question has been solved for approximate entanglement catalysis, as Theorem~\ref{thm:EntropyAppCata} does not require the catalyst state to be pure. Even more, it has been shown in~\cite{Datta2022entanglement} that a catalyst in a mixed state is required to achieve some transitions which are not achievable otherwise. In quantum thermodynamics, most results on catalytic state transformations are limited to specific families of states, e.g., states which are block diagonal in the energy eigenbasis. A complete characterization of catalytic transformations via thermal operations between general quantum states has so far remained open.

Another open question concerns the potential equivalence between asymptotic state transformations without catalysis and single-copy transformations with approximate catalysis. Isolated results demonstrating this equivalence have been presented recently, but a systematic understanding of this phenomenon is still lacking. For Gibbs-preserving operations the equivalence has been proven by \citet{shiraishi2020quantum} showing that for any pair of asymptotically reducible states the transformation can also be achieved via approximate catalysis and vice versa. The equivalence between asymptotic reducibility and approximate catalysis has also been proven for pure-state entanglement in the bipartite setting~\cite{Kondra2102.11136}. For general transformations between multipartite entangled states, it has been shown that asymptotic reducibility implies that the transition is also achievable on the single-copy level via approximate catalysis~\cite{Kondra2102.11136}. However, it remains unclear whether catalysis can offer an advantage over asymptotic setups (without catalysis) in any of these scenarios. This leads to an intriguing question of whether bound entangled states~\cite{HorodeckiPhysRevLett.80.5239} can be distilled into singlets with the aid of catalysis.

In the existing literature, entanglement catalysis has primarily been explored within the confines of bipartite, or tripartite frameworks. However, an emerging and fascinating avenue for research involves the study of entanglement in many-body systems~\cite{frerot2023probing,StreltsovPhysRevLett.125.080502,RevModPhys.80.517}. To effectively explore catalysis in these multipartite configurations, it becomes essential to move beyond the confines of the traditional LOCC framework and embrace more advanced paradigms~\cite{streltsov2023multipartite}. We hope that these problems will be addressed and solved in the near future, thus significantly improving our understanding of quantum catalysis and quantum systems in general. 

Many of the recent results discussed in this article make use of an analogy between asymptotic and catalytic setups, as developed by \citet{shiraishi2020quantum} for quantum thermodynamics. This technique allows to build catalytic protocols in quantum thermodynamics and other quantum resource theories, by resorting to the tools developed for the asymptotic setting. When it comes to asymptotic analysis of resource theories, an important line of research aims to build reversible resource theories from minimal assumptions, and provide quantifiers for the corresponding state transition rates. Indeed, it is believed that under few reasonable assumptions, the conversion rates between quantum states are characterized by the regularized relative entropy of the corresponding resource: \begin{equation}
    R(\rho\rightarrow\sigma)=\frac{E^{\infty}(\rho)}{E^{\infty}(\sigma)}. \label{eq:RateGeneral}
\end{equation}
Here,  $E^{\infty}(\rho)=\lim_{n\rightarrow\infty}E(\rho^{\otimes n})/n$, and $E(\rho)=\inf_{\sigma\in\mathcal{F}}S(\rho||\sigma)$ is the relative entropy of the resource, where $\mathcal F$ is the set of free states of the resource theory under study. 

While~\citet{brandao2015reversible} claimed that Eq.~(\ref{eq:RateGeneral}) holds for a general class of quantum resource theories, the proof of this result presented in \cite{brandao2015reversible} relies on the generalized quantum Stein's lemma~\cite{brandao2010generalization}, which has been recently called into question~\cite{berta2022gap}. It thus remains an important open problem to characterize all quantum resource theories for which Eq.~(\ref{eq:RateGeneral}) holds true. Resolving this problem is also important for the understanding of catalysis in general quantum resource theories. We also note that none of the results reviewed in this article are affected by the problem reported in \cite{brandao2010generalization}, more details are given in the Appendix. 

\section*{Acknowledgements}
We thank Roberto Rubboli for insightful comments on our manuscript. This work was supported by the ``Quantum Optical Technologies'' project, carried out within the International Research Agendas programme of the Foundation for Polish Science co-financed by the European Union under the European Regional Development Fund, the ``Quantum Coherence and Entanglement for Quantum Technology'' project, carried out within the First Team programme of the Foundation for Polish Science co-financed by the European Union under the European Regional Development Fund, and the National Science Centre, Poland, within the QuantERA II Programme (No 2021/03/Y/ST2/00178, acronym ExTRaQT) that has received funding from the European Union's Horizon 2020 research and innovation programme under Grant Agreement No 101017733. CD acknowledges support from the German Federal
Ministry of Education and Research (BMBF) within the funding program ``quantum technologies -- from basic research to market'' in the joint project QSolid (grant
number 13N16163).

\section*{Appendix}
Recently, \citet{berta2022gap} uncovered a gap in the proof of the main result of \cite{brandao2010generalization}, a generalisation of the quantum Stein's lemma.
This puts directly into question the validity of the framework for
general quantum resource theories, 
and the resource theory of quantum entanglement specifically,
presented in \cite{brandao2008entanglement, brandao2010generalization, brandao2010reversible, brandao2015reversible},
which tries to establish those theories as \emph{reversible}.
Briefly speaking, a reversible resource theory is such that it
admits a unique entropy-like quantifier of the resource,
which governs all possible asymptotic transformations between resourceful states \cite{ChitambarRevModPhys.91.025001}.
As of today, it is not known whether the generalised quantum Stein's lemma holds, and hence one has to resort to other particular results, proven independently, 
that assure reversibility of individual resource theories.
For example, 
it is true that the resource theory of coherence is reversible under a relevant class of asymptotically coherence-nonincreasing operations \cite{berta2022gap}.
We scrutinised all publications mentioned in this review and concluded that none of the results presented here are affected by the issue described in \cite{berta2022gap}.

\bibliography{literature}

\end{document}